\documentclass[preprint,12pt,3p]{elsarticle}

\pdfoutput=1 

\usepackage{amsmath}

\biboptions{sort&compress}

\usepackage{amssymb}
\usepackage{color}

\journal{Physics Reports}

\begin{document}

{\it\small
\noindent A corresponding version of this manuscript has been accepted for publication by Physics Reports.\\
Journal Reference: Phys.\ Rep.\ \textbf{554}, 1 (2015)\\
URL: http://www.sciencedirect.com/science/article/pii/S0370157314003871\\
DOI: 10.1016/j.physrep.2014.10.001
}
\vspace{.7cm}

\begin{frontmatter}


\title{Tuned, driven, and active soft matter}

\author{
Andreas M.\ Menzel
}

\address{
Institut f\"ur Theoretische Physik II: Weiche Materie, Heinrich-Heine-Universit\"at D\"usseldorf, Universit\"atsstra{\ss}e 1, D-40225 D\"usseldorf, Germany;\\ phone number: +49-211-81-12056\\[-.4cm]
}

\ead{menzel@thphy.uni-duesseldorf.de}

\begin{abstract}
One characteristic feature of soft matter systems is their strong response to external stimuli. As a consequence they are comparatively easily driven out of their ground state and out of equilibrium, which leads to many of their fascinating properties. Here, we review illustrative examples. This review is structured by an increasing distance from the equilibrium ground state. On each level, examples of increasing degree of complexity are considered. 
In detail, we first consider systems that are quasi-statically tuned or switched to a new state by applying external fields. These are common liquid crystals, liquid crystalline elastomers, or ferrogels and magnetic elastomers. 
Next, we concentrate on systems steadily driven from outside e.g.\ by an imposed flow field. In our case, we review the reaction of nematic liquid crystals, of bulk-filling periodically modulated structures such as block copolymers, and of localized vesicular objects to an imposed shear flow. 
Finally, we focus on systems that are ``active'' and ``self-driven''. Here our range spans from idealized self-propelled point particles, via sterically interacting particles like granular hoppers, via microswimmers such as self-phoretically driven artificial Janus particles or biological microorganisms, via deformable self-propelled particles like droplets, up to the collective behavior of insects, fish, and birds. 
As we emphasize, similarities emerge in the features and behavior of systems that at first glance may not necessarily appear related. We thus hope that our overview will further stimulate the search for basic unifying principles underlying the physics of these soft materials out of their equilibrium ground state. 
\end{abstract}


\end{frontmatter}


\newpage

\tableofcontents

\newpage


\section{Introduction}

It is easy to motivate the study of soft matter systems. Everyday life provides us with the most immediate examples. When we go shopping, we carry home our groceries in plastic bags made from polymers. Most flat screen displays exploit the switching behavior of liquid crystals in an electric field. We use solutions of surfactants when we soap ourselves while taking a shower. And finally, a major part of ourselves is actually soft matter, as almost all biological cells are. Considering the shear amount of soft matter distributed commercially, research activities in this field naturally extend far beyond purely academic purposes. 

It is much less straightforward to give a definite, unambiguous, and precise definition of what the term ``soft matter'' actually means or which states of matter it actually includes. Maybe one should proceed the other way around and explain which materials it \textit{ex}cludes by comparison to common other states of matter. 

On the one hand, we interpret the term ``softness'' as a strong response to a comparatively weak force \cite{doi2013soft}. This becomes most illustrative when we compare the elastic behavior of metal solids and ordinary rubbery materials. The Young modulus of a rubber can for example be ten thousand times lower than that of steel \cite{treloar1975physics}. I.e.\ applying a tensile force to that piece of rubber can stretch it about ten thousand times further than when applying it to a corresponding piece of steel. In general, due to these huge responses to external forces, the observed behavior of soft materials is markedly nonlinear \cite{doi2013soft}. 

On the other hand, the response of soft matter systems to external stimuli is generally comparatively slow \cite{doi2013soft}. The relaxation times in polymers can be hours or days (or even more) \cite{strobl1997physics}, whereas it takes nanoseconds in simple liquids \cite{hansen2006theory}. 
It is therefore easy to drive and maintain the systems out of equilibrium, and non-equilibrium effects play a central role in the study of soft materials \cite{doi2013soft}. 

\begin{figure}
\centerline{\includegraphics[width=\textwidth]{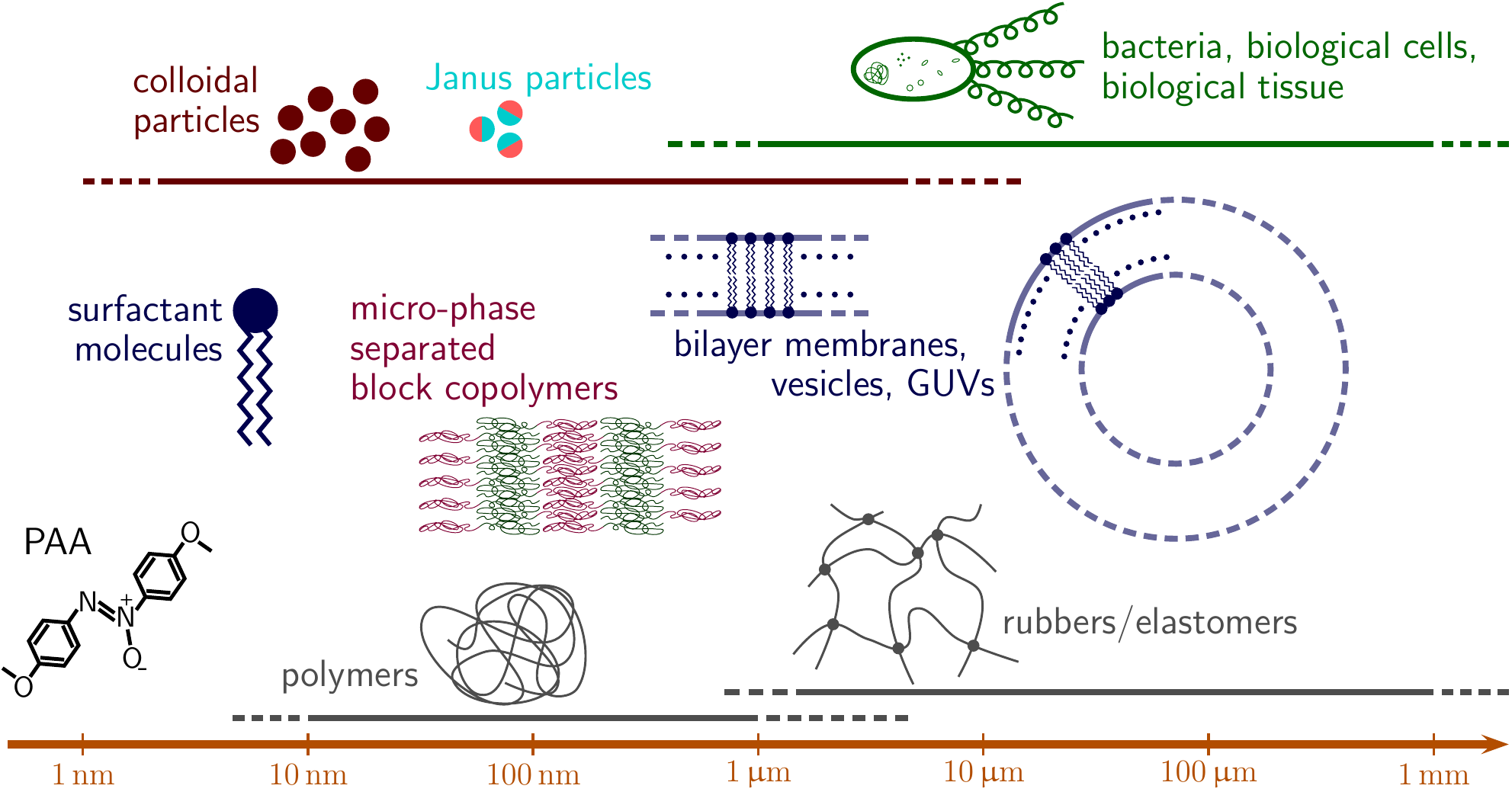}}
\caption{Schematic illustration of typical examples of soft-matter building blocks and materials, together with a rough estimate of their characteristic sizes (horizontal bars indicate a broader range of possible occurring length scales). The rod-like molecule PAA is a common representative of low-molecular-weight liquid crystals \cite{degennes1993physics} that can show an orientationally ordered anisotropic nematic phase. Surfactant molecules combine two antagonistic features on one building block: their head is usually hydrophilic, whereas their tail(s) is (are) mostly hydrophobic \cite{witten2004structured}. This leads, for instance, to the formation of closed bilayer membranes in the form of vesicles \cite{jones2002soft}, with giant unilamellar vesicles (GUVs) as an extreme example (in reality, unlike the schematic, the layer thickness is orders of magnitude smaller than the radius) \cite{hub1982preparation,mueller1983formation,sakuma2011model}. Polymers \cite{strobl1997physics, doi2007theory} in the simplest case can be treated as flexible linear chain-like objects. When covalently chemically crosslinked in the form of rubbers and elastomers (crosslinks are indicated by dots) a macromolecule of macroscopic dimension can be obtained \cite{treloar1975physics,strobl1997physics}. Block copolymers feature chemically different parts (marked by different colors) that can micro-phase separate into regularly ordered structures \cite{hamley1998physics}, here indicated for a lamellar texture. In colloids, the constituting particles are stabilized to form a permanent dispersion \cite{dhont1996introduction,palberg1999crystallization,ivlev2012complex}. Again, different sides of the particles can be functionalized with different chemical properties so that the resulting Janus particles \cite{walther2008janus} form clusters or even regular lattices \cite{chen2012janus,jiang2010janus}. Finally, biological cells such as motile bacteria \cite{harshey2003bacterial,darnton2007torque,kearns2010field} combine a multitude of functionalities. Collective arrangements of biological cells in the form of biological tissue reaches macroscopic dimensions.}
\label{fig_lengths}
\end{figure}

One reason for these properties is the size of the building blocks \cite{doi2013soft}. These sizes typically fall into the range of several nanometers up to about a hundred micrometers. 
Some examples together with a rough classification of the possible associated dimensions are schematically depicted in Fig.~\ref{fig_lengths}. 
Clearly, small rod-like liquid crystalline molecules like PAA with a length of about two nanometers 
are on the lower end 
\cite{degennes1993physics}. Colloids, i.e.\ mesoscopic particles stabilized in a permanent dispersion, cover a broad range of the considered sizes by definition \cite{dhont1996introduction,palberg1999crystallization,ivlev2012complex}. 
On the contrary, a rubber made of sufficiently chemically crosslinked polymer chains can actually be considered as one giant molecule and reach macroscopic dimensions \cite{strobl1997physics}. At this point a strict definition via the size of the constituents becomes more involved. 

Nevertheless, we can often say that soft matter building blocks are large enough so that quantum effects can be neglected, but small enough that thermal fluctuations play a significant role \cite{hamley2007introduction, jones2002soft}. For example, most of the elastic behavior of rubbers must be attributed to entropy, i.e.\ thermal fluctuations \cite{strobl1997physics, doi2007theory}. Entropy also supports the stabilization of colloidal dispersions as a complement to the necessary artificial stabilization mechanisms \cite{dhont1996introduction,ivlev2012complex}. 

Apart from that, the size of the building blocks allows us to use coarse-grained descriptions to characterize their collective physical behavior \cite{jones2002soft}. Chemical details can often be neglected, which leads to a certain degree of ``universality'' in the characterization. This point of view got particularly famous in the field of polymer physics \cite{degennes1979scaling}. 

Finally, the size of the constituents makes it possible to arrange different or even antagonistic properties on one building block. Some examples are depicted in Fig.~\ref{fig_lengths}. These are block copolymers, where chemically different connected parts of the molecules tend to micro-phase separate into 
regular spatial textures \cite{hamley1998physics}; surfactant molecules composed of hydrophobic (``water-fearing'') and hydrophilic (``water-loving'') parts that self-assemble for instance into 
bilayer membranes and vesicles in an aqueous environment \cite{jones2002soft,witten2004structured}; and colloidal Janus particles \cite{walther2008janus} that can form finite clusters or regular lattices \cite{chen2012janus,jiang2010janus}. 

With all these facts in mind, it is not surprising that the rheological behavior of such systems is relatively complex when compared to simple liquids \cite{larson1999structure}. Many soft matter systems are therefore also classified as ``complex fluids''. Sometimes the terms ``soft matter'' and ``complex fluids'' are even used interchangeably \cite{degennes1992soft}. 

In the following, we will concentrate on soft matter systems out of their equilibrium ground state. It is impossible to cover all aspects and areas of this subfield, which was expanding so rapidly over the past decades. Instead, we will mainly focus on typical representatives in each part and embed these examples into the broader context. Roughly, we will increase the degree of activation that leads to non-equilibrium states through the subsequent sections. 

First, we consider the switching or tuning of the states of soft matter systems by external fields. For example, as noted above, in display devices the orientation of the optical axis in liquid crystals is switched by external electric fields. Through a static external field, the system is forced out of its initial equilibrium ground state. However, it can reach a new static equilibrium state in the presence of the external field. \label{mod_intro} We will mainly concentrate on liquid crystalline as well as superparamagnetic or ferromagnetic gels and elastomers, the optical and mechanical properties of which can be tuned by external fields. They can serve as soft actuators, and even their possible role as ``artificial muscles'' was pointed out. 

Next, we focus on systems permanently driven out of equilibrium by steady external shear flows. A brief account of the rheological behavior of liquid crystals and block copolymers is included. Coming back to the introductory cases mentioned above, an example situation from everyday life would be soap or body lotion while we are spreading it on our skin. Related to this topic, we address the progress made on the dynamics of vesicles when driven by an external shear flow. 

The broadest section is devoted to the recently exploding field of ``active'' soft matter systems. We understand the word ``active'' in the sense that individual building blocks have a sort of propulsion mechanism, which leads to motion. Crawling cells or swimming bacteria can run this mechanism for instance by consuming fuel or food, whereas colloidal particles can self-propel via self-phoretic effects. In general, a key point is that the direction of motion is not prescribed from outside. These ingredients lead to new and fascinating properties of the individual constituents and when they collectively act together. There is a significant overlap between the fields of soft matter and biological physics in this area. 

Finally, 
a short summary is appended. We conclude by giving a brief impression of the current state of the field and its evolution.

\newpage
\section{Tuned soft matter}

By ``tuning'' we understand an adjustment of the material properties to a state that serves the current purpose or convenience. The crucial point with the soft matter systems introduced below comes with the non-invasive way of this adjustment. An external field is applied to the closed cell or piece of material to achieve the new setting. 
In general, a static external field can force the system to acquire a new static equilibrium state in the presence of the field. If the external field is increased or decreased in small steps, after each adjustment the system finds itself in a weakly non-equilibrium situation. It has to find the new equilibrium state in the presence of the switched external field. Frequently, this process can be described by simple relaxation dynamics following the minimum of the modified free energy landscape. If the external field is adjusted slowly enough, a quasi-static switching can often be achieved, avoiding genuinely non-equilibrium situations.  
If our focus is on the process of conversion of external field energy to reach a new state of the system, we may speak of ``actuation''.

\subsection{Low-molecular-weight liquid crystals}\label{lmwlc}

As a starting point, let us briefly recall the well-known Fr\'{e}edericksz transition in nematic liquid crystals \cite{degennes1993physics}. This transition can be induced from outside, for instance by applying an external electric field. As will be explained in more detail below, it involves a collective reorientation on the molecular level, changing the optical birefringence properties of a sample cell. In this way, the light transmittance of the cell can be reversibly tuned from outside. 

\begin{figure}[t]
\centerline{\includegraphics[width=\textwidth]{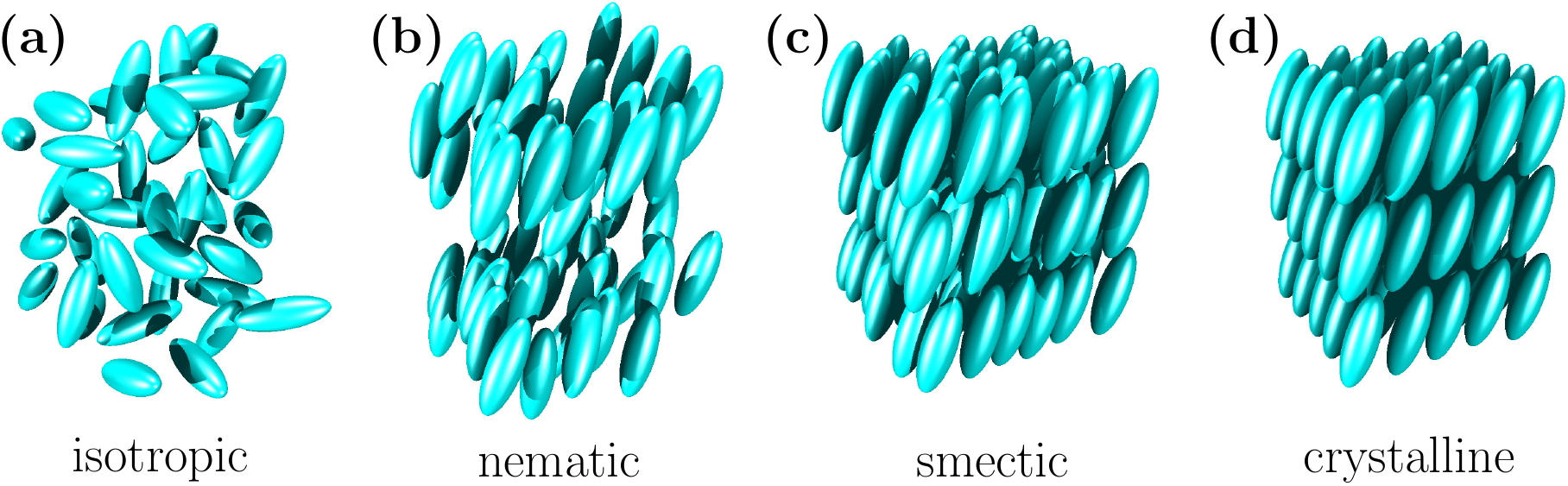}}
\caption{Schematic illustration of typical phases of common low-molecular-weight liquid crystals consisting of elongated rod-like molecules. (a) The disordered isotropic liquid-like state shows neither long-ranged orientational nor long-ranged positional order. (b) In the nematic phase, there is no long-ranged positional order. However, apart from thermal fluctuations, the long axes of the molecules on average orient along a common direction called the director. (c) Smectic phases are characterized by a layer-like positional arrangement of the molecules, while in the depicted smectic-C state there is no long-ranged positional order within each layer. (d) Finally, crystalline solids show the highest degree of order with long-ranged positional order in the form of stacking of the molecules on regular lattice sites and long-ranged orientational order. Transitions between the depicted phases can be induced in corresponding liquid crystalline materials for example by temperature changes.}
\label{fig_lc-phases}
\end{figure}

In the most illustrative case, the liquid crystalline building blocks are small rod-like organic molecules. The molecular structure of one characteristic example, namely PAA, was depicted in Fig.~\ref{fig_lengths}. It shows the typical dimensions, which in this case are about two nanometers in length and about half a nanometer in thickness \cite{degennes1993physics}. 
At high enough temperatures, the material is in an isotropic disordered liquid state as schematically illustrated in Fig.~\ref{fig_lc-phases}~(a). 
The system does not feature any long-ranged positional nor orientational order of the constituents. It behaves in an isotropic way. In contrast to that, we find both positional and orientational order at low enough temperatures when the material forms a crystal. Then the molecules are stacked on regular lattice sites and orient their long axis along a common direction, see Fig.~\ref{fig_lc-phases}~(d). Between these two extremes of a disordered liquid and a crystalline solid, liquid crystals feature further intermediate phases, the so-called mesophases. These liquid crystalline phases show a higher degree of order (corresponding to a lower degree of symmetry) than the fluid but a lower degree of order (corresponding to a higher degree of symmetry) than the crystal. For example, in the nematic state, the one that we will refer to in the following, there is no long-ranged positional order. However, on average, the molecules orient along a common direction, see Fig.~\ref{fig_lc-phases}~(b). In this way, the material becomes anisotropic. Smectic states, which we will come back to later in this review, feature a regular stacking in layers, see Fig.~\ref{fig_lc-phases}~(c). More precisely, they show a quasi-long-ranged positional order in one spatial dimension.

When the system is in the nematic phase, the building blocks tend to collectively order their long axes along a common direction. This average orientation is called the nematic director $\mathbf{\hat{n}}$ \cite{degennes1993physics}. Genuinely nematic phases are non-polar, that is the directions $+\mathbf{\hat{n}}$ and $-\mathbf{\hat{n}}$ cannot be distinguished. The degree of orientational order is measured by a scalar order parameter 
\begin{equation}\label{eq:s}
s = \frac{1}{2}\left\langle(3\cos^2\vartheta-1)\right\rangle,
\end{equation}
where the average $\langle...\rangle$ is taken over all molecules in the sample, and $\vartheta$ for each individual molecule measures the angle between its long axis and the average orientation $\mathbf{\hat{n}}$. We obtain $s=0$ in the disordered isotropic liquid state, and $s=1$ in a perfectly ordered nematic state. A combined symmetric traceless order parameter tensor $\mathbf{S}$ can be introduced, which contains both information, the degree of orientational order and the average orientation $\mathbf{\hat{n}}$,   
\begin{equation}\label{eq:nematicOPtensor}
\mathbf{S} = s\left(\mathbf{\hat{n}}\mathbf{\hat{n}}-\frac{1}{d}\mathbf{I}\right).
\end{equation}
Here, $\mathbf{\hat{n}}\mathbf{\hat{n}}$ denotes a dyadic product, $d$ is the dimension of space (usually $d=3$), and $\mathbf{I}$ represents the unity matrix\footnote{Also another notation is common, in which the letters $S$ and $\mathbf{Q}$ are used instead of $s$ and $\mathbf{S}$, respectively.}. 

On the one hand, since $\mathbf{\hat{n}}$ corresponds to the average orientation of the ordered molecules, it is an axis of anisotropy. For example, the dielectric constants $\epsilon_{\|}$ and $\epsilon_{\bot}$ measured parallel and perpendicular to $\mathbf{\hat{n}}$, respectively, generally differ from each other. Let us consider the case in which the dielectric anisotropy $\epsilon_a=\epsilon_{\|}-\epsilon_{\bot}$ is positive, $\epsilon_a>0$. Then a static external electric field $\mathbf{E}$ tends to orient the director $\mathbf{\hat{n}}$ along itself. This tendency is reflected by a corresponding contribution to the effective thermodynamic potential
\begin{equation}\label{F:elec}
F_{elec} = - \frac{1}{2}\epsilon_a(\mathbf{\hat{n}}\cdot\mathbf{E})^2.
\end{equation}
In this expression, possible constant prefactors due to the system of measure are absorbed into the dielectric anisotropy $\epsilon_a$. 

\begin{figure}[t]
\centerline{\includegraphics[width=\textwidth]{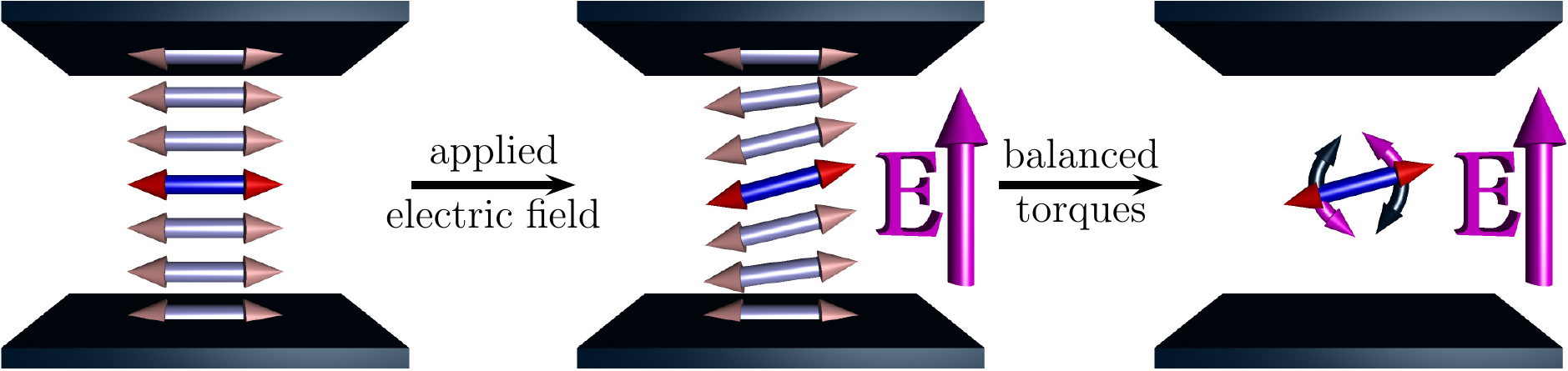}}
\caption{Reorientation of the nematic director induced by an external electric field in a thin cell of prescribed boundary conditions. The nematic director is indicated by the double-headed arrow. At the upper and lower confining plate it is rigidly anchored horizontally. On the one hand, if the sample is thin enough, this leads to a homogeneous horizontal alignment of the director to avoid spatial distortions in the director orientations (left). On the other hand, for positive dielectric anisotropy of the nematic molecules, a vertically applied external electric field $\mathbf{E}$ tends to orient the nematic molecules and the director in vertical direction (center). Both effects -- horizontal alignment due to the boundary conditions and vertical alignment through the external electric field -- compete with each other. Balancing the two resulting torques leads to a finite angle of director reorientation with respect to the horizontal direction (right) above a non-vanishing critical threshold amplitude of the electric field.}
\label{fig_lc_frederiks}
\end{figure}

On the other hand, spatial variations of the director, i.e.\ spatial distortions, increase the thermodynamic potential and are therefore suppressed. One way of quantifying this thermodynamic penalty is by contracting the tensor $\nabla\mathbf{\hat{n}}$,
\begin{equation}\label{F:dist}
F_{dist} =  \frac{1}{2}K\,(\nabla\mathbf{\hat{n}})\!:\!(\nabla\mathbf{\hat{n}}),
\end{equation}
with $K>0$ another material-dependent parameter\footnote{We here use a simplified version of this thermodynamic expression. Due to the anisotropy of the system, the coefficient $K$ in general is replaced by a fourth-rank tensor \cite{degennes1993physics}.}. As an example, the nematic liquid crystal can be confined in a thin cell between two parallel plates as depicted in Fig.~\ref{fig_lc_frederiks}. 
The surfaces of the cell can be prepared such that they anchor the orientation of the director, for instance in a parallel way, as indicated on the left-hand side of Fig.~\ref{fig_lc_frederiks}. This, via Eq.~(\ref{F:dist}), sets a uniform director orientation in the whole cell. 

If now an external electric field is switched on with an orientation perpendicular to the cell surfaces, see the center of Fig.~\ref{fig_lc_frederiks}, it tends to reorient the director and align it parallel to itself as given by Eq.~(\ref{F:elec}). However, via Eq.~(\ref{F:dist}), such a reorientation induces thermodynamically penalized distortions because of the surface anchoring on the plates. The competition between the two effects leads to a critical electric field amplitude that has to be exceeded for director reorientations to become possible. Above this threshold field amplitude, director reorientations are induced. From Eqs.~(\ref{F:elec}) and (\ref{F:dist}) this critical field amplitude follows directly as
\begin{equation}
E_c=\frac{\pi}{d}\sqrt{\frac{K}{\epsilon_a}},
\end{equation}
where $d$ is the thickness of the cell and the impact of possible electric currents has been neglected. In the new static reoriented state, the two effects of the reorienting electric field and the counteracting surface anchoring balance each other. This is indicated on the right-hand side of Fig.~\ref{fig_lc_frederiks}. When the electric field is switched off, the director rotates back into its initial state due to the surface anchoring. 

In this way, the anisotropy properties of the presented sample cell can be reversibly tuned from outside. Naturally, this anisotropy in a real sample is connected to optical birefringence. Thus the optical transmission properties of the cell can be reversibly switched and adjusted by the external electric field. 
This is the underlying principle extensively used to construct optical display devices that are based on liquid crystals \cite{kawamoto2002history}. However, the technical realizations today involve much more complicated geometries than the simple Fr\'{e}edericksz cell outlined above.

\subsection{Liquid crystalline elastomers}

Commonly used liquid crystals have relatively low molecular weight and are classified as complex fluids. As suggested by the name, they typically flow away when not maintained by a containing cell. Another promising material would be one that is stable and self-standing without supporting container walls, but still features the optical transmission properties of common liquid crystals. Such materials were actually realized by stepping towards higher molecular weights in the form of liquid crystalline polymers and elastomers. 

In general, polymers are synthesized by covalently binding together from hundreds up to millions identical molecular repeat units, the so-called monomers. In this way, new huge macromolecules are formed \cite{strobl1997physics}. Different chemical routes for this procedure are available. The number of monomers on the resulting polymeric molecule defines the degree of polymerization. In the simplest case, long linear chain-like molecules are obtained, but several further molecular architectures can be realized. 

\begin{figure}
\centerline{\includegraphics[width=4.5cm]{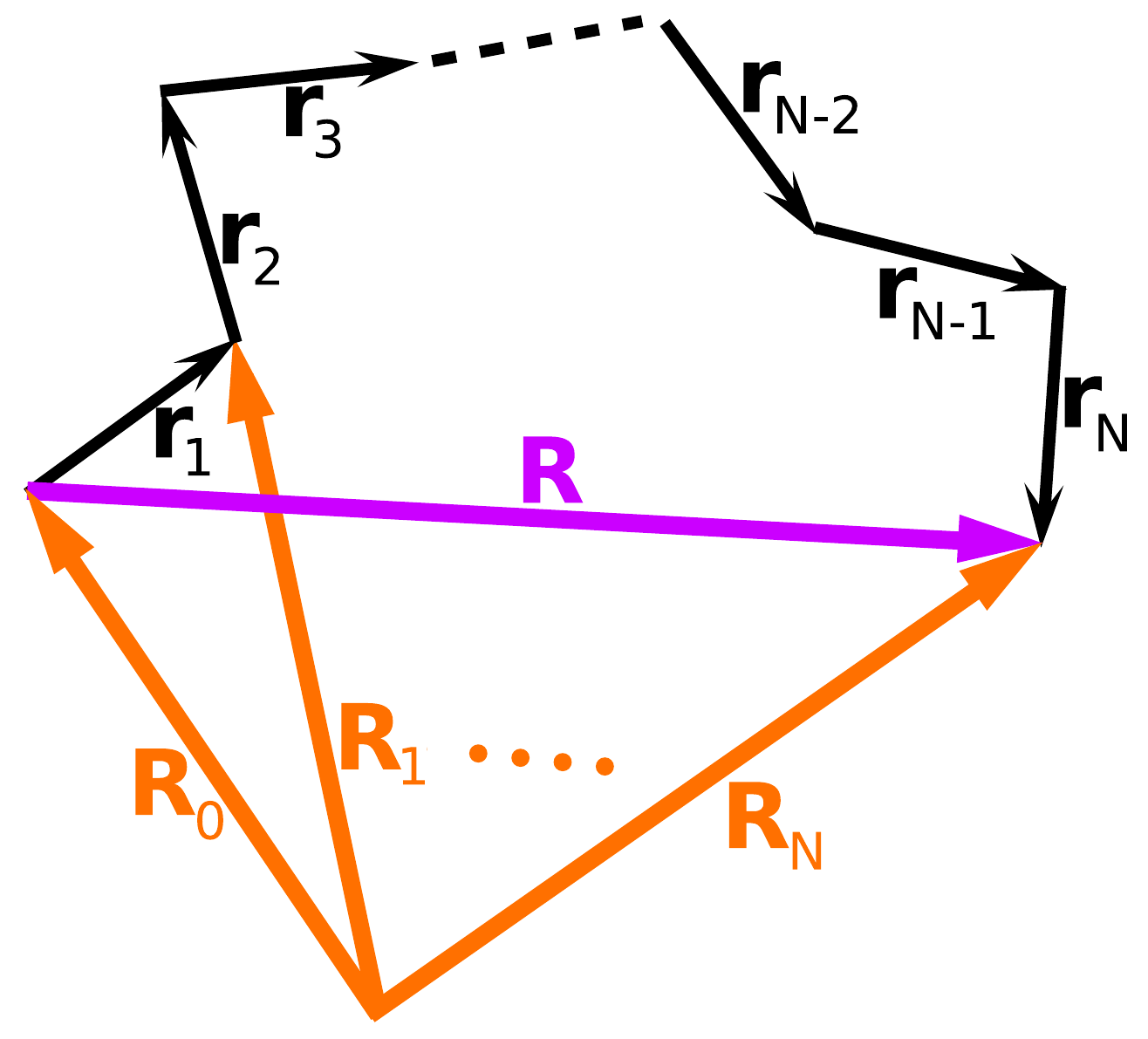}}
\caption{
Illustration of the freely-jointed chain model for a linear polymer chain. The chain is represented by $N$ freely-jointed segments of end-to-end vectors $\mathbf{r}_n=\mathbf{R}_n-\mathbf{R}_{n-1}$ ($n=1,...,N$), where $\mathbf{R}_n$ ($n=1,...,N-1$) give the positions of the joints. All segments are of identical length, $\|\mathbf{r}_n\|=b$ for all $n$, and their orientations are completely independent of each other. $\mathbf{R}=\mathbf{R}_N-\mathbf{R}_0$ is the end-to-end vector of the whole polymer chain.  
}
\label{fig_freely-jointed-chain}
\end{figure}

The simplest theoretical model describing linear polymer chains is the freely-jointed chain model \cite{strobl1997physics}. It already gives us enough qualitative insight to understand the discussion below. In this model, as illustrated in Fig.~\ref{fig_freely-jointed-chain}, the polymeric molecule is reduced to a chain of $N$ straight segments that are linked to each other at their ends. 
Each segment has an identical length $b$ much larger than the size of a single monomer. We denote the positions of the joints between the segments as $\mathbf{R}_n$ for $n=1,...,N-1$, while $\mathbf{R}_0$ and $\mathbf{R}_N$ give the positions of the two ends of the whole resulting chain, see Fig.~\ref{fig_freely-jointed-chain}. The chain is ``freely'' jointed in the sense that each segment orientation is independent of the orientation of all other segments. Thus, for each segment, the probability distribution $\psi(\mathbf{r}_n)$ of its end-to-end vector $\mathbf{r}_n=\mathbf{R}_n-\mathbf{R}_{n-1}$ is given by 
\begin{equation}
\psi(\mathbf{r}_n) = \frac{1}{4\pi b^2}\,\delta\left(\|\mathbf{r}_n\|-b\right) \qquad (n=1,...,N), 
\end{equation} 
where $\delta(\cdot)$ denotes the Dirac $\delta$-function. It serves here to prescribe the length $b$ of the segment. In this minimum model, the probability distribution to find a certain chain configuration is simply given by the product of these individual single-segment probability distributions $\psi(\mathbf{r}_n)$. 

On this basis, it is straightforward to determine the probability distribution for the end-to-end vector $\mathbf{R}=\mathbf{R}_N-\mathbf{R}_0$ of the whole chain. For $N\gg1$, we obtain
\begin{equation}\label{eq:end-to-end}
\psi(\mathbf{R}) = \left(\frac{3}{2\pi Nb^2}\right)^{3/2}\exp\left\{ -\frac{3\mathbf{R}^2}{2Nb^2} \right\}.
\end{equation}
Obviously, this is of Gaussian form. 

From a statistical mechanics point of view, we expect the probability distribution to be of the form $\psi(\mathbf{R})\sim\exp[-F(\mathbf{R})/k_BT]$. Here $F(\mathbf{R})$ denotes the energy of the chain configuration with end-to-end vector $\mathbf{R}$, $k_B$ is Boltzmann's constant, and $T$ the temperature. Comparison with Eq.~(\ref{eq:end-to-end}) shows that
\begin{equation}\label{eq:spring}
F(\mathbf{R}) = \frac{1}{2} \frac{3k_BT}{Nb^2} \mathbf{R}^2.
\end{equation}
Thus the polymer chain behaves like an elastic spring between its two end points $\mathbf{R}_0$ and $\mathbf{R}_N$. In the unstrained state the length of this spring vanishes. We find $3k_BT/Nb^2$ for the magnitude of its elastic constant. 

Based on this simple model, we see that polymeric materials show an elastic response. The important point is that these elastic forces mainly do not origin from an internal energy. They are mostly entropic in nature. Everything that enters Eq.~(\ref{eq:end-to-end}) is the entropy resulting from the single-chain configuration. The elastic constant in Eq.~(\ref{eq:spring}) is set by the temperature $T$. 

As a next step, one would like to obtain an elastic material of macroscopic dimension. For this purpose, many polymer chains are chemically crosslinked, i.e.\ covalently bound, to form one chemical unit. In this way, an elastomer or rubber is obtained. The main source of elasticity is again of entropic origin and referred to as ``rubber elasticity'' \cite{treloar1975physics}. 

\begin{figure}
\centerline{\includegraphics[width=9.cm]{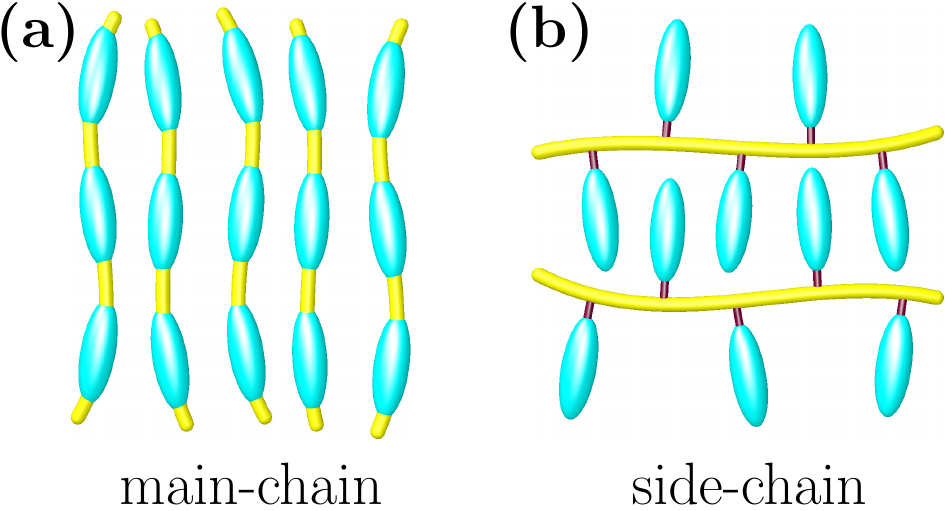}}
\caption{Schematic illustration of two architectures of liquid crystalline polymers. Again the liquid crystalline rod-like components are indicated by ellipsoids as in Fig.~\ref{fig_lc-phases}. Non-liquid-crystalline parts of the polymer ``backbone'' chains are depicted by the bright strings. (a) On the one hand, in main-chain liquid crystalline polymers, the liquid crystalline units are part of the polymer backbones. (b) On the other hand, the liquid crystalline units are attached as side groups to the polymer chains in side-chain liquid crystalline polymers. Usually this is achieved via shorter hydrocarbon chains, so-called spacer groups, which here are indicated in darker color.}
\label{fig_mcsc}
\end{figure}

To make the connection to the field of liquid crystals, first liquid crystalline polymers were synthesized. Two principal architectures are available: on the one hand, liquid crystalline molecules can be part of the polymer backbone chains in liquid crystalline main-chain polymers \cite{percec1991liquid}, see Fig.~\ref{fig_mcsc}~(a); on the other hand, the liquid crystalline molecules can be attached as side-groups to a polymer backbone in liquid crystalline side-chain polymers \cite{finkelmann1978model}, see Fig.~\ref{fig_mcsc}~(b). 
When in a later step such liquid crystalline polymers were chemically crosslinked, rubbery liquid crystalline elastomers were obtained \cite{finkelmann1981liquid,bergmann1997liquid}. In this way, the properties of liquid crystals and those of elastic rubbers could be combined in one material. The coupling provides new interesting features as described below. 

There are different protocols available to obtain different kinds of liquid crystalline elastomers concerning the nematic orientational alignment in the samples. Without special aligning mechanisms 
it is natural for nematic bulk materials in general not to feature a globally aligned nematic director. Already for low-molecular-weight liquid crystals, bulk samples in the nematic phase usually appear turbid. They strongly scatter light \cite{degennes1993physics}. The reason are thermal fluctuations in the director orientations. As a consequence, the nematic director is not macroscopically aligned, but continuously varies over space. 
Additional effort, such as the prescribed alignment on the sample surfaces depicted in Fig.~\ref{fig_lc_frederiks}, is necessary to achieve a global alignment of the director orientation. 

Naturally, thermal orientational fluctuations also occur in polymeric substances. 
Therefore, if during the crosslinking process of synthesizing a liquid crystalline elastomer no special action is taken, polydomain samples are obtained \cite{finkelmann1994liquid}. The director in these polydomain materials is not globally uniform in the nematic state but continuously varies over space. However, protocols to manufacture monodomain samples featuring a macroscopically aligned director are available. For this purpose, during the final crosslinking step of the manufacturing process, the sample must be maintained in the liquid crystalline phase and the director must be homogeneously aligned in the polymer solution. In practice, this was achieved by external magnetic fields \cite{legge1991memory,mitchell1993strain,rogez2011influence}, external electric fields \cite{rogez2011influence}, aligning boundary conditions \cite{urayama2005electrooptical,komp2005versatile,urayama2007stretching}, or stretching a pre-crosslinked sample \cite{kupfer1991nematic,kupfer1994liquid}. A corresponding memory of the director orientation during crosslinking is then imprinted into the material during the crosslinking procedure. When the liquid crystalline molecules were reoriented by some external action, they were observed to rotate back into the memorized orientations after the external influence is switched off again. This was experimentally verified even for polydomain samples crosslinked in the liquid crystalline state \cite{urayama2006slow,urayama2009polydomain}. Apparently the orientations are stored in the architecture of the polymer network. 

The most interesting properties of liquid crystalline elastomers arise from their orient\-ational-deformational coupling. On the one hand, a mechanical deformation like stretching or compression can tune the state of birefringence and the optical properties. As one specific example, liquid crystalline elastomers were used to experimentally demonstrate the possibility to construct mirrorless lasers of a wave-length that is tunable by mechanical deformations \cite{finkelmann2001tunable,schmidtke2005probing}. More in general, it was shown for various different samples that mechanical deformations can reorient the director \cite{kupfer1991nematic,mitchell1993strain,kupfer1994liquid,kundler1995strain,roberts1997single, kundler1998director,urayama2007stretching}. This is indicated on the left-hand side of Fig.~\ref{fig_lc_elastomer}: 
\begin{figure}[t]
\centerline{\includegraphics[width=\textwidth]{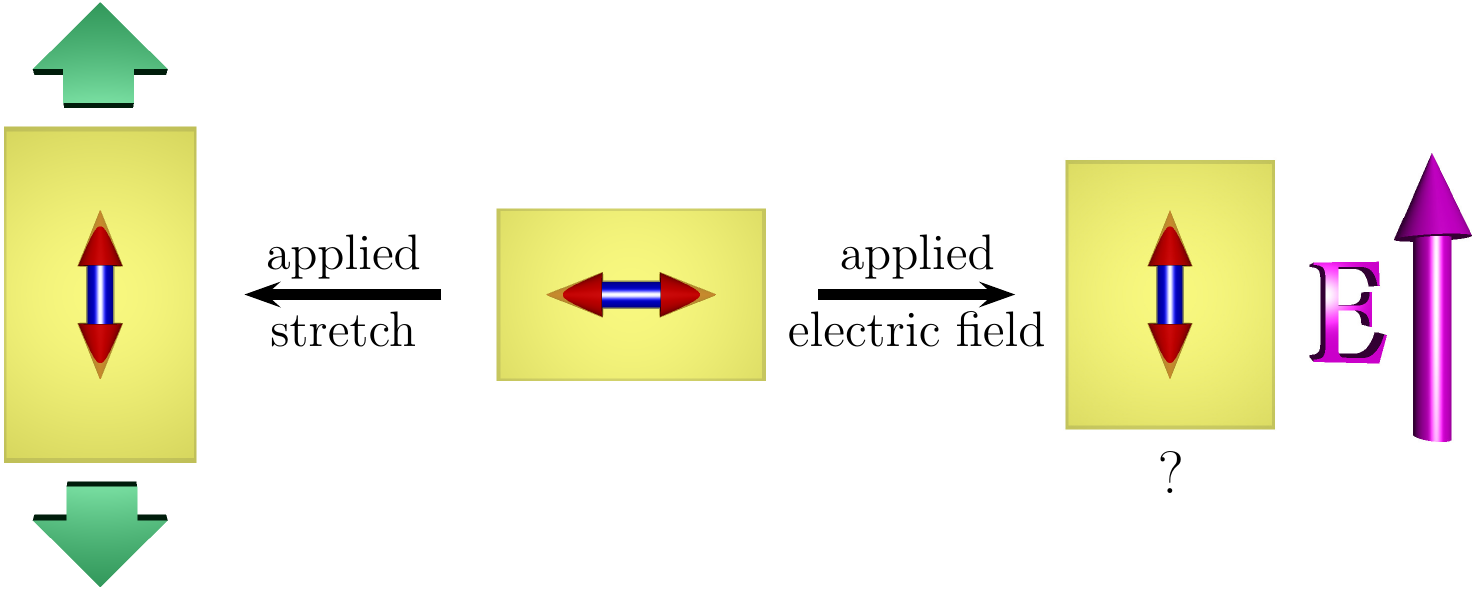}}
\caption{Coupling between director reorientations and elastic deformations in nematic liquid crystal elastomers. The nematic director is indicated by the double-headed arrow, whereas the rectangular box illustrates the deformations of the elastomer. An initial ground state is depicted in the center. When an external stretch (marked by the vertical broad arrows on the left) is applied in the direction perpendicular to the initial director orientation, the coupling to the director can induce its reorientation into the stretching direction. For practical purposes, the inverse effect would be beneficial, i.e.\ mechanical deformations induced by applying an electric field that reorients the director, as illustrated on the right. Unfortunately, the necessary field amplitudes turn out to be too high for practical purposes, except for materials swollen with conventional small-molecule liquid crystals.}
\label{fig_lc_elastomer}
\end{figure}
when a nematic liquid crystalline elastomer is stretched perpendicularly to its initial director orientation, a reorientation of the director into the stretching direction can result. 
The reorientation process involved a pronounced nonlinearity in the corresponding stress-strain curves \cite{kupfer1991nematic,kupfer1994liquid,urayama2007stretching,krause2009nematic}. On the other hand, it was demonstrated that the inverse effect is also within reach experimentally: reorientations of the director can induce mechanical deformations. From an application point of view, this would allow the construction of soft actuators.  
One could use an external electric field to reorient the director, which is transformed into mechanical actuation. Unfortunately, so far the necessary field amplitudes were found to be too high for practical purposes. 
However, for samples swollen with a common low-molecular-weight (small-molecule) liquid crystal, a sizeable amount of so-called electro-mechanical coupling was observed \cite{yusuf2004swelling,urayama2005electrically,yusuf2005low,urayama2005electrooptical, urayama2006swelling,urayama2006deformation,cho2006electrooptical, urayama2007selected,cho2007trifunctionally, fukunaga2008dynamics, hashimoto2008multifunctional}. In this case the director can be reoriented by an external electric field, which can result in deformations of up to the lower double-digit per-cent regime. 

Prestretching a material to the point where director reorientation sets in may allow to observe the electric-field-induced effect also in conventional non-swollen samples, as outlined theoretically \cite{menzel2009response}. Close to this critical point, the reorienting external stretching deformation on the one hand and the imprinted memory that anchors the director on the other hand approximately balance each other. Then a relatively weak additional external mechanical or electric field can result in a comparatively large response in the director orientation. 

Apart from that, other routes of regulation or actuation were demonstrated experimentally, for example photo- \cite{finkelmann2001new,sanchez2011opto} or temperature-induced \cite{yusuf2004hystereses,cho2006thermo,sawa2010thermally,fleischmann2012one} deformations. Possible applications as soft actuators or even ``artificial muscles'' were discussed \cite{hebert1997dynamics,yu2006soft,ohm2010liquid,yang2011micron,jiang2013actuators}. Large length changes of several $100$~\% were observed for loaded and unloaded samples under reversible heating and cooling through the isotropic-nematic transition \cite{wermter2001liquid,krause2009nematic}. For many applications, however, control parameters different from temperature are more desirable. 

On the theoretical side, we mention two principal routes of approach that reproduce the measured nonlinear stress-strain behavior as well as the detected deformations due to director reorientations. The first one is based on the concept of rubber elasticity \cite{treloar1975physics}. It assumes Gaussian statistics for the step lengths along the polymer strands \cite{warner2003liquid}. As a starting point, it further assumes that the main role of the liquid crystalline building blocks is to introduce an anisotropy direction into these Gaussian statistics. Thus, instead of Eq.~(\ref{eq:end-to-end}), the end-to-end vectors of the polymer strands between network points of the polymer mesh are assumed to follow the distorted distribution
\begin{equation}
\psi(\mathbf{R}) \propto \exp\left\{ -\frac{3}{2\tilde{L}^2}\,\mathbf{R}\cdot(\mathbf{I}-r_a\mathbf{\hat{n}}\mathbf{\hat{n}})\cdot\mathbf{R} \right\}, 
\end{equation}
where $\mathbf{\hat{n}}$ again indicates the nematic director, $r_a$ is an anisotropy parameter, $\mathbf{I}$ indicates the unity matrix, and $\tilde{L}$ is an effective length connected to the actual contour lengths of the polymer strands. 
Open issues concerning this picture are how valid the assumption of the distorted Gaussian statistics generally is \cite{rogez2011influence}, and particularly how the orientational memory influences the statistics. 
A different microscopic approach was proposed in the form of randomly crosslinked anisotropically interacting rigid dimers \cite{xing2008nematic}, which reproduces the expressions for the theory of rubber elasticity on a more coarse-grained level. 

\begin{figure}
\centerline{\includegraphics[width=\textwidth]{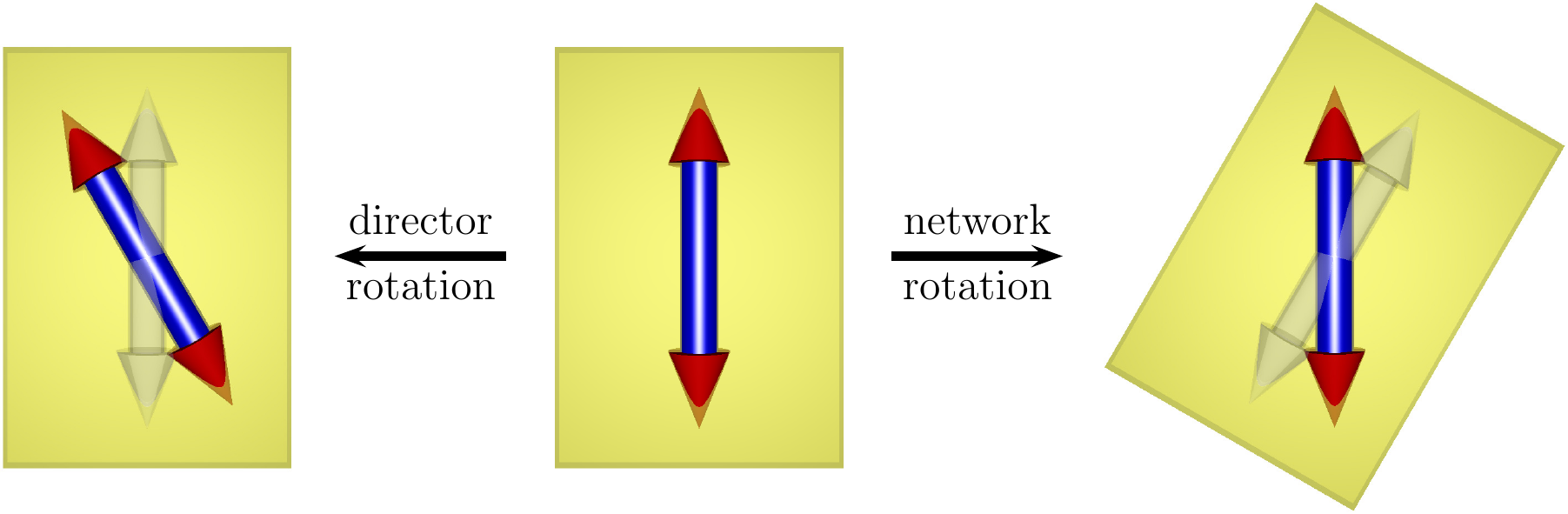}}
\caption{Illustration of the concept of relative rotations in liquid crystalline elastomers. In the ground state (center) the director takes the orientation memorized by the architecture of the polymer network from the time of crosslinking. Relative rotations as a macroscopic variable describe how the director is rotated with respect to that memorized orientation. This can occur by a reorientation of the director relatively to the polymeric network (left), or a rotation of the network together with the imprinted memorized direction with respect to the actual director orientation (right). The variables of relative rotations test whether the results from both processes differ from each other and allow a formal inclusion in the corresponding energetic considerations. In all pictures, the strong-colored double-headed arrow indicates the actual director orientation, and the bright box symbolizes the polymeric network. On the left- and on the right-hand side, the light double-headed arrow represents the imprinted ground-state director orientation memorized by the network structure at the moment of crosslinking.}
\label{fig_relrot}
\end{figure}

The second route is macroscopic and based on symmetry arguments. It couples the continuum descriptions of liquid crystals and elastic materials \cite{degennes1980weak,brand1994electrohydrodynamics,lubensky2002symmetries,muller2005undulation,menzel2006rotatoelectricity, menzel2007cholesteric,menzel2008instabilities,ye2007semisoft,menzel2007nonlinear,menzel2009nonlinear,ye2009phase}. 
We noted above that a memory of the director orientation during synthesis is imprinted into the network structure of the polymer mesh. 
One central ingredient to many of the macroscopic studies is therefore the observation that rotations of the director out of its memorized equilibrium orientation cost energy. Such ``relative rotations'' between the director and the polymer network environment are included as additional variables \cite{degennes1980weak,brand1994electrohydrodynamics,muller2005undulation,menzel2006rotatoelectricity, menzel2007cholesteric,menzel2008instabilities}. First proposed for small deviations from the initial state, this concept was later generalized to the nonlinear regime \cite{menzel2007nonlinear,menzel2009nonlinear}. 
It is schematically illustrated in Fig.~\ref{fig_relrot}. 

\subsection{Ferrogels and magnetic elastomers}

Another example of externally tunable polymeric soft matter systems is given by magnetic gels and elastomers \cite{filipcsei2007magnetic}. Again, these materials combine the features of two different classes of complex fluids: those of ferro- and magnetorheological fluids \cite{rosensweig1985ferrohydrodynamics,odenbach2002ferrofluids,odenbach2003ferrofluids,odenbach2003magnetoviscous, huke2004magnetic,odenbach2004recent,fischer2005brownian,ilg2005structure,klapp2005dipolar,holm2005structure, embs2006measuring,ilg2006structure,gollwitzer2007surface,vicente2011magnetorheological} with those of crosslinked polymers \cite{strobl1997physics}. 

The systems can be classified as composite magnetic hybrid materials \cite{jolivet2002synthesis,garcia2003mesoporous,yuan2011one,sarkar2012polymer,kao2013toward}. They consist of superparamagnetic or ferromagnetic particles that are embedded in a crosslinked polymer matrix. Depending on the degree of swelling of such a magnetic elastomer with a solvent it may rather be called a magnetic gel \cite{jolivet2002synthesis,messing2011cobalt,frickel2011magneto}. In the dry state, the materials can become hard and glass-like \cite{jolivet2002synthesis}. 

Different sizes of the embedded magnetic particles, typically in the range of nano- to micrometers \cite{filipcsei2007magnetic}, can lead to qualitatively different material behavior. If the particles are smaller than ten to fifteen nanometers, the direction of their magnetic moment is not fixed with respect to the particle axes. Instead, thermal fluctuations can lead to a reorientation of the magnetic moment \cite{neel1949theorie}, which is called N\'{e}el mechanism. For larger particles, reorientations of the magnetic moments mainly occur via the Brownian or Debye mechanism \cite{coffey1993ferromagnetic}, i.e.\ rigid rotations together with the whole particle. Still larger particles, from around a hundred nanometers, can feature multidomains of internal magnetization \cite{frenkel1930spontaneous,brown1968fundamental,aharoni1988elongated,seynaeve2001transition,hergt2006magnetic}. 

In the absence of an embedding polymer matrix, suspensions of such magnetic particles in a carrier liquid were stabilized in the form of ferro- and magnetorheological fluids \cite{rosensweig1985ferrohydrodynamics,odenbach2002ferrofluids,odenbach2003ferrofluids,odenbach2003magnetoviscous, huke2004magnetic,odenbach2004recent,fischer2005brownian,ilg2005structure,klapp2005dipolar,holm2005structure, embs2006measuring,ilg2006structure,gollwitzer2007surface,vicente2011magnetorheological}. 
Such fluids show properties of high practical relevance \cite{raj1990commercial,raj1995advances}. In particular, their dynamic flow behavior determined by their macroscopic viscosity can be tuned reversibly through an external magnetic field. This feature, commonly referred to as ``magnetoviscous effect'', was observed experimentally and explained theoretically \cite{rosensweig1969viscosity,mctague1969magnetoviscosity, zubarev2002rheological,thurm2002magnetic,thurm2002magnetoviscous,thurm2003particle, odenbach2003ferrofluids,pop2004microstructure,odenbach2004recent,ilg2005anisotropy,pop2006investigation}. It is attributed on the one hand to a hindrance of particle rotations that would reorient the magnetization axes of the particles away from the aligning external magnetic field direction. On the other hand, it results from the formation of micro-aggregates in the presence of external magnetic fields. Typically, the second contribution, when present, exceeds the first one. The important point is that the adjustment of the viscosity can be achieved in a non-invasive way reversibly from outside. 

We turn back to the presence of an embedding polymer matrix in magnetic gels and elastomers. 
In analogy to the viscous behavior of magnetic fluids, for magnetic gels and elastomers the resistance to elastic deformations, quantified by the elastic moduli \cite{landau1986elasticity}, was experimentally probed. It turned out that the elastic moduli can be tuned reversibly from outside by external magnetic fields \cite{deng2006development,filipcsei2007magnetic,stepanov2007effect, chen2007investigation,bose2009magnetorheological,evans2012highly,borin2013tuning}. 
This is one of the most outstanding properties of these materials. From an application point of view, it is for example interesting for the construction of novel externally tunable damping devices \cite{sun2008study} or vibration absorbers \cite{deng2006development}. 

The analysis of theoretical minimum models demonstrated that the spatial distribution of the magnetic particles plays a central role and qualitatively determines the nature of this effect \cite{ivaneyko2011magneto,ivaneyko2012effects,ivaneyko2014mechanical}. It was found that the tensile elastic modulus decreases or increases with increasing amplitude of an external magnetic field applied along the tension axis, depending on the particle arrangement. For instance, a regular simple-cubic or bcc lattice arrangement of the particle positions implied a decreasing tensile elastic modulus \cite{ivaneyko2011magneto,ivaneyko2012effects}. In contrast to that, the modulus increased for an fcc structure \cite{ivaneyko2012effects}. 

To a certain degree, the particle distribution can in fact be influenced during the synthesis of the materials. When a strong external magnetic field is applied before and during the crosslinking procedure, the magnetic particles were observed to form chain-like aggregates \cite{collin2003frozen,varga2003smart,filipcsei2007magnetic,gunther2012xray, borbath2012xmuct,gundermann2013comparison}. After the crosslinking has been completed, the field can be switched off with the anisotropic particle distribution persisting. The anisotropy has been locked into the materials. 

It has been demonstrated in experiments that this imprinted anisotropy can support the tunability of the elastic modulus by an external magnetic field \cite{filipcsei2007magnetic}. In that case, the largest effect of tunability was found when all three directions were aligned: the imprinted anisotropy direction, the additional external magnetic field to tune the mechanical properties, as well as the direction of mechanical deformation to measure the elastic modulus \cite{filipcsei2007magnetic}. This trend could be reproduced in a numerical study that assumed affine deformations of the material \cite{camp2011modeling,camp2011effects}. Although the degree of anisotropy in this case was significantly lower than that of chain-like aggregates, the effect was clearly observed. 

We remark that the assumption of affine deformations, which maps the macroscopic length changes linearly to all distances within the system, constitutes a major simplification. Yet it is often necessary to allow theoretical progress at all. Nevertheless, the more irregular the distributions of the magnetic particles within the samples become, the more this can lead to increasingly erroneous and even qualitatively incorrect results. Care has to be taken when this approximation is applied to realistic systems as has recently been pointed out on the basis of a simple dipole-spring model 
\cite{pessot2014structural}. 

Allowing for non-affine deformations, an explicit numerical investigation on the impact of chain-like particle arrangements was performed using finite element simulations \cite{han2013field}. It showed that deviations from a straight chain architecture in the form of a zig-zag structure can qualitatively influence the behavior under external magnetic fields: the zig-zag angle determines whether the compressive elastic modulus decreases or increases with increasing amplitude of the external magnetic field. Even non-monotonic behavior is possible. 

The behavior within one linear chain was studied in a minimum dipole-spring model of hard spheres connected by elastic springs \cite{annunziata2013hardening}. When the magnetic moment of the spheres is increased by an external magnetic field such that the hard spheres attract each other, they will come closer until they finally touch. Neglecting thermal fluctuations of the spheres, this process has the signature of a phase transition. It is of second order when the initial separation is small. If the initial separation is large, it is of first order, with the spheres clashing together in the final part of approach. A critical point separates both scenarios. Since the compressive elastic modulus along the chain diverges when the spheres touch each other, the process was called ``hardening transition''. At the critical point, just before the spheres touch each other, the compressive elastic modulus along the chain vanishes. 
Also the behavior of free-standing filamental chains was studied numerically \cite{sanchez2013filaments,cerda2013phase}. In this case, various different states like compact, helicoidal, partially collapsed, simply closed, and extended open configurations were detected. 

\begin{figure}[t]
\centerline{\includegraphics[width=\textwidth]{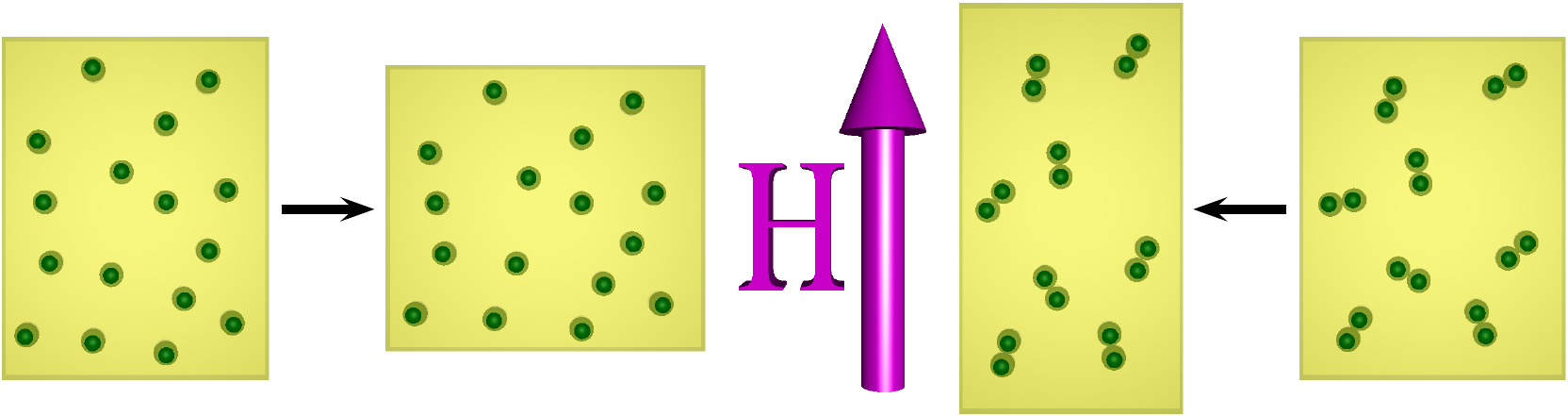}}
\caption{In magnetic gels and elastomers the spatial distribution of the magnetic particles can qualitatively influence the macroscopic behavior. Here, this is illustrated for the magnetostrictive behavior of a two-dimensional sheet in a homogeneous external magnetic field (see Ref.~\cite{stolbov2011modelling}). Left-hand side: for a random ``gas-like'' distribution of magnetic particles switching on the external magnetic field was observed to result in a contraction along the field direction \cite{stolbov2011modelling}. Right-hand side: in contrast to that, for randomly oriented doublets of magnetic particles, the same set-up was found to lead to an elongation along the field direction \cite{stolbov2011modelling}. 
}
\label{fig_ferrogel}
\end{figure}

A second fascinating property of magnetic elastomers and gels arises from their reversible deformations that can be induced by external magnetic fields. The spatial arrangement of the magnetic particles can qualitatively impact whether elongation or contraction occurs along the field direction, as pointed out by theory and simulations \cite{stolbov2011modelling,zubarev2013effect}. An example is illustrated in Fig.~\ref{fig_ferrogel}. 
Apart from the magnetostrictive behavior \cite{guan2008magnetostrictive,stolbov2011modelling,gong2012full, zubarev2012theory,zubarev2013magnetodeformation,allahyarov2014magnetomechanical}, i.e.\ deformations induced by homogeneous external magnetic fields, particularly the ``actuation'' by inhomogeneous external magnetic fields \cite{zrinyi1996deformation,zrinyi1997direct} was investigated in experiments and by modeling. To minimize the free energy, the materials are drawn into regions of higher magnetic field. This can lead to elastic deformations. After switching off the external field, the elastic energy is released and the probe can reversibly switch back to its initial state. It was demonstrated for a real sample that frequencies around $40\mbox{~Hz}$ can easily be followed \cite{filipcsei2007magnetic}. Consequently these materials again are ideal candidates for the construction of soft actuators \cite{zimmermann2006modelling,filipcsei2007magnetic} or ``artificial muscles'' \cite{zrinyi2000intelligent}. 
These applications involve dynamic properties, which so far have been addressed in only a few studies on the theoretical side \cite{jarkova2003hydrodynamics,bohlius2004macroscopic,tarama2014tunable}, requiring further investigation in the future. 
Apart from that, the use of magnetic gels as sensors indicating magnetic fields and field gradients by deformation was experimentally outlined \cite{szabo1998shape,ramanujan2006mechanical, liu2006magnetic}. Other studies discuss their application to control by external magnetic fields the amount and rate of drug release \cite{liu2006magnetic,brazel2009magnetothermally}. A further subject is the combat against cancer cells. For this purpose, magnetic particles are embedded into the corresponding cell tissue. A quickly alternating external magnetic field can be applied that continuously remagnetizes the particles. Through hysteretic losses during the remagnetization cycles, heat is generated. Such hyperthermic treatment  \cite{jordan1999magnetic,babincova2001superparamagnetic, lao2004magnetic,hergt2006magnetic} can destroy the cancer cells. 

\begin{figure}
\centerline{\includegraphics[width=9.cm]{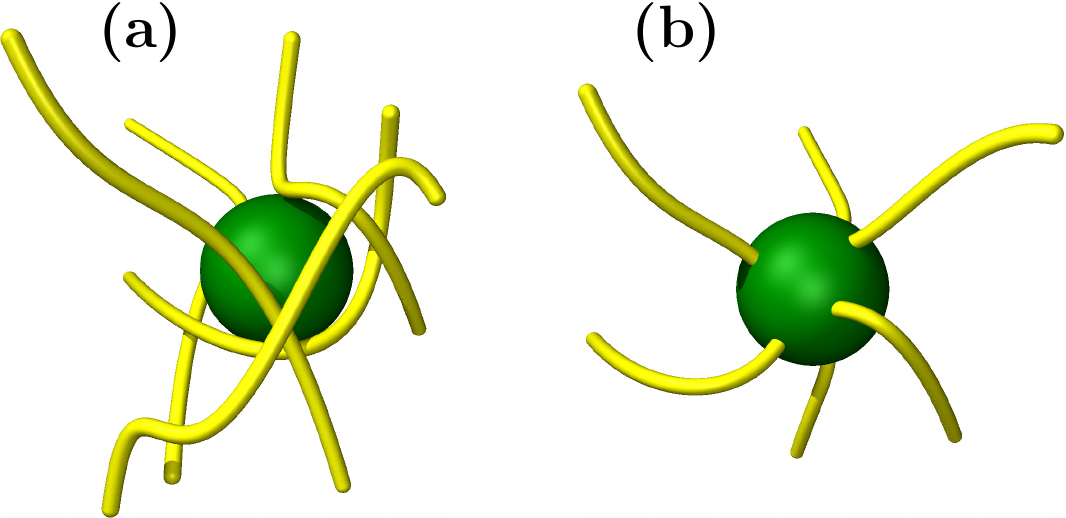}}
\caption{Schematic illustration of two qualitatively different kinds of ferrogels implying different degrees of magneto-mechanical coupling. (a) In the classical materials the magnetic particles are simply embedded in the surrounding mesh of polymer chains that form the polymer matrix. For spherical particles, there may not be any permanent restoring torque induced by particle rotations. (b) In a new class of ferrogels, the polymer chains are covalently bound to the surface of the magnetic particles \cite{frickel2011magneto,messing2011cobalt}. Thus the magnetic particles serve as crosslinkers. Rotations of the spherical particles inevitably induce deformations of the surrounding polymer mesh, leading to permanent restoring torques. In both cases, the density of polymer chains is significantly higher than indicated in the simplified schematics.}
\label{fig_kinds-ferrogels}
\end{figure}

For all these purposes, the degree of the so-called magneto-mechanical coupling should be maximized. A recent attempt in this direction was to not only embed the magnetic particles in the polymer matrix, but to directly crosslink them into the polymer network \cite{frickel2009functional,frickel2011magneto,messing2011cobalt}, see Fig.~\ref{fig_kinds-ferrogels}. 
During sample preparation, this was achieved via surface-functionalization of the magnetic particles by attaching reactive groups on their surfaces. As a consequence, the magnetic particles themselves can serve as crosslinkers. Rotating the particles, the polymer chains attached to their surfaces get ``rolled up'' around the particles as investigated by a simulation study \cite{weeber2012deformation}. This ``rolling-up'' leads to restoring torques. 
An orientational memory of the particle orientations arises, implying an energetic penalty when the particles rotate relatively to their environment \cite{annunziata2013hardening}. It can have a significant impact on the appearance of the system. For example, depending on the orientational memory, qualitatively different energetic ground states were obtained for a linear chain of ferromagnetic particles in a dipole-spring model \cite{annunziata2013hardening}: ferromagnetic, with the magnetic moments aligned along the chain axis; antiferromagnetic-like, with neighboring dipoles rotated by 180 degrees around the chain axis; or a spiral-like magnetization, with neighboring dipolar moments rotated by an intermediate angle around the chain axis.  
Even stronger magneto-mechanical coupling might be achieved through the use of elongated instead of spherical magnetic particles \cite{roeder2012shear}.

\subsection{Links between liquid crystalline and magnetic elastomers}

At the end of this section, we include a brief qualitative comparison between the two classes of materials considered above, i.e.\ liquid crystalline and magnetic elastomers and gels. We only concentrate on a few selected points that may be important when a mutual support of the two fields is considered. With the macroscopic theories for the two classes of material being partially mappable onto each other, also their observable behavior in analogous experiments may be so. 

Both materials combine the elastic behavior of polymeric substances with the anisotropy arising from an orientational ordering of their constituents. In the case of liquid crystalline elastomers, this anisotropy results from the average orientation of the liquid crystalline molecules along the director. For magnetic elastomers and gels, it is identified with the orientation of the magnetic moments in the equilibrium configuration. 

Depending on the procedure of synthesis, a more or less pronounced memory of the initial orientations at the crosslinking time is locked into the samples. Then relative rotations between this locked-in memory direction and the actual orientation 
cost energy. We outlined above the use of such relative rotations as a variable in the macroscopic characterization of liquid crystalline elastomers \cite{degennes1980weak,brand1994electrohydrodynamics, menzel2007nonlinear,menzel2009nonlinear,menzel2009response}, see Fig.~\ref{fig_relrot}, and in a mesoscopic model of magnetic gels and elastomers \cite{annunziata2013hardening}. Furthermore, relative rotations have already been considered in a macroscopic characterization of anisotropic magnetic gels \cite{bohlius2004macroscopic}. 
A next step will be to connect the mesoscopic and the macroscopic descriptions of magnetic gels by an appropriate coarse-graining procedure.  
Similarities between the two classes of materials, i.e.\ between liquid crystalline elastomers and magnetic gels, were pointed out on the macroscopic scale \cite{brand2011physical}. 

Generally, it is an interesting question, how the memory actually gets locked into the architecture on the molecular level. In other words, what structural changes occur for a polymer network in comparison to a conventional material without memory. The differences between the two classes of materials are of course large, already in view of the corresponding length scales. On the one hand, the liquid crystalline molecules are relatively small and can even be part of the polymer backbones in main-chain liquid crystalline elastomers. Therefore one would expect that an orientational memory should be reflected on the molecular level by the structure of the crosslinked polymer network. On the other hand, magnetic elastomers are real composite materials of mesoscopic colloidal magnetic particles embedded in a polymer matrix. Here, the spatial arrangement of the mesoscopic magnetic particles within the elastic matrix can be the source of the orientational memory. So the mechanism of memorizing may be quite different in the two cases. 

Nevertheless, based on the similarities in the macroscopic theories, it would be worthwhile to test the materials in those experiments in which the respective other material class features its outstanding properties. The experience acquired for one of the two material classes may help to guide the way also for the other material class. On the one hand, this means to clarify the question, whether an orientational coupling between strain deformations and the anisotropy direction also exists within anisotropic magnetic gels. In a next step, this would lead to the question, whether this property is likewise connected to a marked nonlinear stress-strain behavior as observed for liquid crystalline elastomers. On the other hand, in analogy to the elastic moduli that are reversibly tunable by external magnetic fields in the case of magnetic gels, a similar effect may be observable for liquid crystalline elastomers and gels. Here, rather an externally applied electric field would serve to tune the elastic properties. It would be satisfying to observe a mutual benefit between the two fields of liquid crystalline elastomers and magnetic gels. 

Moreover, instead of only tuning their properties, the materials could also be utilized as model systems for driven and active soft matter in genuinely non-equilibrium situations. We mentioned that both liquid crystalline elastomers and magnetic gels can be used as soft actuators by switching an appropriate external field \cite{yusuf2005low,urayama2006deformation,urayama2007selected,cho2007trifunctionally, fukunaga2008dynamics,hashimoto2008multifunctional,urayama2005electrooptical,urayama2005electrically, finkelmann2001new,sanchez2011opto,cho2006thermo,sawa2010thermally,fleischmann2012one, hebert1997dynamics,yu2006soft,ohm2010liquid,yang2011micron,jiang2013actuators,filipcsei2007magnetic, stolbov2011modelling,guan2008magnetostrictive,gong2012full,zubarev2012theory,zubarev2013magnetodeformation, allahyarov2014magnetomechanical,zrinyi1996deformation,zrinyi1997direct,zimmermann2006modelling, zrinyi2000intelligent}. In a next step, these materials could be steadily driven by external fields so that they cannot reach an equilibrium state any more. Examples of driven soft matter will be addressed in the subsequent part of this review. In another step, they can be continuously ``activated'' and then be categorized as active soft materials. For instance, the potential of liquid crystalline elastomers for the construction of light-activated swimmers was experimentally outlined \cite{camacho2004fast}. Similarly, an efficient production of light-activatable microscopic artificial cilia was demonstrated via inkjet printing \cite{vanoosten2009printed}; induced deformations of these slender artificial fibers \cite{vanoosten2009printed} mimic the deformation cycle of natural cilia that are found on bacterial surfaces and serve for self-propulsion of the microorganisms \cite{brennen1977fluid}. Likewise, magnetically activated elastic filaments were used to construct propelling artificial microswimmers \cite{dreyfus2005microscopic}. We will come back to the migration of self-propelled particles and microswimmers in the later part of this review. 
Apart from that, on the theoretical side, the macroscopic characterizations of gels featuring orientational degrees of freedom can be extended to include active stresses that result for example from chemical reactions. Such approaches were used to identify and describe possible mechanisms for the motility of crawling biological cells \cite{kruse2004asters,kruse2005generic,juelicher2007active, carlsson2011mechanisms,recho2013contraction,recho2014optimality}.

\newpage
\section{Driven soft matter}

Naturally, in each system thermal fluctuations lead to a sort of ``internal drive''. For instance, colloidal particles are subject to Brownian motion, which adds to prevent them from sedimentation in a gravitational field and keeps them suspended \cite{dhont1996introduction}. Another example are experiments on colloidal particles thermally driven along the top of aligned membrane tubes that adhere to a substrate \cite{wang2009anomalous,wang2012brownian}. An interesting crossover between exponential and Gaussian forms of the displacement statistics was observed in the latter context. Generally, when asymmetric ratchet-like conditions direct the internal thermal drive, net motion can arise \cite{peskin1993cellular,faucheux1995optical}. 

\begin{figure}
\centerline{\includegraphics[width=4.5cm]{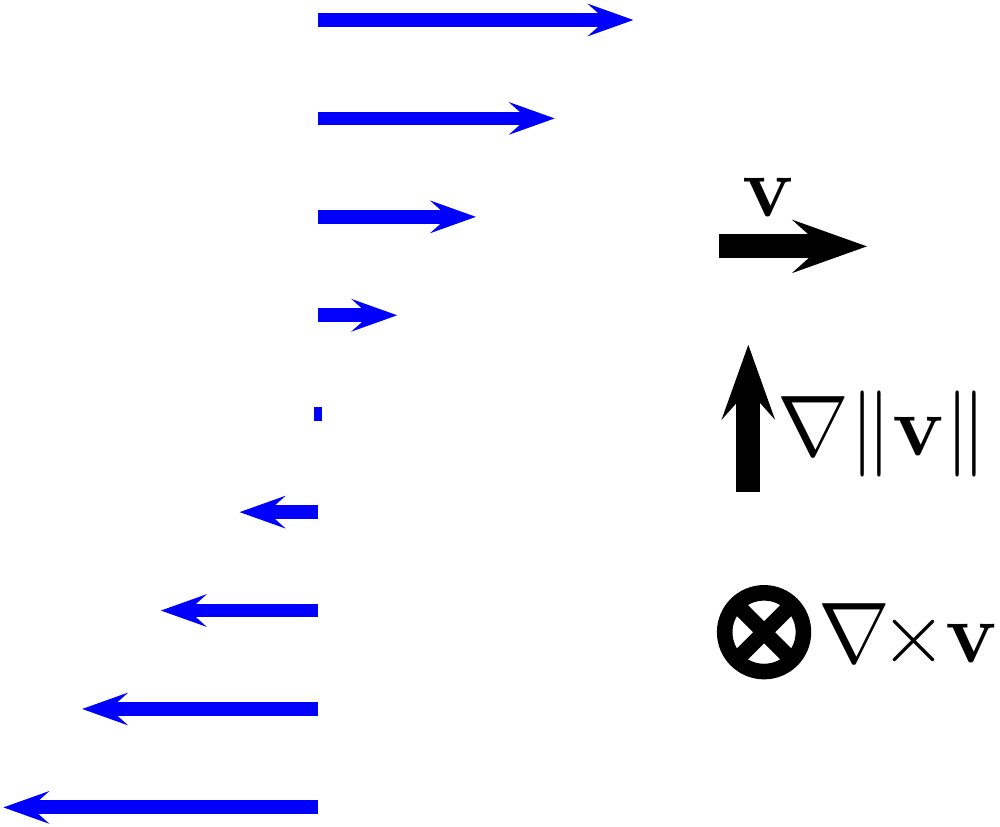}}
\caption{The flow profile corresponding to a planar linear shear flow is depicted on the left. It reveals the three different directions indicated on the right: the flow direction oriented along the flow velocity $\mathbf{v}$ (horizontal); the direction of the shear gradient $\mathbf{\nabla}\|\mathbf{v}\|$ (vertical); and the vorticity direction along $\mathbf{\nabla}\times\mathbf{v}$, here oriented perpendicular to the shear plane.}
\label{fig_linear-shear-flow}
\end{figure}

In contrast to that, we here focus on soft matter systems driven from outside. In our case this is achieved by externally imposed shear flows. The simplest situation is the one of an imposed planar linear shear profile as the one schematically indicated in Fig.~\ref{fig_linear-shear-flow}. This geometry is characterized by three different directions: the flow direction, here horizontally oriented; the direction of the shear gradient, here vertically oriented; and the vorticity direction perpendicular to the shear plane, i.e.\ here pointing into the plane of the figure. 
In practice, the strategy to approximately realize this flow field in simple situations is to horizontally drive the top and/or bottom plate(s) of the confining cell with constant speed. 

We start by repeating simple effects that emerge for bulk-filling low-molecular-weight nematic liquid crystals. Increasing the complexity, we then briefly review aspects occurring for bulk-filling periodically modulated phases such as they appear in block copolymer systems. Finally we address localized structures in the form of vesicles.

\subsection{Flow alignment and director tumbling in nematic liquid crystals}

Similar to the previous chapter, we first turn to conventional low-molecular-weight liquid crystals deep in the nematic phase. Locally, the liquid crystalline molecules align on average along an ordering direction, the so-called director $\mathbf{\hat{n}}$ \cite{degennes1993physics}, see also Sec.~\ref{lmwlc}. In the hydrodynamic equations describing the macroscopic dynamic behavior of a nematic liquid crystal, the director orientation $\mathbf{\hat{n}}$ couples to the hydrodynamic flow field $\mathbf{v}$ \cite{degennes1993physics,pleiner1996pattern}. 
For an imposed flow field as the one depicted in Fig.~\ref{fig_linear-shear-flow}, we now consider a bulk state of vanishing director distortions $\nabla\mathbf{\hat{n}}=\mathbf{0}$. 
Then the hydrodynamic equation for the director orientation becomes \cite{pleiner1996pattern,rienacker1999orientational}
\begin{equation}\label{eq:dyn-n}
\frac{\partial\mathbf{\hat{n}}}{\partial t} = 
\mathbf{\Omega}\cdot\mathbf{\hat{n}} + 
\lambda \left[ 
\mathbf{A}\cdot\mathbf{\hat{n}}-(\mathbf{\hat{n}}\cdot\mathbf{A}\cdot\mathbf{\hat{n}})\,\mathbf{\hat{n}} 
\right].
\end{equation}
In general, the two gradient tensors $\mathbf{\Omega}=[(\nabla\mathbf{v})^T-(\nabla\mathbf{v})]/2$ and $\mathbf{A}=[(\nabla\mathbf{v})^T+(\nabla\mathbf{v})]/2$
contain the rotational and elongational effect of the flow field $\mathbf{v}$, respectively. Here ${\nabla}\mathbf{v}$ again denotes a dyadic product and the superscript $^T$ its transpose. A convective contribution $\mathbf{v}\cdot\nabla\mathbf{\hat{n}}$ vanishes in our case due to the assumed spatially homogeneous director orientation $\nabla\mathbf{\hat{n}}=\mathbf{0}$. 

\begin{figure}[t]
\centerline{\includegraphics[width=\textwidth]{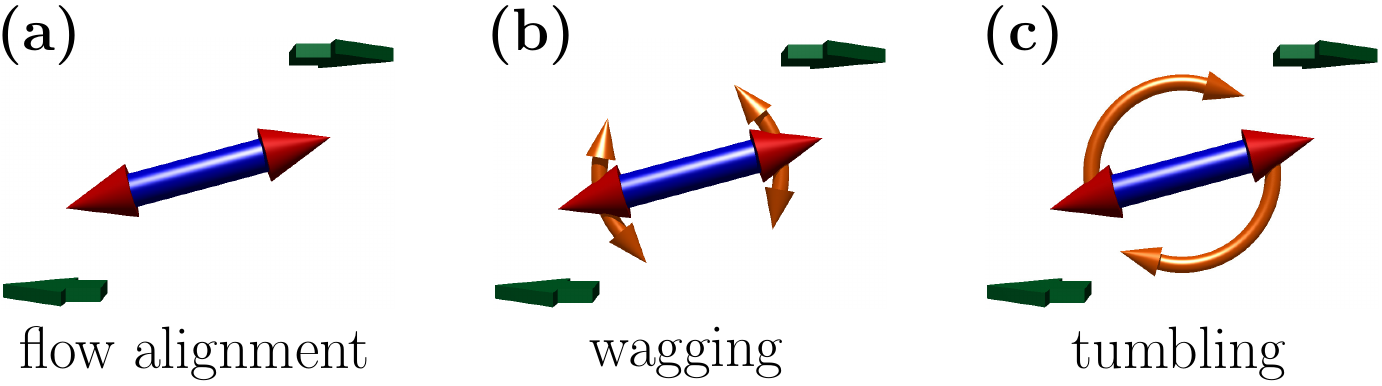}}
\caption{Illustration of three dynamic modes of a nematic liquid crystal under an imposed planar shear. The nematic director is indicated by the double-headed arrow, whereas the orientation of the shear is given by the two arrows in the top right and bottom left in each panel. (a) In the case of ``flow alignment'' or ``shear alignment'' the director takes a steady orientation within the shear plane and with a finite tilt angle with respect to the flow direction. (b) When ``wagging'', the director oscillates within the shear plane around an orientation of usually non-vanishing tilt with respect to the flow direction. (c) During ``tumbling'' motions the director makes full rotations in the shear plane.}
\label{fig_lc_shear}
\end{figure}

We can see from Eq.~(\ref{eq:dyn-n}) that, besides the shear rate set by the imposed flow field, only one parameter $\lambda$ determines the dynamic behavior. It can be written as the ratio between two viscosities and thus represents material properties \cite{degennes1993physics,rienacker1999orientational}. 
Depending on the magnitude of this parameter, the director can show qualitatively different types of behavior in the shear flow. On the one hand, for $|\lambda|\geq1$, the director reaches a steady-state orientation within the shear plane \cite{leslie1968some,degennes1993physics,gahwiller1972temperature,pieranski1974two}. This situation is typically called ``flow alignment'' or ``shear alignment'' and is depicted in Fig.~\ref{fig_lc_shear}~(a). 
The orientation angle with respect to the flow direction, i.e.\ the ``flow alignment angle'', remains constant and is determined by the parameter $\lambda$. On the other hand, the director does not find a steady-state orientation, but continuously rotates in the shear plane, for $|\lambda|<1$. This dynamic state is usually referred to as ``tumbling'' \cite{gahwiller1972temperature,pieranski1974two}, see Fig.~\ref{fig_lc_shear}~(c), and consequently $\lambda$ is often referred to as the tumbling parameter. 
Using the Fokker-Planck approach, these two dynamic regimes were connected to properties on the molecular level \cite{risken1996fokker,zwanzig2001nonequilibrium,archer1995molecular,kroger1995viscosity}. 
The above considerations imply that the shear rate of the imposed flow field is low enough so that the degree of nematic ordering, see Eq.~(\ref{eq:s}), is not significantly altered from its equilibrium value. 

Director tumbling was also observed and discussed for mixtures of liquid crystals \cite{ternet1999flow} and for liquid crystalline polymers \cite{burghardt1991role,rienacker1999orientational}. For certain parameter values, a transition was found with increasing shear rate from tumbling to flow alignment via an intermediate third dynamic state called ``wagging'' \cite{larson1990arrested,rienacker1999orientational}. In this third state, which is included in Fig.~\ref{fig_lc_shear}~(b), the director oscillates back and forth instead of performing full tumbling rotations. Still further dynamic modes were identified later, including chaotic types of motion \cite{rienacker2002chaotic}.

\subsection{Alignment and reorientations in periodically modulated phases}

At our next level of complexity, we concentrate on materials that in their ground states cannot be considered as spatially homogeneous any more. Instead, they feature regular periodic modulations in their density or in the concentrations of their constituents. An illustrative example are block copolymer melts or solutions. 

To understand the situation in block copolymer melts, it is instructive to recall the problem of mixing of two polymers of a different kind. This scenario is addressed by the Flory-Huggins theory \cite{strobl1997physics}. For low-molecular-weight liquids, a spontaneous mixing is usually driven by a gain in translational entropy: each molecule is now provided with the extended volume of the whole mixture, not only with the one of its initial single component liquid. In polymers, however, this effect is strongly reduced. The monomers are chemically trapped in the polymer chains. Only each polymer chain as a whole can increase its translational entropy. Therefore the gain in translational entropy under mixing is significantly lower than for low-molecular-weight liquids. Instead, the inter-species interactions between the monomers of different kinds in comparison to the intra-species interactions between monomers of the same kind determine whether the polymers mix or form two separated phases. In most cases, this competition favors the separated state. Most pairs of polymers do not form homogeneous mixtures at ambient temperatures \cite{strobl1997physics}. 

\begin{figure}
\centerline{\includegraphics[width=\textwidth]{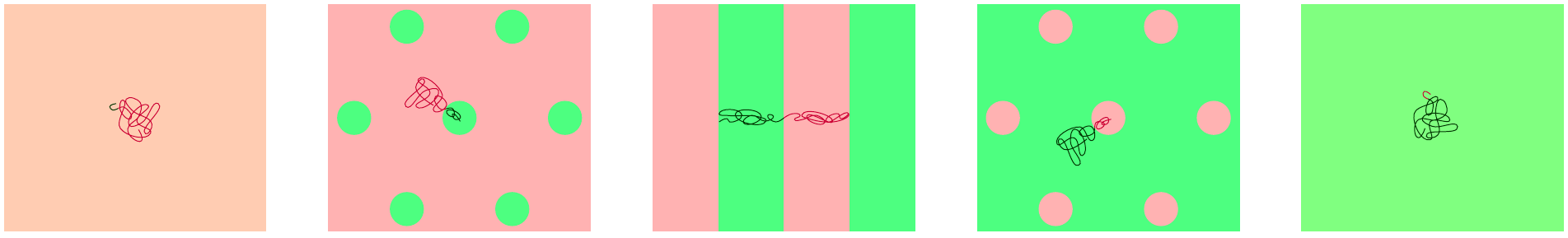}}
\caption{Schematic illustration of the two-dimensional patterns shown by micro-phase separated linear diblock copolymers. The temperature must be low enough so that micro-phase separation can take place. In the series from left to right, the fraction of one kind of polymer on the diblock copolymer chain with respect to the other one increases. This is depicted by a typical chain in each panel. If one species strongly dominates by its volume fraction, the state is disordered, i.e.\ not micro-phase separated, as in the most left and right examples. When the volume fractions become more balanced but still one species dominates, hexagonal textures can be observed as in the second examples from left and right. In three dimensions these would correspond to hexagonally arranged cylinders elongated perpendicular to the plane of the picture. Lamellar structures (center) appear for approximately equal volume fractions of the two blocks and are likewise obtained in three dimensions. 
}
\label{fig_microphase}
\end{figure}
The situation becomes interesting in block copolymers. Here, the different polymer chains that tend to demix under the previous conditions are combined into one molecule. 
For example, a diblock copolymer chain is composed of two chains of chemically different polymers that are covalently bound to each other at one of their ends. 
Due to the covalent bonds between the blocks, a macroscopic demixing is not possible for block copolymers. Consequently, demixing can only occur on the length scale of a single block in the form of a micro-phase separation. These demixed blocks can then arrange into spatially periodic patterns. Depending mostly on temperature and the volume fractions of the two different blocks, lamellar phases, hexagonally arranged cylinders, bcc or fcc (in solutions) structures, and more complicated textures are obtained \cite{meier1969theory, leibler1980theory,ohta1986equilibrium,fredrickson1987fluctuation,bates1990block, fredrickson1996dynamics,strobl1997physics,hamley1998physics,hamley2001structure, teramoto2002double,nonomura2003formation,yamada2004kinetics}. 
For approximately equal volume fractions of the two blocks, a lamellar micro-phase separated state emerges. It features a layered arrangement of the different blocks and was depicted in Fig.~\ref{fig_lengths}. Patterns expected for two-dimensional arrangements of linear diblock copolymers are depicted in Fig.~\ref{fig_microphase}. 

When such materials are exposed to shear, an orientation of the periodic structures with respect to the shear velocity and its gradient has been observed in numerous experimental studies \cite{koppi1992lamellae,winey1993interdependence,balsara1994shear, balsara1994insitu,patel1995shear,zhang1995frequency,zhang1995symmetric, zhang1996annealing,fredrickson1996dynamics,zhang1997symmetric,maring1997threshold, wiesner1997lamellar,chen1997pathways,chen1998flow,pople1999shear,wang1999ordering, leist1999double,zipfel1999shear,wang2000birefringence,hamley2000effect, hamley2001structure,langela2002microphase}; see Fig.~\ref{fig_linear-shear-flow} for the shear geometry. The most illustrative example is the one of lamellar layered textures \cite{koppi1992lamellae,winey1993interdependence,balsara1994shear, balsara1994insitu,patel1995shear,zhang1995frequency,zhang1995symmetric, zhang1996annealing,fredrickson1996dynamics,zhang1997symmetric,maring1997threshold, wiesner1997lamellar,chen1997pathways,chen1998flow,pople1999shear,leist1999double, zipfel1999shear,hamley2000effect,hamley2001structure,langela2002microphase}. 
For a simple planar linear shear flow, three principal orientations of the lamellae are possible as illustrated in Fig.~\ref{fig_bcp_shear}: 
\begin{figure}[t]
\centerline{\includegraphics[width=\textwidth]{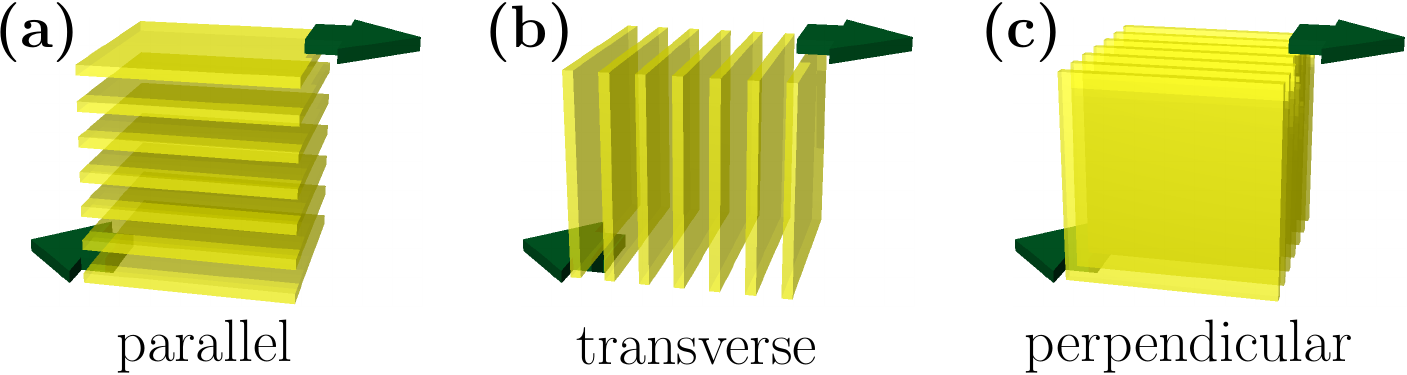}}
\caption{Illustration of the three principal orientations of a lamellar sheet-like structure under an imposed planar shear: (a) parallel, (b) transverse, and (c) perpendicular.}
\label{fig_bcp_shear}
\end{figure}
a ``parallel'' orientation with the lamellae normals parallel to the direction of the shear gradient; a ``transverse'' orientation with the lamellae normals parallel to the direction of the shear flow; and a ``perpendicular'' orientation with the lamellae normals perpendicular to the two previous cases, i.e.\ parallel to the vorticity direction. 

Interestingly, when a steady linear shear flow 
was applied to a lamellar block copolymer sample, the signature of the transverse orientation was observed at low shear rates \cite{pople1999shear}. This is surprising because it leads to a significant distortion of the lamellar layers. Switching to higher shear rates, a reorientation process occurred with the layer normals now lying in the plane spanned by the directions of the shear gradient and the vorticity \cite{pople1999shear}. This implies a combination between the parallel and the perpendicular alignment, which was also observed in other studies \cite{balsara1994shear,balsara1994insitu}. However, the steady shear could not induce a perfect reorientation. Even more, the steady shear led to defects in previously well-aligned lamellar samples for example in the form of focal conical textures \cite{winey1993interdependence}. It turned out that large-amplitude oscillatory shear was much more effective to align these samples than steady shear flow. Thus the orientational effects of oscillatory shear were studied extensively.  

For large-amplitude oscillatory shear, reorientation transitions in lamellar phases were observed by varying the shear frequency or the shear amplitude \cite{koppi1992lamellae,balsara1994insitu,patel1995shear,zhang1995frequency,zhang1995symmetric, zhang1996annealing,zhang1997symmetric,maring1997threshold, wiesner1997lamellar,chen1997pathways,chen1998flow,leist1999double, zipfel1999shear,hamley2001structure,langela2002microphase}. Even a ``double-flip'' reorientation process can be found for increasing shear frequency. At low frequencies, a parallel orientation is often obtained \cite{balsara1994shear,balsara1994insitu,zhang1995frequency,zhang1996annealing,zhang1997symmetric, maring1997threshold,wiesner1997lamellar,leist1999double,zipfel1999shear}. An increase in the shear frequency can lead to a flip to the perpendicular orientation \cite{zhang1995frequency,zhang1996annealing,zhang1997symmetric, maring1997threshold,wiesner1997lamellar,leist1999double,zipfel1999shear}. Very high frequencies, however, can induce a second flip back into the parallel geometry \cite{patel1995shear,zhang1995frequency,maring1997threshold,wiesner1997lamellar,leist1999double}. Despite the significant distortion of the lamellae, the transverse orientation was also observed in the case of oscillatory shear \cite{zhang1995symmetric,wiesner1997lamellar,chen1997pathways}. 

\begin{figure}
\centerline{\includegraphics[width=9.cm]{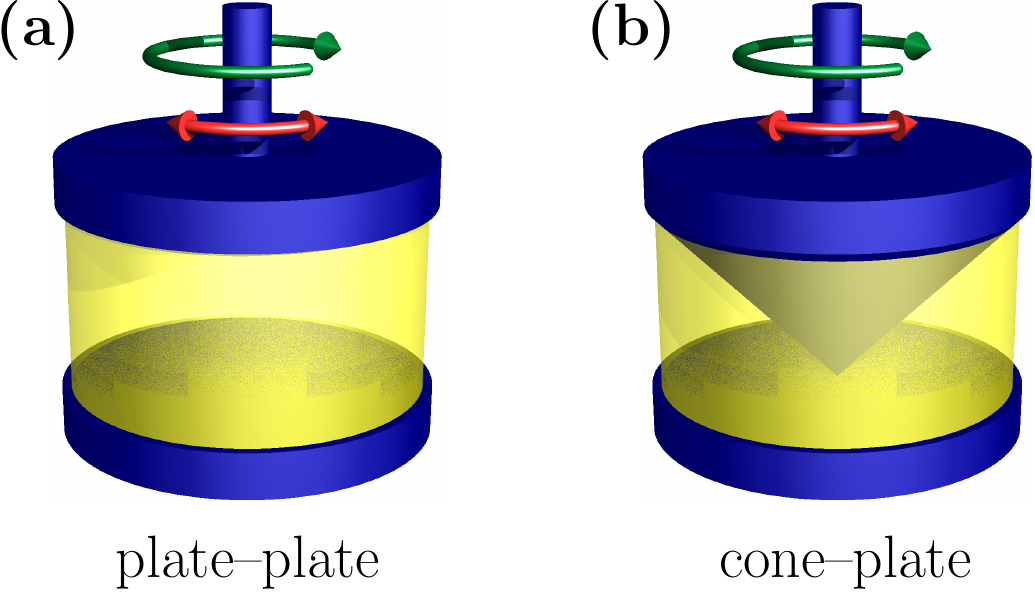}}
\caption{Schematic illustration of two principal types of shear rheometer. (a) In the plate-plate geometry, the sample is sandwiched between two parallel plates, one of them rotating relatively with respect to the other around their common axis. The displacements on the plate surface and thus the shear amplitude increase linearly from the rotation axis in the center to the outside. (b) A cone-plate geometry counteracts this effect because the sample thickness likewise increases linearly from the central rotation axis to the outside. In both cases, steady shear (darker single-headed arrow) or oscillatory shear (brighter double-headed arrow) can be applied. The arrows also coincide with the local direction of the shear velocity, see Fig.~\ref{fig_linear-shear-flow}. 
}
\label{fig_rheometer}
\end{figure}

Experimentally, the investigations can be performed in a Couette cell or using rheometric devices as displayed in Fig.~\ref{fig_rheometer}. 
In the plate-plate geometry, the sample is sandwiched between two parallel plates that rotate relatively with respect to each other around their common axis. At the surface of the rotating plate, the displacement increases linearly with increasing distance from the rotation axis in the center. Thus effects of different shear amplitude, defined here as displacement versus sample thickness, can be investigated simultaneously. A cone-plate geometry counteracts this effect. Here, one plate is replaced by a cone, leading to a linearly increasing sample thickness from the rotation axis in the center to the outside. Generally, the principal lamellar orientation in the sample cell can for instance be determined by small-angle x-ray or neutron scattering \cite{koppi1992lamellae,winey1993interdependence,balsara1994shear, balsara1994insitu,patel1995shear,zhang1995frequency,zhang1995symmetric, zhang1996annealing,fredrickson1996dynamics,zhang1997symmetric,maring1997threshold, wiesner1997lamellar,chen1997pathways,chen1998flow,wang1999ordering, leist1999double,zipfel1999shear,pople1999shear,hamley2000effect, hamley2001structure,langela2002microphase}. From the results of different beam directions, conclusions can be drawn about the orientational state. 

Understanding the underlying reasons for the orientational behavior under shear is a complicated task and also material dependent. Several different effects connected to various length scales play together and determine the overall behavior. For example, the boundary of the sample cavity can have an aligning effect on the structure \cite{boker2002large}; the micro-phase separated textures are distorted by shear deformations, leading to elastic strains \cite{kawasaki1986phase,yamada2006elastic,tamate2008structural}; structural fluctuations can couple to shear deformations \cite{bruinsma1992shear}; viscosity contrasts are present in each sample due to the stacking of the molecules and due to the material properties of the different blocks \cite{patel1995shear,zhang1995frequency}; the molecules are elongated along the layer normal, which sets a preferred path of molecular motion in particular for entangled polymers at higher molecular weights \cite{zhang1995symmetric}; shear can induce a tilting of the elongated chains away from the layer normal, leading to a restoring torque \cite{chen1997pathways,auernhammer2000undulation,auernhammer2002shear,soddemann2004shear}; and as we have seen in the previous section on low-molecular-weight liquid crystals, elongated molecules by themselves can already show various types of dynamic alignment behavior. Apparently, different frequencies and amplitudes of the applied shear can be used to preferably address some of these ingredients. In this way, their relative impact on the collective dynamic response can be modified and the overall appearance of the sample can be tuned. 

Other materials like common smectic (layered) low-molecular-weight liquid crystals \cite{safinya1991nematic,bruinsma1992shear,larson1993rheology,panizza1995effects}, see Fig.~\ref{fig_lc-phases}~(c), or lamellar phases of surfactant solutions \cite{bruinsma1992shear,berghausen1998shear,zipfel1999influence, zipfel2001cylindrical,mortensen2001structural,richtering2001rheology} can similarly feature orientational effects due to their ``sheet-like'' structure. 
Reorientations under shear can also be observed in structural phases different from the lamellar one \cite{fredrickson1996dynamics,daniel2000effect,wang2000birefringence,hamley2000effect, hamley2001structure,angelescu2004macroscopic}. Besides, for hexagonally arranged cylinders as well as for spheres ordered in bcc- and fcc-textures, a sliding of whole structural sheets over each other was suggested \cite{doi1993anomalous,ohta1993anomalous,koppi1994epitaxial, berret1996shear,molino1998identification,daniel2000effect, daniel2001nonlinear,hamley2001structure}. 

Applying shear to block copolymer systems is an issue of high practical relevance. Due to the reorientation effects, it can serve to heal multi-domain textures and obtain a monodomain structure \cite{winey1993morphology,winey1993interdependence,larson1993rheology, molino1998identification,chen1998flow,wang1999ordering}. Apart from shear, alignments can also be induced by other routes, for example by applying external electrical fields to materials that show a dielectric contrast between their blocks \cite{amundson1991effect,amundson1993alignment,amundson1994alignment, gurovich1994microphase,gurovich1995why,boker2002microscopic,boker2002large, kyrylyuk2002lamellar,zvelindovsky2003comment,boker2003electric, xu2005electric,olszowka2006large,lyakhova2006kinetic,olszowka2008control,liedel2012beyond}. From a technological point of view, block copolymers possess a high potential for applications in the nano-sciences \cite{hamley2003nanostructure,park2003enabling,darling2007directing}: they provide regular periodic arrays of tunable material properties on the molecular length scales of the different blocks. This can be exploited for nano-lithographic processes, to produce regular arrays of nano-dots or nano-wires, or to synthesize nanoporous materials. Monodomain structures are often beneficial for this purpose. Therefore understanding and clarifying the possible routes to generate them is a task of central importance.

\subsection{Vesicles in shear flow}

In the above cases, we considered the properties of bulk-filling phases under shear flow. This is different from the following examples, where we focus our attention on localized finitely-sized objects such as 
closed membranes and vesicles. In contrast to a bulk-filling periodic phase, complete rotations of a limited object in a liquid environment is easily possible. Together with the deformability of the considered entities, this leads to new dynamic states. 

\begin{figure}[t]
\centerline{\includegraphics[width=\textwidth]{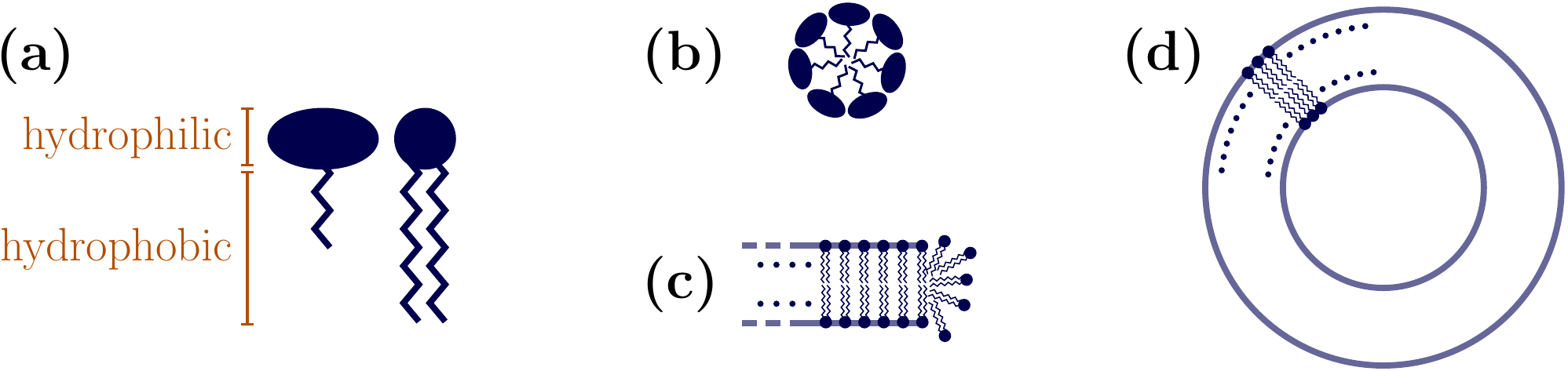}}
\caption{Schematic examples of amphiphilic (surfactant) molecules and illustration of some of the aggregates that they can form in an aqueous solution. (a) Two extreme examples of typical structures of amphiphilic molecules are displayed. Both have a hydrophilic (``water-loving'') head group, typically in the form of some polar or charged molecular groups. The tails are usually based on hydrocarbon chains and hydrophobic (``water-fearing''). In the first example, the head group is relatively bulky and there is only one shorter tail. In the second example, there are two tails of approximately the same diameter as the head group. The formation of aggregates in an aqueous environment in both cases is possible above a certain concentration. It leads to a screening of the hydrophobic tails from the aqueous environment by the hydrophilic head groups. (b) Due to their geometric properties, the molecules of the first kind favor the formation of micellar aggregates of high surface curvature. (c) The molecules of the second kind, due to their geometric shape, are suitable to generate flat bilayer membranes. Thermal fluctuations are of course present but not indicated here. To close the membranes at their outer ends, some arrangements of high surface curvature and thus higher energy are necessary. (d) For large membranes, this can be avoided by forming closed surfaces in the form of vesicles. The ratio of their radius versus the membrane thickness in reality can be much larger than depicted here, so that locally the surface curvature of the vesicle becomes negligible. See also the indicated sizes in Fig.~\ref{fig_lengths}.}
\label{fig_micellar-aggregates}
\end{figure}

The systems that we focus our attention on are based on amphiphilic molecules in an aqueous environment \cite{lipowsky1991conformation,seifert1997configurations}. Such molecules typically feature a hydrophilic  (``water-loving'') head group and one or more hydrophobic (``water-fearing'') tails, usually based on hydrocarbon  chains; see Figs.~\ref{fig_lengths} and \ref{fig_micellar-aggregates}~(a). 

To reduce the contact with the surrounding water molecules, the hydrophobic chains above a critical concentration tend to form intermolecular aggregates. This is again a sort of micro-phase separation. A simple example is depicted in Fig.~\ref{fig_micellar-aggregates}~(b) in the form of a micelle. In this aggregate, the hydrophobic chains can hide away from the water molecules and form the core of the micelle. They are screened from the water molecules by the outward-pointing hydrophilic head groups that search the contact to the aqueous environment. 

Both, the concentration and the geometry of the single amphiphilic molecules determine which sort of intermolecular aggregates form. 
If the head groups are very bulky and if there is only one short chain per molecule, the formation of spherical micelles as depicted in Fig.~\ref{fig_micellar-aggregates}~(b) is supported. They have a high curvature of their surface. 
In contrast to that, if the hydrophilic head and the hydrophobic tail groups have approximately the same diameter, the formation of aggregates with low curvature of the surface is beneficial. This situation can for example be observed for bilayer membranes. In this case, the amphiphilic molecules are arranged in two adjacent layers stacked on each other as illustrated in Figs.~\ref{fig_lengths} and \ref{fig_micellar-aggregates}~(c). Here, the hydrophobic chains point to the inside of the two stacked layers and expel the water molecules. The hydrophilic heads point to the outside and screen the inner region from the aqueous environment. 

At its lateral rims, the bilayer membrane would either have to expose some of the hydrophobic chains to the aqueous environment or show an unnaturally high curvature of its surface. 
To avoid such high-energy regions at the lateral rims, the bilayer can bend on length scales large compared to its thickness and form a closed surface. The resulting closed bilayer membrane is called a vesicle and also indicated in Figs.~\ref{fig_lengths} and \ref{fig_micellar-aggregates}~(d). Typically, an aqueous solution is found on both sides of the bilayer membrane, i.e.\ outside the vesicle and also enclosed on the inside. 

Interesting dynamic states arise due to the typical constraints following from the architecture of vesicles. On experimental time scales, their inner volume is generally conserved \cite{deuling1976curvature,seifert1997configurations}. Most importantly, also the surface area of vesicles remains practically constant \cite{deuling1976curvature,seifert1997configurations}. This is in contrast to simple liquid drops and qualitatively affects their shape and dynamics \cite{danker2007rheology,danker2008rheology}. Apart from that, a possible viscosity contrast between the inner encapsulated fluid and the outer fluid impacts the dynamic behavior 
\cite{biben2002advected,biben2003tumbling,beaucourt2004steady,rioual2004analytical,biben2005phase, kantsler2005orientation,kantsler2006transition,mader2006dynamics,  misbah2006vacillating,noguchi2007swinging,lebedev2007dynamics, danker2007dynamics,danker2007rheology,mader2007coupling, finken2008two,kessler2009elastic,messlinger2009dynamical,deschamps2009phase, noguchi2009swinging,ghigliotti2010rheology, finken2011micro,farutin2011symmetry,biben2011three,zabusky2011dynamics, farutin2012squaring,gires2012hydrodynamic,farutin2013analytical,lamura2013dynamics}. 
The same is true for a possible friction between the two layers of amphiphilic molecules that form the bilayer membrane and may slide over each other during certain dynamic modes \cite{seifert1997configurations}. 

\begin{figure}
\centerline{\includegraphics[width=6.cm]{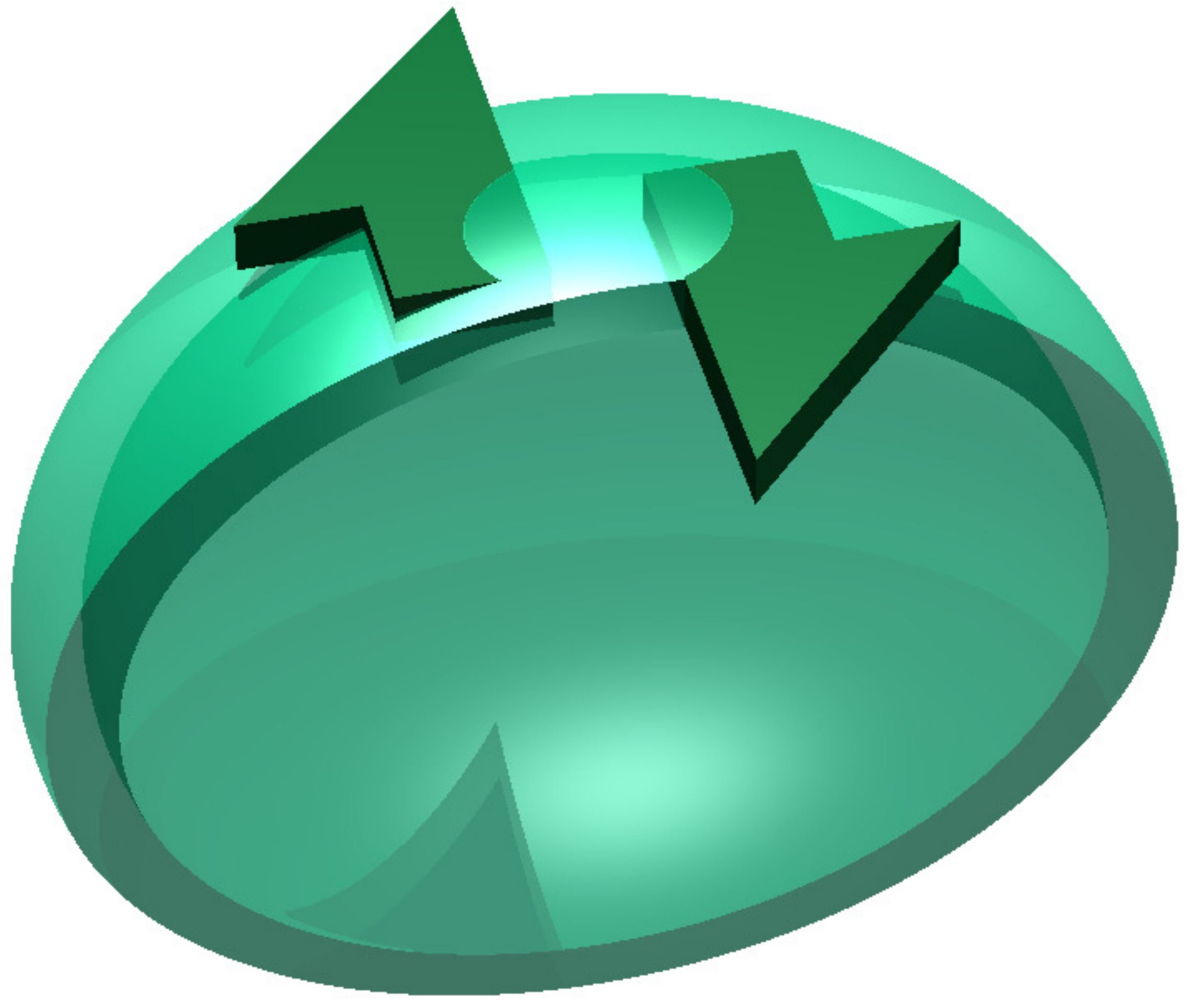}}
\caption{A schematic vesicle is half-cut to show the membrane location. Within the plane of the membrane, a shear stress is applied to the membrane as indicated by the two arrows. If the membrane does not sustain an in-plane shear stress but behaves like a two-dimensional liquid, the vesicle is called a ``fluid vesicle''. Likewise, it is referred to as a ``viscous vesicle'' when the effect of the intra-membrane viscosity is emphasized. In contrast to that, if the membrane does sustain in-plane shear stresses, the vesicle is called an ``elastic vesicle''.
}
\label{fig_inplaneshear}
\end{figure}

\begin{figure}[t]
\centerline{\includegraphics[width=\textwidth]{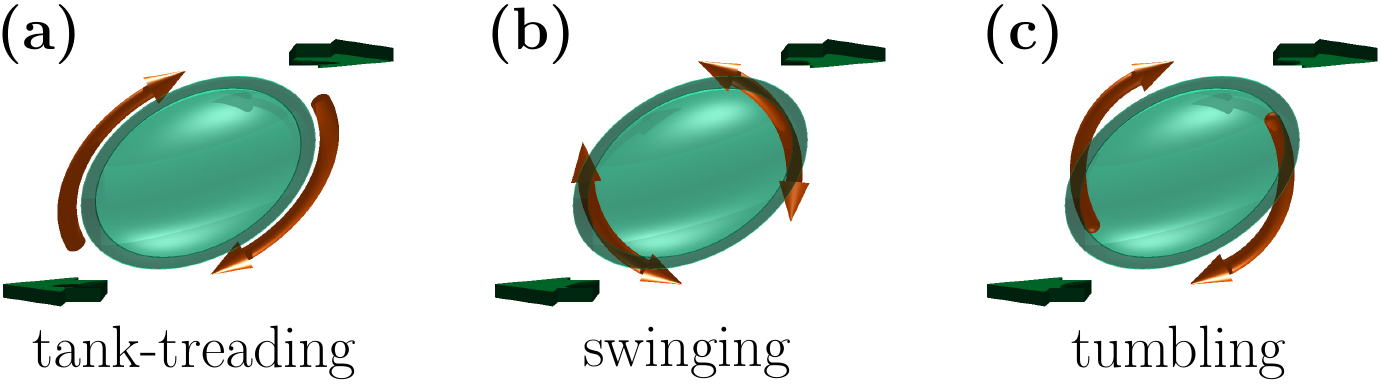}}
\caption{Illustration of three dynamic modes of a vesicle under an imposed planar shear flow. The vesicle is indicated by the ellipsoid. (a) While ``tank-treading'', the vesicle takes a steady finitely inclined orientation with respect to the flow direction within the shear plane. Only the vesicle membrane performs a continuous rotational motion along its contour. (b) During ``swinging'' motions (also ``trembling'' or ``vacillating breathing'') the long axis of the vesicle oscillates within the shear plane around an orientation that is usually finitely inclined with respect to the flow direction. (c) When ``tumbling'' the vesicle performs full rotations in the shear plane.}
\label{fig_vesicle_shear}
\end{figure}

On the one hand, the membrane may correspond to a two-dimensional liquid that does not sustain in-plane shear stresses, see Fig.~\ref{fig_inplaneshear}. 
Then the vesicles are called ``fluid vesicles'' \cite{lipowsky1991conformation,seifert1997configurations}. We concentrate on the dynamics of such objects in simple linear planar shear flow as introduced in Fig.~\ref{fig_linear-shear-flow}. 
If there are no viscosity contrasts between the liquid on the inside and on the outside of the vesicle, a dynamic state called ``tank-treading'' is forecast by modeling and observed experimentally \cite{kraus1996fluid,haas1997deformation,biben2002advected, biben2003tumbling,beaucourt2004steady,kantsler2005orientation,mader2006dynamics, noguchi2007swinging,lebedev2007dynamics,danker2007dynamics,danker2008rheology, messlinger2009dynamical,deschamps2009phase,biben2011three,zabusky2011dynamics}. It is depicted in Fig.~\ref{fig_vesicle_shear}~(a): due to the elongational part of the shear flow the vesicle becomes oriented within the shear plane, but with a constant inclination angle to the flow direction; the rotational part of the shear flow induces a rotational motion of the membrane around the vesicle interior. 
Increasing the viscosity contrast at not too high shear rates, a bifurcation to a different type of motion called ``tumbling'' is predicted from modeling and observed experimentally \cite{biben2002advected,biben2003tumbling,beaucourt2004steady, rioual2004analytical,kantsler2006transition,mader2006dynamics, noguchi2007swinging,lebedev2007dynamics,danker2007dynamics,danker2008rheology, messlinger2009dynamical,deschamps2009phase,biben2011three,zabusky2011dynamics}. It is illustrated in Fig.~\ref{fig_vesicle_shear}~(c). Here, the vesicle as a whole rotates in the shear flow, instead of only its membrane tank-treading around its interior. 
Thermal fluctuations can induce tumbling already for vanishing viscosity contrast at low shear rates \cite{abreu2012effect}. 
At higher shear rates, ``swinging'' (also ``trembling'' or ``vacillating breathing'') was observed instead of pure tumbling at least for intermediate viscosity contrasts \cite{kantsler2006transition,misbah2006vacillating,noguchi2007swinging, danker2007dynamics,lebedev2007dynamics,danker2008rheology, messlinger2009dynamical,deschamps2009phase,biben2011three, zabusky2011dynamics}. In this state, the long axis of the vesicle does not perform full rotations as in the tumbling mode, but only oscillates up and down as indicated in Fig.~\ref{fig_vesicle_shear}~(b). 
Summarizing, a tendency towards the right-hand side in the series of panels (a)--(c) in Fig.~\ref{fig_vesicle_shear} is generally supported by an increasing viscosity contrast between the outside and the inside of the vesicle and a decreasing shear rate. Yet, the situation can be more involved in each individual case. 
More complex dynamic modes were identified recently \cite{biben2011three,farutin2012squaring}, and the qualitative influence that thermal fluctuations can have were pointed out \cite{deschamps2009dynamics,zabusky2011dynamics,levant2012amplification,abreu2013noisy}. 

Apart from that, the impact of an explicitly viscous membrane was investigated \cite{noguchi2004fluid,noguchi2005dynamics,noguchi2005vesicle,noguchi2007swinging}. These ``viscous vesicles'' generally show a reduced inclination angle during tank-treading. Increasing the membrane viscosity can further induce a transition from tank-treading to tumbling or swinging.

On the other hand, the membrane may for example be polymerized so that it  does sustain in-plane shear stresses \cite{lipowsky1991conformation,seifert1997configurations}, see Fig.~\ref{fig_inplaneshear}. Such objects are referred to as ``elastic vesicles'' \cite{noguchi2005shape}. 
The basic features of the dynamics of fluid vesicles in linear shear flow are recovered both from modeling and experiments: a tumbling motion at low shear rate and a tank-treading motion at high shear rate emerge \cite{abkarian2007swinging,kessler2008swinging,kessler2009elastic, noguchi2009swinging,abreu2012effect,koleva2012deformation}. However, during the tank-treading motion, the orientation angle with respect to the flow direction oscillates due to the shape-memory \cite{abkarian2007swinging,kessler2008swinging,kessler2009elastic, noguchi2009swinging,abreu2012effect,koleva2012deformation}. This is referred to as ``swinging''. Furthermore, if the viscosity inside becomes large compared to the outside viscosity, an intermittent type of motion is observed as a combination between the swinging tank-treading and the tumbling dynamics \cite{abkarian2007swinging,kessler2008swinging,kessler2009elastic, noguchi2009swinging,abreu2012effect}. 

\begin{figure}
\centerline{\includegraphics[width=\textwidth]{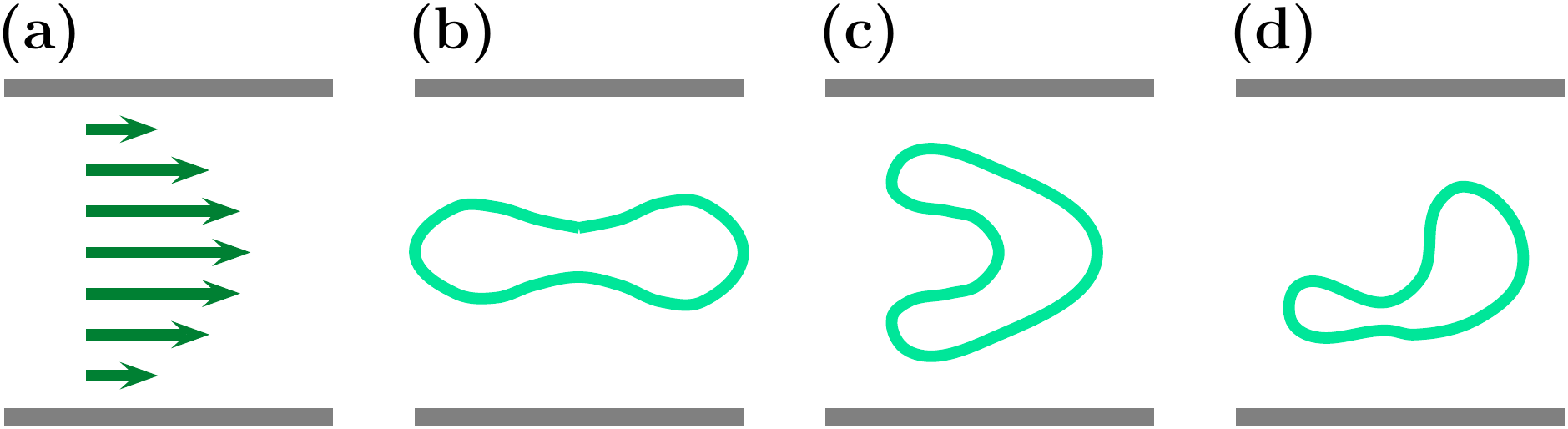}}
\caption{Schematic illustration of different possible shapes adopted by vesicles in a Poiseuille flow. The snapshots have to be considered as two-dimensional cross-sectional views through the center of a tube-like channel. (a) Profile of the Poiseuille flow across the channel. (b) Elongation of a discoidal vesicle along the fluid flow. (c) Parachute-like shape and orientation. (d) Slipper-like configuration. Gray bars indicate the channel walls.}
\label{fig_poiseuille}
\end{figure}

From a biological point of view, elastic vesicles form an important topic because red blood cells fall into this category \cite{abkarian2007swinging,dupire2010chaotic}. To understand the transport of red blood cells in blood vessels, the dynamics of elastic vesicles subjected to Poiseuille flows in cylindrical tubes, see Fig.~\ref{fig_poiseuille}~(a), is studied extensively. 
Whereas fluid vesicles of initially discoidal shape typically elongate in the Poiseuille flow, see Fig.~\ref{fig_poiseuille}~(b), or show axisymmetric ``parachute'' shapes in strong flow fields, see Fig.~\ref{fig_poiseuille}~(c), elastic vesicles usually adopt a parachute-like deformation or an asymmetric ``slipper''-like shape as indicated in Fig.~\ref{fig_poiseuille}~(d) \cite{noguchi2005shape,kaoui2009why}. Many vesicles together in a tube were found to adopt disordered distributions with discoidal shapes, ordered states of aligned parachutes, as well as slippers arranged in a zigzag-like fashion \cite{mcwhirter2009flow,mcwhirter2011deformation,mcwhirter2012ordering} depending on the flow velocity and packing fraction. Hydrodynamic lift forces keep the vesicles away from the walls \cite{seifert1999hydrodynamic,cantat1999lift,lorz2000weakly,sukumaran2001influence, abkarian2002tank,callens2008hydrodynamic,messlinger2009dynamical}. 
In addition to that, further dynamic effects have been discovered, for example a wrinkling deformation of the elastic membrane in shear flows \cite{walter2001shear,finken2006wrinkling,finken2011micro}. 

Analytical approaches included the above-mentioned constraints following from the vesicle architecture and were able to analyze the resulting dynamic properties \cite{keller1982motion,seifert1999hydrodynamic,rioual2004analytical, finken2006wrinkling,misbah2006vacillating, noguchi2007swinging,lebedev2007dynamics,skotheim2007red,mader2007coupling, danker2007dynamics,danker2007rheology,finken2008two,danker2008rheology, kessler2009elastic,noguchi2009swinging,danker2009vesicles,noguchi2009swinging, farutin2011symmetry,finken2011micro,abreu2012effect, gires2012hydrodynamic,abreu2013noisy}. 
Due to the curved vesicle surfaces, elements from differential geometry are necessary to perform these calculations. Apart from the analytical approaches, two simulation routes were particularly successful to investigate the dynamic behavior of vesicles in a liquid environment. One of them is particle-based and referred to as multi-particle collision dynamics \cite{noguchi2004fluid,noguchi2005dynamics,noguchi2005shape,noguchi2005vesicle, noguchi2006dynamics,finken2008two,messlinger2009dynamical,mcwhirter2009flow, noguchi2010dynamics,mcwhirter2011deformation, mcwhirter2012ordering,lamura2013dynamics}. It includes the surrounding and encapsulated fluid particles, modeling the many-particle interactions in an effective way. Still, it correctly obeys the physical conservation laws. The second route is a phase field model \cite{biben2002advected,biben2003tumbling,beaucourt2004steady,biben2005phase, jamet2007towards,jamet2008toward,ghigliotti2010rheology}. It is based on an auxiliary continuous scalar order parameter field. Two different constant values of this scalar order parameter are used to distinguish between the inside and the outside of the vesicle, for example $-1$ and $+1$, respectively. At the location of the membrane, the order parameter field rapidly but continuously changes between these two values. In this way, the boundary of the vesicle can easily be identified. It can be effectively tracked by simply advecting the order parameter field with the hydrodynamic flow of the system. 

A third route to the problem was outlined in the form of a density-field approach \cite{menzel2011density}. For periodic bulk-filling textures, such a procedure had for example been introduced for block copolymer structures \cite{ohta1986equilibrium,kawasaki1986phase,bahiana1990cell, ohta1993anomalous,doi1993anomalous,drolet1999lamellae,nonomura2001growth, chen2002lamellar,yamada2006elastic,yamada2007interface, tamate2008structural} or for crystalline materials via the phase field crystal approach \cite{elder2002modeling,elder2004modeling,berry2006diffusive, athreya2007adaptive,provatas2007using,tupper2008phase, chan2009molecular,chan2010plasticity}. 
These kinds of formalism are based on a single scalar real order parameter naturally given by the density field. They apply on diffusive time scales. 
A corresponding free energy density to obtain localized nonperiodic structures was introduced in a phenomenological way and the resulting density-field description was coupled to hydrodynamic flow fields \cite{menzel2011density}. 
In this way, in two spatial dimensions, characteristics of tank-treading were observed in linear shear flow, and folding into the parachute form as well as an elongation in a Poiseuille channel flow were obtained \cite{menzel2011density}. 
A challenging task for the future is to derive such density-field approaches starting from a more microscopic and particle-resolved basis.

\newpage
\section{Active soft matter}

So far we have been concerned with ``passive'' systems that at most were driven from outside. We now turn to ``active'' or ``self-driven'' materials. 
It is not always straightforward to clearly distinguish between such ``active'' soft matter \cite{ebbens2010pursuit,ramaswamy2010mechanics,romanczuk2012active, cates2012diffusive,marchetti2013hydrodynamics} and ``externally driven'' soft matter as considered in the previous section. 
The term ``active'' generally implies that at least some components of the system feature an individual ``internal drive''. 
Unfortunately, this definition requires a sort of restricted point of view. There is always a residual coupling of the internal driving mechanism to the outside world. We understand this statement by considering examples of active self-propelled particles. 

One case of particularly well studied artificial self-propelled microswimmers
is given by dispersed colloidal Janus particles \cite{paxton2004catalytic,howse2007self,jiang2010active,volpe2011microswimmers,buttinoni2012active, theurkauff2012dynamic,buttinoni2013dynamical}. 
In general, colloids are made of mesoscopic particles or droplets of sizes between $1$~nm and $10-100$~$\mu$m, see also Fig.~\ref{fig_lengths}, dispersed in another substance \cite{dhont1996introduction,palberg1999crystallization,ivlev2012complex}. Everyday examples are ink, paint, and milk. To stabilize colloidal suspensions, coagulation of the particles must be hindered. For this purpose, the refractive index of the colloidal particles and their environment should be matched to each other to reduce the van-der-Waals interaction. Furthermore, electric charges on the particles can lead to stabilization due to repulsion. Apart from that, also steric stabilization is possible, when the colloidal particles are covered by polymer brushes: close contact between the particles would restrict the possible configurations of the polymer chains, which induces an effective entropic short-range repulsion. Micron-sized colloidal particles offer the advantage that they can easily be tracked by optical methods. 

There are different ways to cover one side of the surface of a colloidal particle by a material of significantly different properties than the other side \cite{hong2006simple,walther2008janus,jiang2010janus,chen2012janus}. Such colloidal particles are then called Janus particles. This can be exploited, e.g., to selectively heat the stronger absorbing side by laser light irradiated from outside \cite{jiang2010active}. As a result, a self-propelled motion can be induced via thermophoretic effects: the particle via its asymmetric light absorption generates a temperature gradient around itself; this leads to hydrodynamic stress gradients that can set the surrounding fluid into motion, which effectively leads to particle motion \cite{golestanian2007designing}. A further possibility is to enforce demixing of a binary fluid on the heated side \cite{volpe2011microswimmers,buttinoni2012active,buttinoni2013dynamical} or to selectively catalyze chemical reactions only on the surface of one side of the particles \cite{howse2007self,theurkauff2012dynamic}. The arising concentration gradients can lead to self-propulsion via diffusiophoretic effects analogous to the case of thermophoresis \cite{golestanian2007designing}. However, the fuel to drive these chemical reactions needs to be provided from outside, as must be the irradiated laser light in the case of thermophoretic motion. 

Granular hoppers on a vibrating plate form another model system to study self-propelled motions \cite{narayan2007long,kudrolli2008swarming,deseigne2010collective,deseigne2012vibrated}. Here, typically the motion in the plane parallel to the surface of the plate is observed. However, this motion is initiated and driven from outside by the externally tuned vibration. 
Finally, even such units as swimming bacteria \cite{berg1972chemotaxis,wada2007model,suematsu2011localized,drescher2011fluid}, crawling cells and amoebae \cite{rappel1999self,szabo2006phase,peruani2012collective}, or molecular motors that can set filaments into motion \cite{schaller2010polar,sanchez2012spontaneous} are not completely independent of their surroundings. At some point they need to take up the food or fuel provided by their environment. 

We thus further narrow our present definition of active motion. The internal drive of the active units may be induced from outside. However, their motion must break a symmetry of the system. In other words, there must be a degree of freedom in the direction of induced motion that is not prescribed by the external driving. For example, a vertical vibration of the substrate can induce the motion of granular hoppers on a horizontal plate. However, the direction of motion within the plane of the plate surface is not specified. In that sense, the active particles are free to choose their migration direction. 

In the following, we are mainly interested in collective effects that arise when many active and self-propelled particles act together. Several aspects of this complex behavior seem to have general character and can already be understood from minimal model systems. Therefore, we increase the level of complexity of the systems under investigation. 
We start from the simplest case of point-like self-propelled particles featuring a constant self-propulsion velocity. Although this is an idealized situation, basic principles of the collective behavior are revealed. In a next step, we include systems where steric interactions occur, e.g.\ dry granular systems. After that, we outline the consequences of replacing the constant self-propulsion velocity by an active driving force. Artificial microswimmers like self-propelling colloidal Janus particles and biological swimmers in the form of bacteria are addressed. The influence of hydrodynamic interactions on their behavior is discussed. We then outline the role of particle deformability. Finally, we give examples of studies dedicated to the collective behavior of animals.

\subsection{Point-like self-propelled particles}

\begin{figure}[t]
\centerline{\includegraphics[width=.9\textwidth]{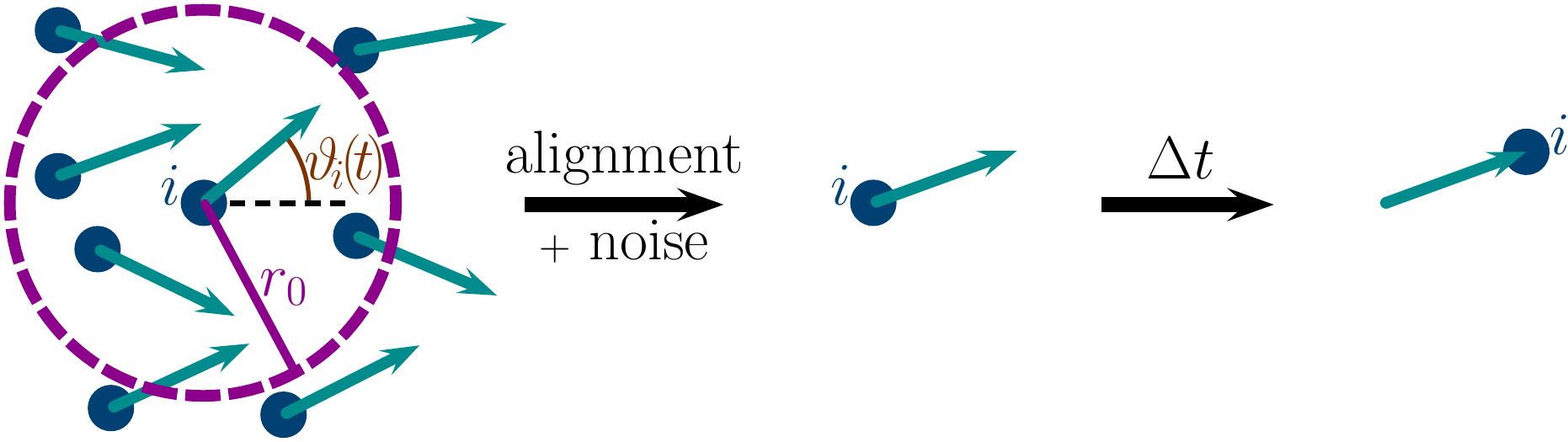}}
\caption{Illustration of the Vicsek model \cite{vicsek1995novel} in two spatial dimensions. All particles, marked by the dark bullets, self-propel with the same speed $v_0$. This is indicated by the velocity vectors attached to the bullets that all have the same magnitude. At each discrete numerical time $t$, the velocity orientation $\vartheta_i(t)$ of each particle $i$ is updated: it is set to the average velocity orientation of all  particles located within a distance $r_0$ from the $i$th particle; furthermore some orientational noise is added. During each time step $\Delta t$, the particles move along their velocity orientations by a discrete distance $v_0\Delta t$, here shown for a unit time step $\Delta t=1$.}
\label{vicsekmodel}
\end{figure}

The most prominent and fundamental model introduced in the context of idealized point-like particles is the one by Vicsek et al.\ \cite{vicsek1995novel}, which is illustrated in Fig.~\ref{vicsekmodel} and explained in the following. 
In this model, $N$ particles self-propel in a two-dimensional plane of periodic boundary conditions. As a key ingredient and major simplification, the magnitude of the individual velocities of all particles, i.e.\ their speed, is assumed to be equal to a constant $v_0$. It remains identical and unchanged for all times. The current orientation of the velocity vector of the $i$th particle can be parameterized by an angle $\vartheta_i$, $i=1,...,N$. This orientational angle $\vartheta_i$ is adjusted at each time step to the mean of the velocity orientations of all particles $j$ that are located within a spherical environment of radius $r_0$ around the $i$th particle:  
\begin{equation}\label{vicsek_vartheta}
\vartheta_i(t+\Delta t) = \langle \vartheta_j(t) \rangle_{r_0} + \eta_i(t).
\end{equation}
Here, $\langle...\rangle_{r_0}$ describes this average over all particles within the sphere of radius $r_0$ around the $i$th particle. $t$ marks the time and $\Delta t$ sets the discrete time step. Furthermore, the angular noise $\eta_i(t)$ is originally taken as a random number with uniform probability out of a centered interval at each time step. After each time step $\Delta t$, the particle positions $\mathbf{r}_i$ are updated as
\begin{equation}\label{eq:vicsek_rupdate}
\mathbf{r}_i(t+\Delta t) = \mathbf{r}_i(t)+\Delta t\, v_0 
\left(\begin{array}{c} \cos\left[\vartheta_i(t)\right] \\ \sin\left[\vartheta_i(t)\right] \end{array}\right). 
\end{equation}

\begin{figure}
\centerline{\includegraphics[width=9.cm]{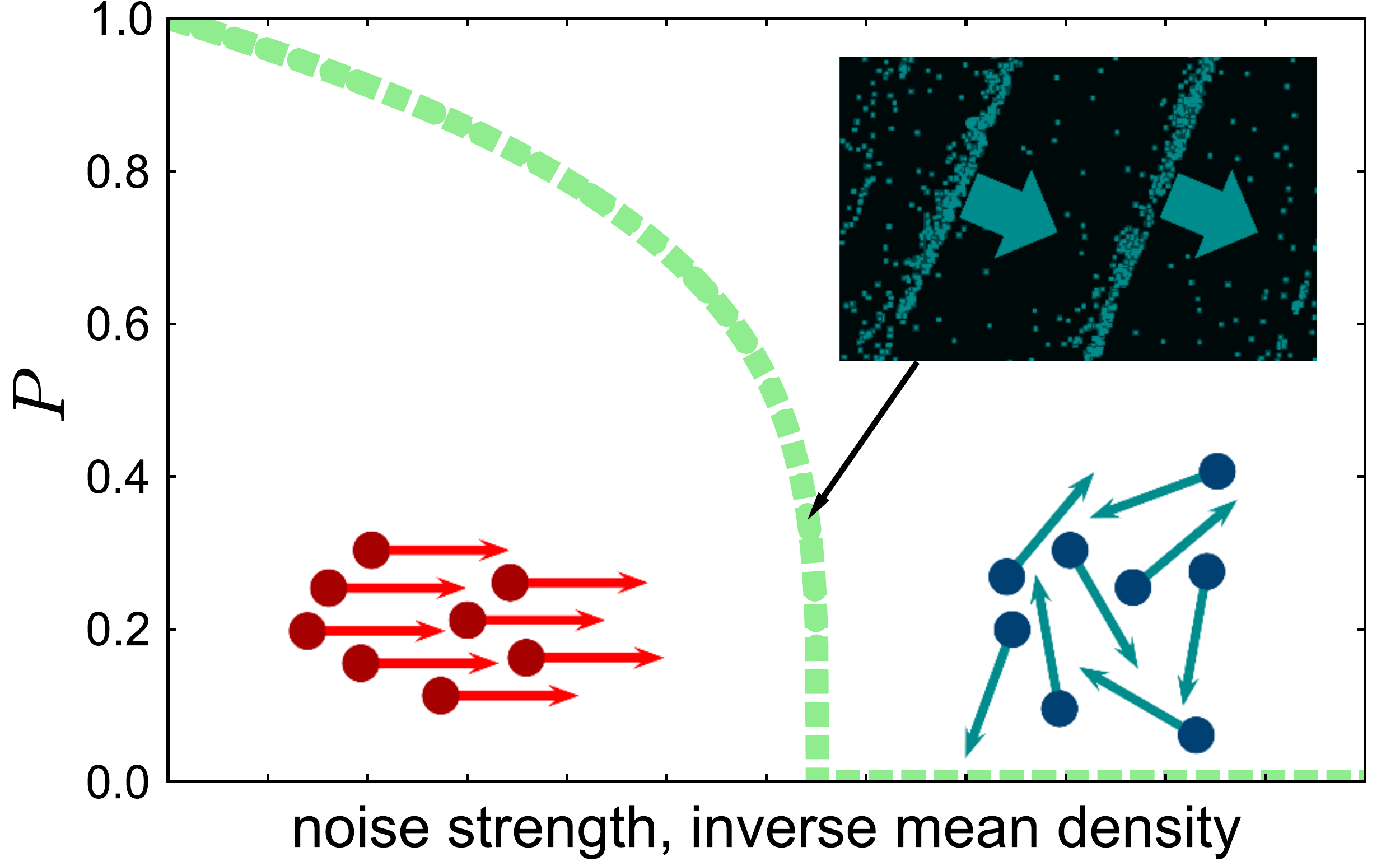}}
\caption{Schematic illustration of the order-disorder transition in the Vicsek model and its variants. The transition can be induced by increasing the characteristic noise amplitude or by decreasing the mean particle density. In the ordered state (bottom left inset) the particles on average migrate collectively into a common direction, whereas in the disordered state (bottom right inset) coherent particle motion does not occur. The order parameter $P$ is calculated from the magnitude of the sample-averaged velocity orientations, see Eq.~(\ref{eq:Pvicsek}). Close to the transition, spatial inhomogeneities are usually observed in the form of density bands that travel perpendicularly to their elongation direction through a diluted disordered background (top right inset).}
\label{fig_do-phase-transition}
\end{figure}

As a result of the competition between the alignment with the local environment and the stochastic noise, a phase transition is observed. This is a transition between a phase of disordered motion of zero net particle flux on the one hand and a phase of ordered collective motion on the other hand. 
In the second case, all self-propelled particles migrate on average collectively into the same direction. The two phases are schematically indicated in the bottom insets of Fig.~\ref{fig_do-phase-transition}. 

An order parameter to quantify this order-disorder transition is given by the magnitude of the sample-averaged velocity orientations, 
\begin{equation}\label{eq:Pvicsek}
P(t) 
  =  \left\|\frac{1}{N}\sum_{i=1}^N \left(\begin{array}{c} \cos\left[\vartheta_i(t)\right] \\ \sin\left[\vartheta_i(t)\right] \end{array}\right) \right\| ,
\end{equation}
where the velocity orientations are parameterized
by the orientation angles $\vartheta_i$ ($i=1,...,N$) [see Eqs.~(\ref{vicsek_vartheta}) and (\ref{eq:vicsek_rupdate}) as well as Fig.~\ref{vicsekmodel}]. The magnitude of the order parameter is $P=1$ in a completely ordered state of all particles collectively migrating into the same direction, and $P=0$ in the disordered state. As indicated in Fig.~\ref{fig_do-phase-transition}, the phase transition from the ordered to the disordered state of motion can be induced by
decreasing the mean particle density or increasing the characteristic noise amplitude. 

This transition was studied in detail over the past few years \cite{bertin2006boltzmann,chate2008collective,bertin2009hydrodynamic}. In simulations of large systems, it was found that the transition is of first order \cite{gregoire2004onset,bertin2006boltzmann,chate2008collective,bertin2009hydrodynamic}. The discontinuity in the transition is apparently related to spatial inhomogeneities that arise in the particle density around the transition point. Density bands emerge that tend to collectively migrate perpendicularly to their elongation \cite{chate2008collective,bertin2009hydrodynamic,mishra2010fluctuations,peruani2011polar} 
as indicated by the top inset in Fig.~\ref{fig_do-phase-transition}. Due to this directed collective migration, they can pick up further particles from the environment of disordered motion. Such a process further increases the density within the band. In this sense, a kind of self-supporting mechanism develops \cite{gopinath2012dynamical}. Also in real experiments, traveling density bands have been observed \cite{schaller2010polar}. 

The nature of these traveling density bands was discussed in the framework of solitons \cite{chate2008collective,bertin2009hydrodynamic}. Typically the density bands feature a sharp front and an extended tail. Recently, their behavior under head-on collisions is increasingly investigated for point- and non-point-like particles \cite{ihle2013invasion,yamanaka2014formation,tarama2014individual}. Penetration of colliding density bands and recovery after collision have been observed \cite{ihle2013invasion,yamanaka2014formation}. It has been demonstrated that they are obtained as different propagating solutions of conventional continuum models for self-propelled particle crowds \cite{toner1995long,toner1998flocks} in the form of multiple parallel density bands, single solitary bands, and single active droplets \cite{caussin2014emergent}. 

Several variants of the Vicsek model were pointed out and analyzed \cite{chate2008modeling}. For example, the effects of metric-free alignment interactions \cite{ginelli2010relevance} were discussed. In three spatial dimensions, the traveling density bands were recovered in the form of migrating density planes \cite{chate2008collective}. Other studies replaced the discrete nature of the dynamic equations Eqs.~(\ref{vicsek_vartheta}) and (\ref{eq:vicsek_rupdate}) by differential equations. In particular, the discrete averaging process $\langle...\rangle_{r_0}$ in Eq.~(\ref{vicsek_vartheta}) was replaced by a more continuous functional form \cite{peruani2008mean,peruani2010cluster,menzel2012collective}:
\begin{equation}
\frac{d\vartheta_i(t)}{dt} = -\frac{\partial U}{\partial\vartheta_i}+\Gamma_i(t), \qquad \frac{d\mathbf{r}_i}{dt}=v_0\left(\begin{array}{c} \cos\left[\vartheta_i(t)\right] \\ \sin\left[\vartheta_i(t)\right] \end{array}\right), \qquad i=1,\dots,N.
\end{equation}
For simplicity, the orientational noise $\Gamma_i(t)$ is assumed to result from a Gaussian white process. The continuous function $U$ is based on pairwise alignment interactions between the particles. For example, the functional form
\begin{equation}\label{eq_U}
U(\mathbf{r}_1,\dots,\mathbf{r}_N,\vartheta_1,\dots,\vartheta_N) = -\sum_{\substack{i,j=1\\ i<j}}^N \Theta\left(r_0-\|\mathbf{r}_i-\mathbf{r}_j\|\right)\cos(\vartheta_i-\vartheta_j)
\end{equation}
again leads to pairwise velocity alignment for particles closer to each other than the distance $r_0$. Here, $\Theta$ represents the Heaviside step function. 
The snapshot in the top inset of Fig.~\ref{fig_do-phase-transition} was obtained from a numerical calculation following a procedure along these lines. 

Using such an approach, the situation of a binary mixture was considered for different rules of the inter-species velocity alignment \cite{menzel2012collective}. Starting from the particle picture, continuum equations for the one-particle probability densities were derived \cite{menzel2012collective} within the Fokker-Planck framework \cite{risken1996fokker,zwanzig2001nonequilibrium,lee2010fluctuation,savel2003controlling}. 
Also macroscopic hydrodynamic-like continuum equations for the macroscopic order parameters were obtained and analyzed \cite{menzel2012collective}. 
Interestingly, when the above-mentioned density bands appear in one species, they can induce spatial heterogeneities in the other species via the inter-species coupling \cite{menzel2012collective}. 
These results may be interesting for the dynamics of biofilm formation. Biofilms are surface- or interface-attached communities of microorganisms \cite{otoole2000biofilm,stoodley2002biofilms}, in nature usually composed of more than one species \cite{an2006quorum,elias2012multi}. Often at least part of the microorganisms in a film adopts a motile state \cite{kearns2005cell,an2006quorum,vlamakis2008control,veening2010gene}. Their collective behavior should be influenced by inter-species interactions. 

\begin{figure}
\centerline{\includegraphics[width=11.cm]{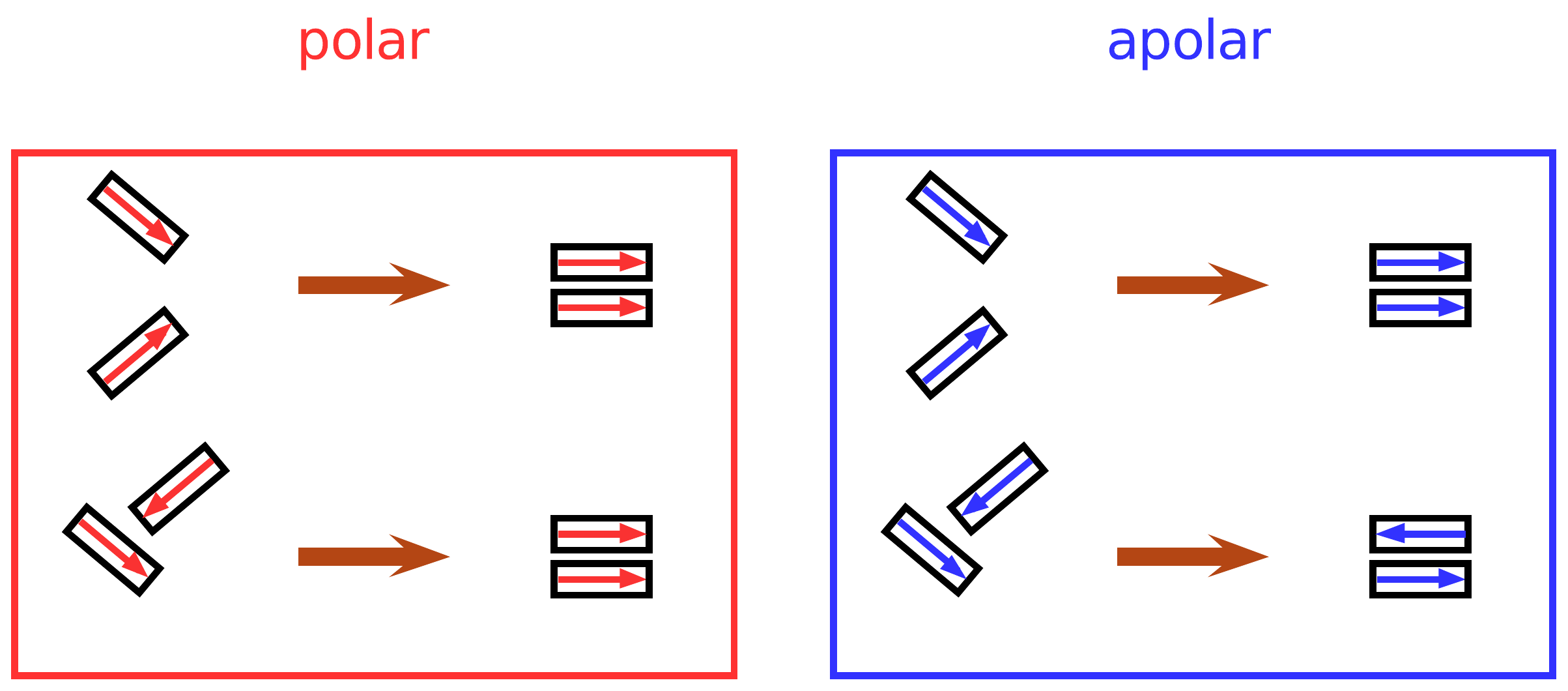}}
\caption{
Schematic illustration of polar alignment rules on the left and apolar alignment rules on the right. Self-propelled particles are indicated by rectangular boxes, while the arrows inside the boxes mark the direction of active drive. For polar alignment rules, the particles always tend to align their velocity orientations into the same direction. In the case of apolar alignment rules, which may be induced, for example, by steric interactions of elongated particles, an antiparallel orientation of the velocity vectors is equally preferred.
}
\label{fig_apolar}
\end{figure}

Other variants of the Vicsek model, such as polar particles of apolar alignment interactions \cite{chate2008modeling,peruani2008mean,ginelli2010large,peruani2010cluster,peruani2011polar} or nematic particles that randomly reverse their migration direction \cite{chate2006simple,chate2008modeling} were investigated. 
For the first kind of systems, density bands with particle migration along the contour of the band were observed \cite{chate2008modeling,ginelli2010large,peruani2011polar}, in contrast to the above-mentioned density bands that migrate perpendicularly to their elongation direction. An example of apolar alignment interaction is depicted in Fig.~\ref{fig_apolar}. 

A further characteristic feature of self-propelled particle systems has been outlined for the active nematic case \cite{ramaswamy2003active}. We denote the average number of active particles in a subregion of the whole system by $N$ and the fluctuations of this number by $\delta\! N$. Typically a relation of the form ${\langle(\delta\! N)^2\rangle}^{1/2}\propto N^{\zeta}$ is obtained, with $\zeta=0.5$ in equilibrium systems. In the two-dimensional case of active nematics, an exponent $\zeta=1$ was predicted \cite{ramaswamy2003active,toner2005hydrodynamics} and later confirmed through simulations \cite{chate2006simple}. These large density fluctuations were termed ``giant number fluctuations''. Likewise, in the case of active polar particles an elevated exponent $\zeta>0.5$ was predicted \cite{toner1995long,toner1998flocks} and confirmed in simulations \cite{chate2008modeling,ginelli2010large,peruani2011polar}. Such large number fluctuations were also observed in experiments \cite{narayan2007long,deseigne2010collective,deseigne2012vibrated}. As mentioned below, in systems featuring steric interactions, large number fluctuations can result from a cluster transition during which self-propelled particles mutually block their active migration \cite{fily2012athermal,redner2013structure,buttinoni2013dynamical, bialke2013microscopic,fily2014freezing,stenhammar2013continuum,stenhammar2014phase}.

Finally, we note that two-dimensional systems of self-propelled particles can feature a long-ranged orientational order of their migration directions \cite{toner1995long,toner1998flocks,toner2005hydrodynamics}. This is in marked contrast to equilibrium systems \cite{mermin1966absence}. Long-ranged order can emerge in the active case because information about the orientations is additionally propagated by the self-propelled motion of the particles.

\subsection[Self-propelled particles featuring steric interactions]{\hspace{-.1cm}Self-propelled particles featuring steric interactions}

Originally, the point-like self-propelled particles in the Vicsek model \cite{vicsek1995novel} only interact via the alignment of their velocity orientations. In principle, this can lead to artifacts such as an unbounded growth of the local particle density. 

As a first step to solve this problem, modifications of the Vicsek approach were introduced. For example, the migration directions of particles that come too close to each other can be reversed to bound the density \cite{czirok1996formation}. Or a pairwise alignment of the velocity vectors into opposite directions is induced for particles that are not well separated \cite{romenskyy2013statistical,menzel2013unidirectional,weber2014defect}. Another route adjusts the migration direction according to an inter-particle potential \cite{gregoire2003moving}. If a finite intermediate distance is preferred, crystal-like arrangements can arise within the particle swarm \cite{gregoire2003moving}. Such a situation can be interpreted as a combination of steric repulsive and cohesive attractive interactions \cite{gregoire2004onset}. 

\begin{figure}
\centerline{\includegraphics[width=\textwidth]{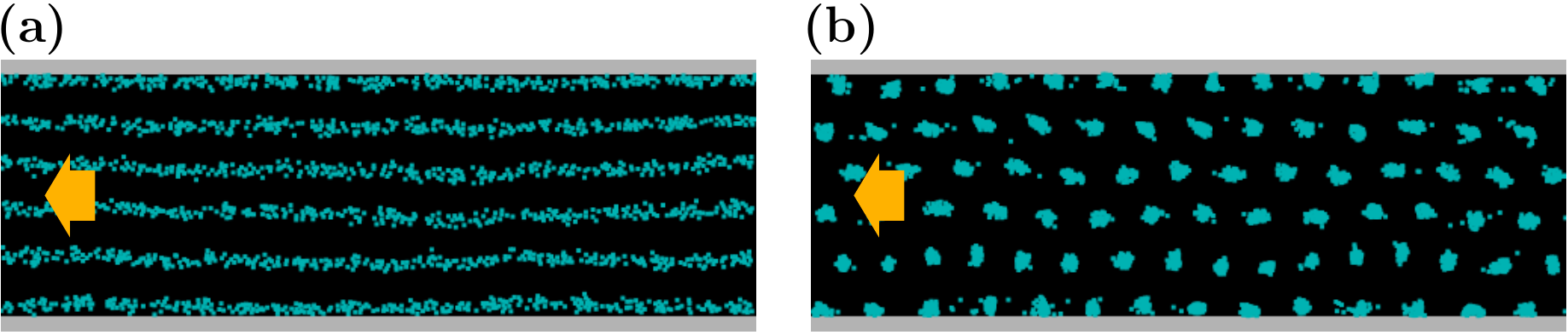}}
\caption{
Pattern formation in a modified Vicsek model of self-propelled particles confined in a channel between two parallel horizontal plates \cite{menzel2013unidirectional}. The patterns appear at intermediate overall particle densities. As indicated by the arrows, the structures in both cases collectively migrate along the channel direction. (a) Increasing the overall particle density from below, first a state of {unidirectional} laning appears \cite{menzel2013unidirectional}: all particles on average migrate into the \textit{same} direction. (b) At slightly higher particle densities, migrating cluster crystals emerge \cite{menzel2013unidirectional}. Each lattice point in this state is occupied by several self-propelled particles. The mechanism leading to the pattern formation is based on discrete migration steps of the self-propelled particles and the possibility of an overreaction in their velocity alignment. Channel walls are indicated by the gray bars and periodic boundary conditions are applied in the horizontal direction. 
}
\label{fig_unidirectional}
\end{figure}

In addition to steric interactions between the particles, also steric interactions with confining walls can be considered. An example is a channel geometry made by two parallel confining plates \cite{menzel2013unidirectional}. Using a functional form for the alignment interactions as in Eq.~(\ref{eq_U}) in combination with discrete time steps, pattern formation was observed at intermediate overall particle densities in the channel \cite{menzel2013unidirectional}. On the one hand, lanes emerge along the channel direction that are either directly supported by the walls or ``free-standing'' in the interior \cite{menzel2013unidirectional}, see Fig.~\ref{fig_unidirectional}~(a). These lanes are ``unidirectional'' in the sense that all particles in the channel self-propel on average into the same direction. On the other hand, at slightly higher particle densities, active cluster crystals form that collectively migrate along the channel \cite{menzel2013unidirectional}, see Fig.~\ref{fig_unidirectional}~(b). The mechanism behind the formation of these structures results from a combination of two ingredients: first discrete migration steps of the self-propelled particles, and second the possibility for overreactions in the velocity alignment. 

From an experimental point of view, dry systems of manifestly sterically interacting self-propelled particles can be realized by granular hoppers \cite{blair2003vortices,volfson2004anisotropy,galanis2006spontaneous,narayan2007long, aranson2007swirling,kudrolli2008swarming,deseigne2010collective,deseigne2012vibrated}. For this purpose, the granular particles are typically set into motion by a vertical vibration of the horizontal substrate. Only the in-plane displacements are recorded. Both, the migrations and ordering of apolar \cite{blair2003vortices,galanis2006spontaneous,narayan2007long,aranson2007swirling} and polar \cite{kudrolli2008swarming} rod-like objects were investigated, as well as the motion of polar disks \cite{deseigne2010collective,deseigne2012vibrated,weber2013long}. In the latter case, the steric interactions between the particles are isotropic due to the disk-like shape and do not explicitly provide an alignment interaction. Interestingly, the collective behavior of these vibrated polar disks could nevertheless remarkably well be mapped onto the Vicsek model \cite{deseigne2012vibrated}. 

A polar granular hopper features an anisotropic density distribution along its body to break the forward-backward symmetry. Under vibration it thus shows an imposed preferred migration direction \cite{kudrolli2008swarming}. In contrast to that, for apolar hoppers the symmetry must be broken from outside by an additional horizontal mode of vibration \cite{aranson2007swirling}, or the symmetry must be broken spontaneously to perform steps of propagation. Vibrated apolar rods were observed to incline with respect to the substrate surface and to preferentially migrate into the inclination direction \cite{blair2003vortices,volfson2004anisotropy}. The mechanism of spontaneous symmetry breaking was investigated in more detail for a dimer model particle \cite{dorbolo2005dynamics}. 
In the case of vibrated polar disks, the polarity results from breaking the symmetry of contact with the vibrating substrate by attaching two asymmetric legs \cite{deseigne2012vibrated}. 

As revealed by these studies \cite{volfson2004anisotropy,dorbolo2005dynamics}, the ``dry'' friction between the particles and the substrate plays a major role for the propagation mechanism. The displacement statistics of ``dry'' objects on vibrated substrates were analyzed experimentally, also as a function of the substrate inclination \cite{goohpattader2009experimental,goohpattader2010diffusive,goohpattader2011stochastic}. Interestingly, the displacement distribution functions appear to have exponential tails \cite{goohpattader2009experimental,goohpattader2010diffusive,goohpattader2011stochastic}. Similar results were obtained for vibrated water droplets \cite{goohpattader2009experimental,mettu2010stochastic}.

Typically the interactions with the substrate are modeled by a ``dry'' friction term of the Coulomb type \cite{persson2000sliding}. Generally, a simple model equation for the one-dimensional motion of a single granular particle exposed to dry friction with the substrate and an additional dynamic (viscous) friction with the remaining environment can be written in the form of a Langevin equation as \cite{gennes2005brownian,hayakawa2005langevin,baule2010path,baule2011stick, menzel2011effect,baule2013rectification}
\begin{equation}\label{Coulomb}
m\frac{dv}{dt} = -m\frac{v}{\tau}-\sigma(v)\Delta+\gamma(t), \qquad\frac{dx}{dt}=v. 
\end{equation}
Here, $m$ is the effective mass of the particle, $v$ its (generally non-constant) velocity, and $x$ its position. The first term on the right-hand side of the velocity equation denotes the dynamic (viscous) friction with a relaxation time $\tau$. $\sigma(v)$ returns the sign of $v$, i.e.\ $\sigma(v>0)=+1$, $\sigma(v=0)=0$, and $\sigma(v<0)=-1$. Thus the second term on the right-hand side includes the dry Coulomb friction of constant strength $\Delta$. It implies that a moving particle is always slowed down by the same amount of deceleration, independently of its current speed. Finally, $\gamma(t)$ describes a stochastic force, which may arise from fluctuations of the environment and stochastic noise in the vibration mechanism. 
Transferring Eq.~(\ref{Coulomb}) to the Fokker-Planck framework \cite{risken1996fokker,zwanzig2001nonequilibrium, kawarada2004non,gennes2005brownian,hayakawa2005langevin,menzel2011effect}, one finds that the steady-state velocity distribution function features a cusp at $v=0$ due to Coulomb friction \cite{kawarada2004non,hayakawa2005langevin,gennes2005brownian, baule2010path,baule2011stick,menzel2011effect,baule2012singular}. Interestingly, the irreversible part of the resulting Fokker-Planck equation can be mapped onto the case of a quantum mechanical harmonic oscillator with a pinning delta-potential at its center \cite{avakian1987spectroscopy,janke1988statistical,patil2006harmonic,touchette2010brownian,menzel2011effect}. 
Starting from a sharp initial state, the spatial particle distributions develop non-Gaussian tails on intermediate time-scales \cite{menzel2011effect}. Multiple time-scales emerge in the evolution of the distribution functions due to Coulomb friction \cite{menzel2011effect}. Nevertheless, despite such a non-Gaussian behavior, the mean-square displacement of the particles still grows linearly in time \cite{wang2009anomalous,menzel2011effect,wang2012brownian}. 
It will be an interesting problem for the future to study in more detail the effect of aspects of the friction mechanism on the collective behavior of granular hoppers.

\subsection{Active driving force instead of constant velocity}
\label{sec-active-driving-force}

In the previous sections, we reported about the success of the Vicsek model, which, despite its simplicity, could even reflect the collective behavior in an example system of granular hoppers \cite{deseigne2012vibrated}. One of the key assumptions in the Vicsek model is the constant self-propulsion speed of each individual particle \cite{vicsek1995novel}. This is an adequate simplification in a dilute system. However, it becomes problematic in dense systems of interacting particles. In particular, jammed situations where particles mutually block their ways cannot be described adequately in this model. 

It is then reasonable to switch to the picture of an active driving force \cite{hagen2011brownian,henkes2011active,zottl2012nonlinear,fily2012athermal,bialke2012crystallization, wittkowski2012self,redner2013structure,ni2013pushing,zottl2013periodic,ferrante2013collective,ferrante2013elasticity, abkenar2013collective,bialke2013microscopic,buttinoni2013dynamical,hagen2014gravitaxis} instead of a constant self-propulsion velocity. In the simplest general case, the particles are now characterized by their position $\mathbf{r}$, by their velocity $\mathbf{v}$, and by a polar unit vector $\mathbf{\hat{p}}$ that gives the current orientation of the active driving force. Here, situations can arise, in which the direction of the velocity $\mathbf{v}$ and the orientation $\mathbf{\hat{p}}$ of the driving force are not parallel. The magnitude of the velocity $\mathbf{v}$ can change over time, whereas the strength of the active driving force $F_{d}$ is usually kept constant for simplicity. 

In many cases, the motion of the particles is overdamped. This is true for instance for colloidal Janus particles \cite{paxton2004catalytic,howse2007self,jiang2010active,volpe2011microswimmers,buttinoni2012active, theurkauff2012dynamic,buttinoni2013dynamical} or bacteria \cite{berg1972chemotaxis,wada2007model,suematsu2011localized,drescher2011fluid} that self-propel in an aqueous environment at low Reynolds numbers \cite{purcell1977life}. Overdamped Brownian dynamics is appropriate to describe this kind of motion and the velocity variable becomes redundant. The structure of the rescaled dynamic equations for an isotropic self-propelled particle of constant strength $F_d$ of its active driving force can be written as   
\begin{equation}
\frac{d\mathbf{r}}{dt}=-\nabla U+F_{d}\,\mathbf{\hat{p}}+\mbox{\boldmath$\xi$\unboldmath}, 
\qquad\frac{d\mathbf{\hat{p}}}{dt}=(\mbox{\boldmath$\omega$\unboldmath}+\mbox{\boldmath$\eta$\unboldmath})\times\mathbf{\hat{p}}.
\label{constantdriving}
\end{equation}
Here, the potential $U$ contains for example the influence of external fields like gravity, 
the steric interactions with other particles, or the steric interactions with confining walls. The angular velocity $\mbox{\boldmath$\omega$\unboldmath}$ results from the torque on the particle for example due to external alignment by a magnetic field. Both, $\mbox{\boldmath$\xi$\unboldmath}$ and $\mbox{\boldmath$\eta$\unboldmath}$, include the impact of stochastic fluctuations. 
So far, most of the corresponding studies have been restricted to two spatial dimensions. 

First, the stochastic motion of single isolated self-propelled particles was analyzed \cite{howse2007self,hagen2011brownian}. In the long-time limit, the stochastic noise determines the statistical properties. Consequently, in the long-time regime, the motion appears diffusive. However, the magnitude of the corresponding overall diffusion constant can be significantly increased by self-propulsion when compared to the case of an analogous passive particle \cite{howse2007self,hagen2011brownian,zheng2013non}. In contrast to that, self-propulsion significantly alters an intermediate time regime. For example, non-Gaussian displacement statistics can emerge at intermediate times in spite of a Gaussian form of the stochastic noise \cite{hagen2011brownian}. These results were confirmed by comparison to experimental systems of self-propelling colloidal Janus particles \cite{zheng2013non}. Also the impact of an external flow field was studied \cite{zottl2012nonlinear,zottl2013periodic}. 

Apart from that, Eqs.~(\ref{constantdriving}) were generalized and investigated for the case of rod-like particles, where the diffusion and friction matrices become uniaxially anisotropic \cite{teeffelen2009clockwise}. For general particle shapes, aligning torques can arise due to steric interactions between anisotropic particles or because of interactions with confining boundaries. These aligning torques enter the equations via the angular velocity $\mbox{\boldmath$\omega$\unboldmath}$. Furthermore, a torque that leads to a continuous reorientation of the polar direction $\mathbf{\hat{p}}$ can arise from the mechanism of self-propulsion in combination with the particle shape or be actively generated. The stochastic motions of such circle swimmers were analyzed \cite{teeffelen2008dynamics,teeffelen2009clockwise} and compared with experiments \cite{kummel2013circular}. Self-organization in array-like vortex textures was found \cite{kaiser2013vortex}. Moreover, it was demonstrated that in three spatial dimensions the asymmetry of biaxial self-propelled particles can lead to helical and even to superhelical trajectories \cite{wittkowski2012self}. 

Concerning the collective behavior of many interacting particles, we first focus on the case of spherical objects \cite{henkes2011active,fily2012athermal,bialke2012crystallization, bialke2013microscopic,buttinoni2013dynamical,ni2013pushing}. Due to their isotropic shape, they do not feature a steric alignment interaction. This promotes the emergence of new truly non-equilibrium effects. A spherical particle that self-propels towards a hard wall is blocked in its motion and slowed down \cite{elgeti2013wall}. It can only escape again when the direction of active drive has reoriented away from the wall. The time scale for this process is set by rotational diffusion and thus indirectly by the temperature of the system. During this time, other particles can hit and additionally get blocked, if the density and self-propulsion speed are high enough. 
\begin{figure}
\centerline{\includegraphics[width=9.cm]{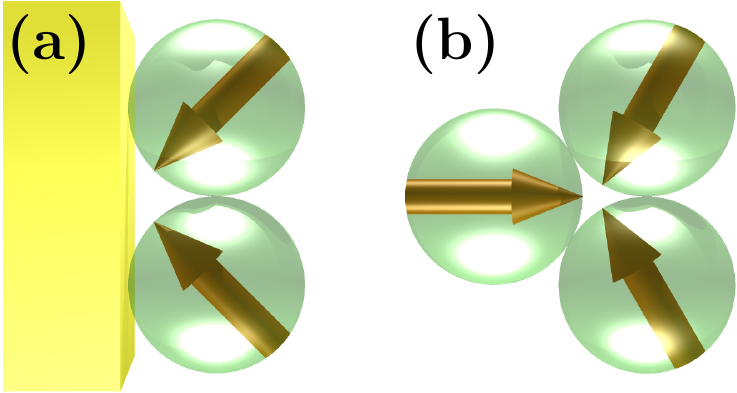}}
\caption{
Schematic illustration of the blocking mechanism of isotropically sterically interacting self-propelled particles. The self-driven particles are of spherical shape. Arrows indicate the direction of active drive which can reorient by rotational diffusion. A two-dimensional motion of the particles within the plane of the figure is assumed. (a) Two particles block each other, supported by a rigid wall. (b) Self-supported blocking can also occur in the bulk. If rotational diffusion is slow enough in comparison to the active drive and the density is high enough, both situations can serve as seeds for the formation of larger clusters. 
}
\label{fig_trapped}
\end{figure}
Through such a mutual blocking, see Fig.~\ref{fig_trapped}~(a), wall-supported clusters can emerge\footnote{This was demonstrated for rod-like particles \cite{wensink2008aggregation}. In that case, after a certain time, the clusters can slip off the wall, supported by the reorienting steric interactions with the wall \cite{elgeti2013wall}. On the contrary, due to their isotropic steric interactions, self-propelled spheres are more effectively blocked by walls than rod-like particles \cite{elgeti2013wall}.}.
Interestingly, the analogous process is also observed in a self-supported way in the bulk \cite{fily2012athermal,redner2013structure} in two-dimensional systems, see Fig.~\ref{fig_trapped}~(b). Here, the blocking mechanism was analyzed in more detail \cite{bialke2013microscopic} and also investigated experimentally \cite{buttinoni2013dynamical}. A real phase separation into a clustered and a gas-like phase can be obtained as a steady state of the system \cite{redner2013structure,cates2013active,stenhammar2013continuum}. For clarity, it is stressed that these clusters form in the absence of any attractive forces and despite purely repulsive interactions between the particles. The clusters emerge solely from mutual steric blocking of the migration due to the non-equilibrium active drive. 

Continuum equations were derived that trace the phase separation into the gas-like and clustered phases back to a force imbalance arising from the mutual steric blocking \cite{bialke2013microscopic}, and a weakly nonlinear stability analysis \cite{cross1993pattern} was carried out \cite{speck2014effective}. In this way, the nature of the transition of phase separation could be investigated in more detail. Interestingly, the onset of the clustering transition can be mapped onto the phase separation of passive particles described by the famous Cahn-Hilliard model \cite{cahn1958free}. In particular, this implies the existence of an effective free energy density that can characterize the non-equilibrium clustering transition \cite{speck2014effective}. 
The analysis suggests a change of the clustering transition from continuous at high particle densities to discontinuous at lower particle densities. This change in the nature of the clustering transition was confirmed by numerical investigations \cite{speck2014effective}. 
Naturally, the clusters imply large fluctuations of the local particle density \cite{fily2012athermal}. A reentrance of the fluid phase is observed as a function of the driving force \cite{bialke2013microscopic}. 

At high densities, the active systems become jammed \cite{henkes2011active,ni2013pushing} or crystallize \cite{bialke2012crystallization,redner2013structure}. Active crystals were also observed experimentally \cite{palacci2013living}, with attractive particle interactions typically being involved in the crystallization process \cite{palacci2013living,mognetti2013living}. 

On the one hand, active crystals have been studied numerically by particle simulations \cite{bialke2012crystallization,redner2013structure,ferrante2013collective,ferrante2013elasticity, weber2014defect}. The self-organization of purely repulsively interacting actively driven particles into crystal-like structures was investigated \cite{bialke2012crystallization}. Furthermore, the collective behavior of self-propelled particles that are ordered in crystal-like arrangements and interact by harmonic-spring potentials has been analyzed in detail \cite{ferrante2013collective,ferrante2013elasticity}. In the latter case, there is no explicit alignment interaction between neighboring particles. Nevertheless, the elastic interactions between them channel the individual attempts of self-propulsion towards one global orientation of motion. Finally, the whole structure collectively migrates into one common direction. 

\begin{figure}
\centerline{\includegraphics[width=\textwidth]{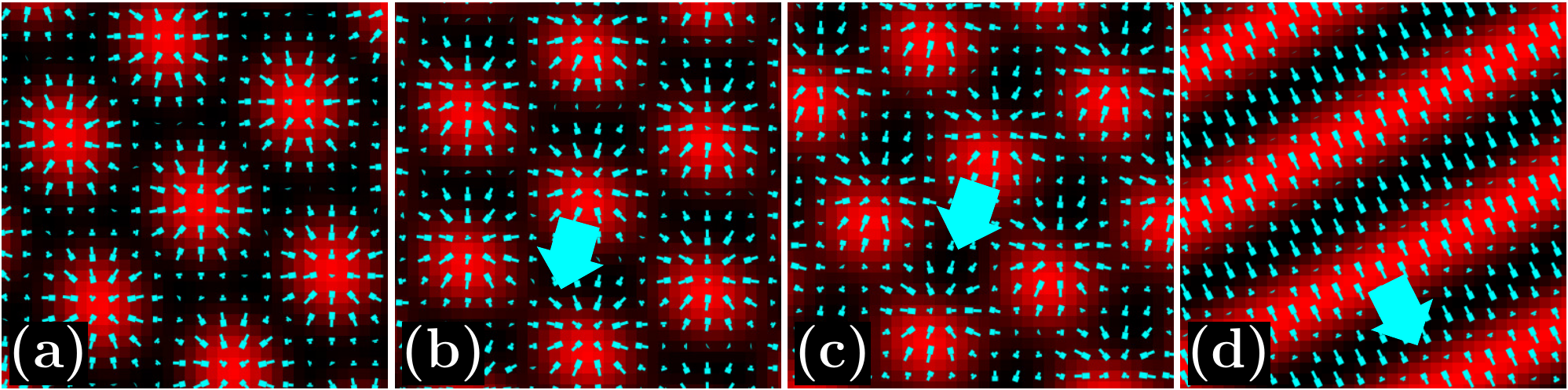}}
\caption{Active crystals as modeled and displayed in Refs.~\cite{menzel2013traveling,menzel2014active}. The particle-resolved density field $\psi(\mathbf{r},t)$ is encoded by the color, where brighter color means higher density. Thin needles, pointing from the thicker to the thinner ends, indicate the local orientation and magnitude of the polar orientational order parameter field $\mathbf{P}(\mathbf{r},t)$. Thick arrows mark directions of collective motion. Panels from left to right were obtained for an increasing strength of the active drive. Here, explicit alignment mechanisms for the self-propulsion directions were not included. Therefore the hexagonal active crystal in panel (a) at low active drive remains at rest. Beyond a threshold magnitude of the active drive, however, this structure can nevertheless start to migrate, as shown for the traveling hexagonal crystal in panel (b). Increasing the active drive further, structural transitions to traveling rhombic and quadratic (c) as well as traveling lamellar textures (d) were observed.
}
\label{fig_activecrystal}
\end{figure}

On the other hand, a field-theoretic approach was introduced to study the collective behavior of active crystals \cite{menzel2013traveling,menzel2014active}. It can be derived microscopically from classical dynamical density functional theory \cite{marconi1999dynamic,archer2004dynamical, elder2007phase,teeffelen2009derivation,tegze2009diffusion,jaatinen2009thermodynamics, lowen2010phase,wittkowski2010derivation,wittkowski2011polar} applying appropriate assumptions. 
Two microscopic order parameter fields were used to characterize the state of the system composed of many self-driven particles. 
The first one is a particle-resolved density field $\psi(\mathbf{r},t)$; 
the second one is a polar order parameter field $\mathbf{P}(\mathbf{r},t)$ that characterizes the locally preferred orientation of the active driving direction $\mathbf{\hat{p}}$, see Eqs.~(\ref{constantdriving}). Coupled dynamic equations describe the time evolution of these order parameter fields \cite{menzel2013traveling,menzel2014active}. 
Resulting periodic structures are displayed in Fig.~\ref{fig_activecrystal}. Concepts from two prominent continuum descriptions were unified in this approach. The first one is the phase field crystal model characterizing periodic modulations in the density field. 
This model was successfully used to reproduce solidification and crystallization phenomena in conventional passive crystalline structures on particle-resolved length and diffusive time scales \cite{elder2002modeling,elder2004modeling,stefanovic2006phase,elder2007phase, teeffelen2009derivation,tegze2009diffusion,jaatinen2009thermodynamics, stefanovic2009phase,chan2010plasticity,ramos2010dynamical,tegze2011faceting}. 
The second prominent approach contained here stems from the macroscopic Toner-Tu model for the collective motion of non-crystallized self-propelled particles \cite{toner1995long,toner1998flocks}. 
It can distinguish between systems that feature spontaneous directed self-propulsion, for example emerging from explicit alignment interactions, and systems without spontaneous alignment of their self-propulsion directions. 
Starting from initially disordered active systems, the field-theoretic approach shows that an active single crystal can form via the coarsening of multi-domain structures \cite{menzel2013traveling}. If a local alignment mechanism for the active driving directions of the particles exists, 
the crystal finally collectively travels into one common migration direction. If there is no such alignment, 
the active crystal remains at rest for small active drive, see Fig.~\ref{fig_activecrystal}~(a). However, beyond a threshold value of the active drive, the crystal still finds a common migration direction and starts to collectively travel. This is similar to the above-mentioned results from the particle-resolved simulations of active crystalline structures \cite{ferrante2013collective,ferrante2013elasticity}. With increasing active drive, a transition from traveling hexagonal to traveling rhombic, quadratic, and lamellar structures was observed \cite{menzel2013traveling}. Such a transition series is displayed in Fig.~\ref{fig_activecrystal}~(b)--(d).  

In addition to that, using the active phase field crystal theory \cite{menzel2013traveling}, a linear stability analysis of the traveling active single crystalline structures was performed \cite{menzel2014active}. This analysis is complicated by the fact that the density field in the crystalline ground state is already spatially modulated. It was found that the investigated collectively traveling active single crystals are linearly stable. The impact of hydrodynamic interactions on the stability of such active crystals was considered and is discussed in the next section. 

\begin{figure}
\centerline{\includegraphics[width=7.5cm]{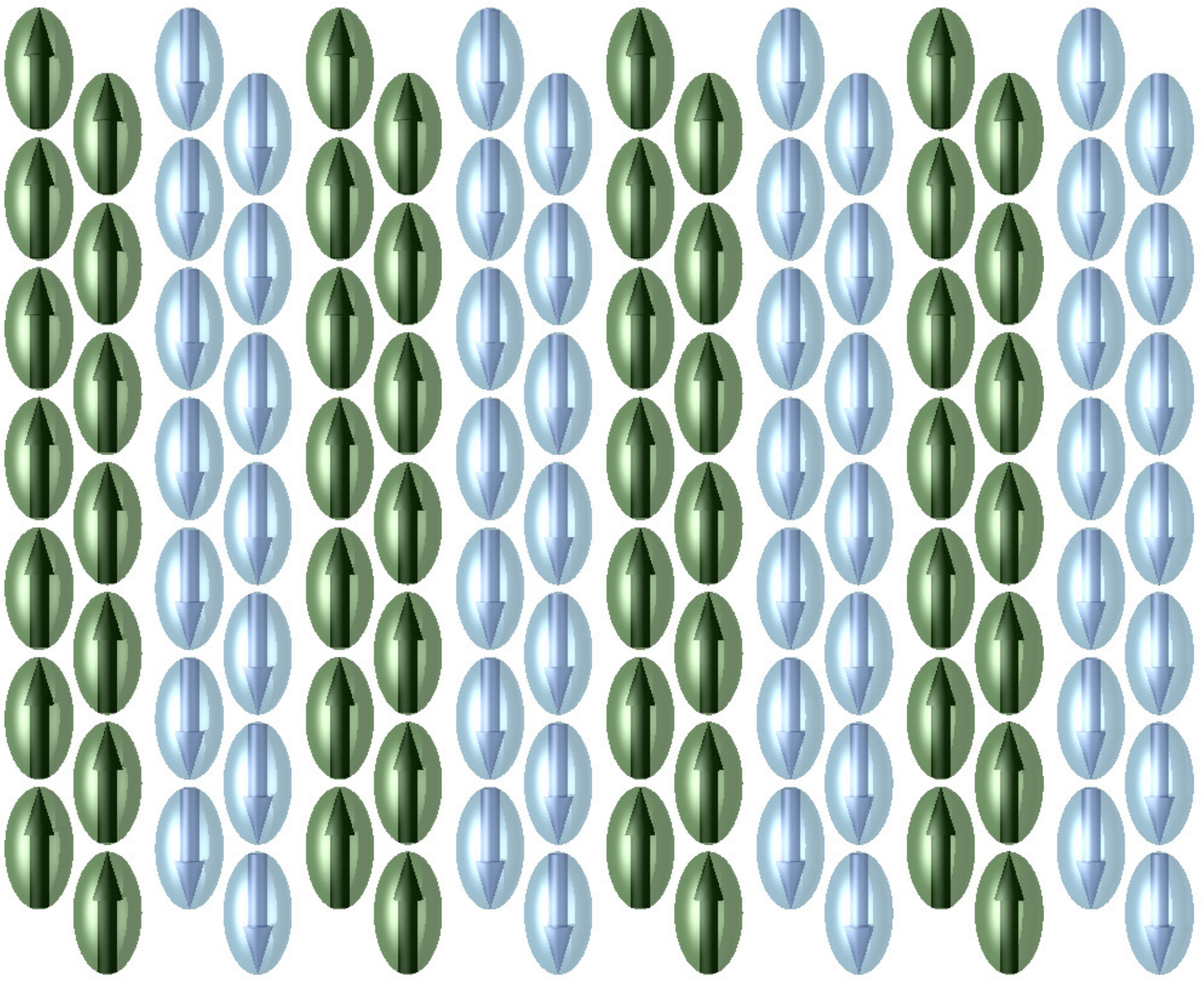}}
\caption{
Schematic illustration of a laning state. Darker particles move upwards, whereas brighter particles migrate downwards. The particles organize themselves into neighboring lanes of opposite migration directions. Either the opposite migration directions of the individual particles are explicitly imposed from outside for particles driven by an external force; or the opposite migration directions emerge due to self-organization in systems of identical self-driven particles. In reality the lane thickness is not necessarily as regular as depicted here and can comprise different numbers of particles. 
}
\label{fig_laning}
\end{figure}

Turning now to non-spherical self-propelled particles, a well-studied example is given by self-propelled rods actively driven parallel to their long axis \cite{peruani2006nonequilibrium,wensink2008aggregation,baskaran2008enhanced, baskaran2008hydrodynamics,elgeti2009self,yang2010swarm, wensink2012meso,baskaran2010nonequilibrium,wensink2012emergent, mccandlish2012spontaneous,kaiser2012how,abkenar2013collective,kaiser2013vortex,kaiser2013capturing}. The biological motivation arises for instance from experimental investigations on elongated bacteria 
crawling or gliding on a substrate \cite{harshey2003bacterial,peruani2006nonequilibrium,aranson2007model,kearns2010field,peruani2012collective}. Another biological example are filaments actively driven by molecular motors. These motors can be fixed on a substrate at one end \cite{schaller2010polar,schaller2011frozen} or directly work between the filamental rods \cite{sanchez2011cilia,sanchez2012spontaneous}. Non-biological artificial experimental systems can in particular be realized by vibrating polar granular rods \cite{kudrolli2008swarming}. In all these cases, the anisotropy of the steric interactions provides a mechanism of particle alignment. It turns out that the self-propulsion supports nematic ordering of rod-like particles by lowering the transition density from the isotropic to the nematic state \cite{baskaran2008enhanced}. This is in line with the prediction of long-ranged orientational order in two-dimensional self-propelled particle systems \cite{toner1995long,toner1998flocks,toner2005hydrodynamics}. Apart from that, various different collective states of migration have been revealed for interacting self-propelled rods \cite{peruani2006nonequilibrium,wensink2008aggregation,yang2010swarm,wensink2012meso,wensink2012emergent,mccandlish2012spontaneous,abkenar2013collective}. Among those are disordered, jammed, clustered,
and swarming states, where in the latter case the rods organize in localized packets of common migration direction
\footnote{The notation in this context is not unambiguous. Sometimes polar swarms are also denoted as clusters, whereas sometimes non-polar clusters are denoted as swarms.}. 
Laning states were identified \cite{wensink2012meso,wensink2012emergent,mccandlish2012spontaneous,abkenar2013collective} that feature parallel stripes or lanes within which all particles propel into the same direction. However, neighboring lanes in these systems show antiparallel migration directions, see Fig.~\ref{fig_laning} for a schematic illustration. 
This is in contrast to the above-mentioned case of unidirectional laning \cite{menzel2013unidirectional}. Laning as in Fig.~\ref{fig_laning} was previously observed in systems of externally driven particles, where for different particles a driving force into opposite directions was imposed from outside 
\cite{dzubiella2002lane,chakrabarti2004reentrance,rex2008influence,wysocki2009oscillatory, lowen2010particle,vissers2011lane,ikeda2012instabilities,glanz2012nature}. 
In the case of self-propelled particles, laning appears spontaneously without explicitly superimposing the opposite migration directions. Finally, even turbulent states were discovered in simulations, theory, and experiments on swimming bacteria \cite{cisneros2007fluid,wensink2012meso,dunkel2013fluid}, although these organisms swim in an environment of low Reynolds numbers. The source of this  turbulence at low Reynolds numbers is the continuous energy input on the length scale of the individual self-driven particles. 

Apart from that, the behavior of active particles featuring various shapes more complicated than rod-like was studied \cite{wensink2014controlling,nguyen2014emergent}. In particular, forward-backward symmetries can be broken through the particle shape \cite{wensink2014controlling}. Identical shapes, but differently applied active driving forces can lead to phase separation \cite{wensink2014controlling}. Depending on the relative orientation of the self-propulsion direction with respect to the particle shape, concave particles can self-organize into micro-rotors \cite{wensink2014controlling}. Finally, for binary mixtures of non-spherical active rotors, a phase separation into domains of clockwise and counter-clockwise rotations was identified \cite{nguyen2014emergent}.

\subsection{Role of hydrodynamic interactions}

Many realizations of self-propelled particles exist in a liquid environment. Examples are 
artificial microswimmers as for instance the colloidal Janus particles addressed above \cite{paxton2004catalytic,howse2007self,jiang2010active,volpe2011microswimmers,buttinoni2012active, theurkauff2012dynamic,buttinoni2013dynamical}, and swimming bacteria \cite{berg1972chemotaxis,wada2007model,suematsu2011localized,drescher2011fluid}. 
Any self-propelling swimmer sets its surrounding fluid into motion. The motion of the other swimmers, which are suspended in the fluid, is naturally influenced by this fluid flow. In turn, also the motion of all the other swimmers acts back onto the first one. These fluid-mediated interactions are referred to as hydrodynamic interactions. They are long-ranged. 

To include hydrodynamic interactions, we must take into account the fluid flow. It is typically a very good approximation to consider the -- often aqueous -- surrounding liquid as incompressible, i.e.\ to set its mass density $\rho$ equal to a constant. Then the central equation of motion in fluid dynamics that determines the fluid flow, the Navier-Stokes equation \cite{landau1987fluid}, can be written in the form 
\begin{equation}\label{eq:NS}
\rho\,\frac{\partial\mathbf{v}(\mathbf{r},t)}{\partial t} + \rho \left[\mathbf{v}(\mathbf{r},t)\cdot\nabla\right]\mathbf{v}(\mathbf{r},t) = -\nabla p(\mathbf{r},t) + \eta\nabla^2\mathbf{v}(\mathbf{r},t) + \mathbf{f}(\mathbf{r},t). 
\end{equation}
It describes the influence of volume force densities acting on a fluid element at position $\mathbf{r}$ and time $t$. $\mathbf{v}(\mathbf{r},t)$ is the resulting flow velocity field of the fluid at position $\mathbf{r}$ and time $t$. The left-hand side of Eq.~(\ref{eq:NS}) gives the inertial contribution;  
here the second term expresses the fact that the actual motion of the fluid elements must be taken into account to correctly determine the change of momentum at a spatial position $\mathbf{r}$. 
On the right-hand side, the first term denotes the force density due to gradients in the pressure field $p(\mathbf{r},t)$, the second term includes dissipation with $\eta$ the viscosity of the fluid, and the last term takes into account any additional force density field $\mathbf{f}(\mathbf{r},t)$ acting on the fluid. In our case, $\mathbf{f}(\mathbf{r},t)$ contains the force densities that each swimmer exerts on its liquid environment. 

Rescaling all lengths by a characteristic length scale $L$, all velocities by a characteristic velocity $V$, time by $L/V$, pressure by $\rho V^2$, and force density by $\rho V^2/L$, Eq.~(\ref{eq:NS}) takes the form
\begin{equation}\label{eq:NSresc}
\frac{\partial\mathbf{v}'}{\partial t'} + \left[\mathbf{v}'\cdot\nabla'\right]\mathbf{v}' = -\nabla' p' + \mbox{Re}^{-1}\nabla'^2\mathbf{v}' + \mathbf{f}',
\end{equation}
with space and time dependencies not marked explicitly any more. We find 
\begin{equation}\label{eq:Re}
\mbox{Re}=\frac{\rho L V}{\eta}
\end{equation}
for the famous dimensionless Reynolds number. 
Typical microswimmers in the form of colloidal Janus particles and swimming bacteria have dimensions $L$ in the range between $10$~nm and $10$~$\mu$m. A characteristic swimming speed would for example be $10$~$\mu$m$/$s. Thus, we can see from Eq.~(\ref{eq:Re}) that their swimming in an aqueous environment usually occurs at low Reynolds numbers $\mbox{Re}\ll1$ \cite{purcell1977life}. 

We can infer the consequences of this estimate from Eq.~(\ref{eq:NSresc}). For low Reynolds numbers, the inertial terms on the left-hand side can be neglected when compared to the dissipative contribution on the right-hand side. 
Hydrodynamics at low Reynolds numbers is thus described by Stokes' equation \cite{happel1983low}: 
\begin{equation}\label{eq:Stokes}
\mathbf{0} = -\nabla' p' + \mbox{Re}^{-1}\nabla'^2\mathbf{v}' + \mathbf{f}'.
\end{equation}
This equation leads to some peculiarities for swimming at low Reynolds numbers that are markedly different from our every-day experiences. In particular, the equation is linear in the velocity field and instant in time. Consequently, for balanced pressure gradients, flow immediately stops when no further force density $\mathbf{f}'$ is applied to the fluid. This has qualitative consequences when we compare to the situation at higher Reynolds numbers.  

\begin{figure}
\centerline{\includegraphics[width=.8\textwidth]{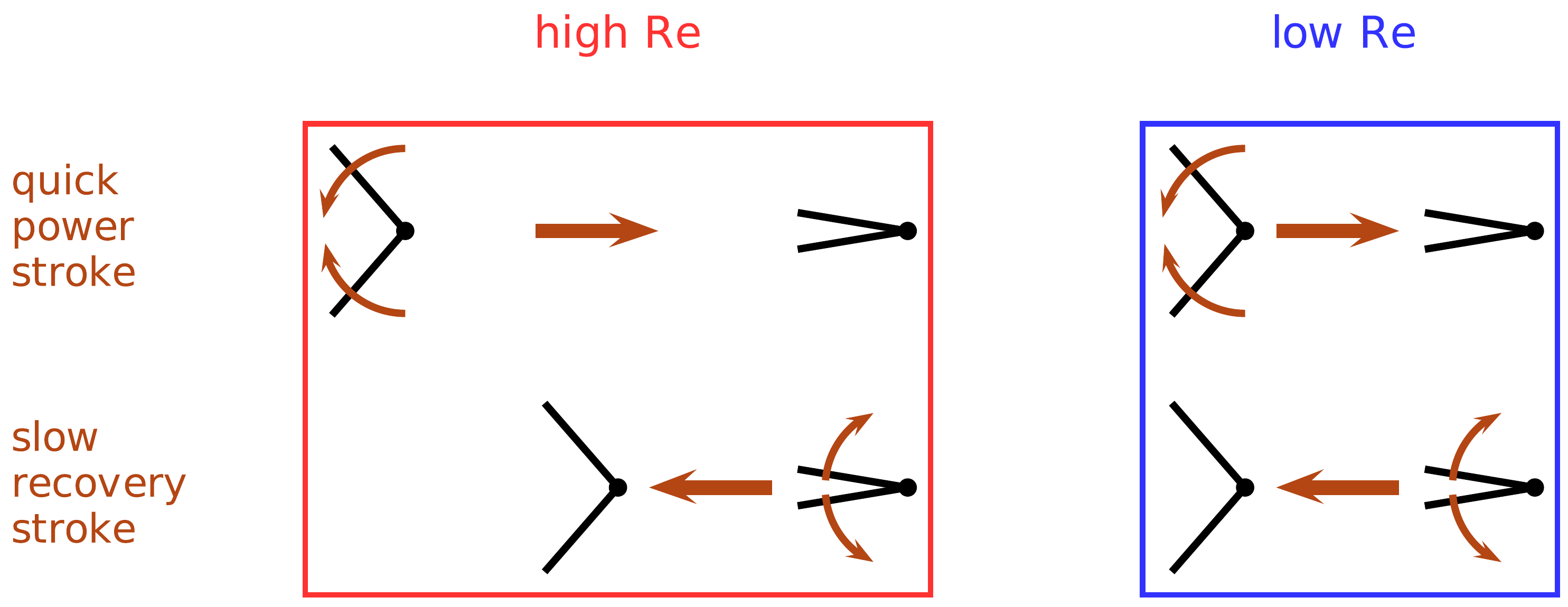}}
\caption{Illustration of the qualitatively different situation for swimming at high and low Reynolds number (``scallop theorem'') \cite{purcell1977life}. The simple swimmer (black) performs a reciprocal shape change, i.e.\ the shape change during the power stroke is just the inverse of the shape change during the recovery stroke. Only the speed of shape change is significantly higher during the power stroke than during the recovery stroke. At high Reynolds numbers this leads to a net displacement after the cycle is completed due to inertial effects. At low Reynolds number, inertial effects are negligible and a net displacement cannot be achieved in this way. }
\label{figure_reciprocal}
\end{figure}

At high enough Reynolds numbers, a swimmer can move by reciprocal shape changes, see also Fig.~\ref{figure_reciprocal}: first a quick shape change is performed in the form of a power stroke; then a slow recovery stroke follows in the exactly inverse way of the power stroke, only that it is performed significantly more slowly. 
So the central difference between the two strokes lies in the speed of execution. In this situation, the inertial terms on the left-hand side of Eq.~(\ref{eq:NS}) can lead to a symmetry breaking and net motion.  
However, reciprocal shape changes of the swimmer cannot lead to such net motion at low Reynolds numbers. 
The symmetry-breaking inertial terms are missing in the Stokes equation Eq.~(\ref{eq:Stokes}). 
These facts are often referred to as the ``scallop theorem'' \cite{purcell1977life}. 

Furthermore, in the absence of pressure gradients, the swimmer motion completely stops together with the fluid flow $\mathbf{v}'$ as soon as the swimmer does not apply any force density $\mathbf{f}'$ to the fluid any longer. Apart from that, the sum of all instant forces exerted by a low Reynolds number swimmer on the surrounding fluid vanishes because each stroke that it performs is balanced by a counter-acting drag force due to the motion of its body. It can be shown that, with increasing distance from such a ``force-free'' swimmer, the induced flow field decays significantly more quickly than for an object that exerts a net force on the surrounding fluid \cite{yeomans2014introduction}. Furthermore, the swimmer must provide a suitable mechanism to break the symmetry and achieve a net forward motion \cite{lauga2011life}. 

\begin{figure}[t]
\centerline{\includegraphics[width=\textwidth]{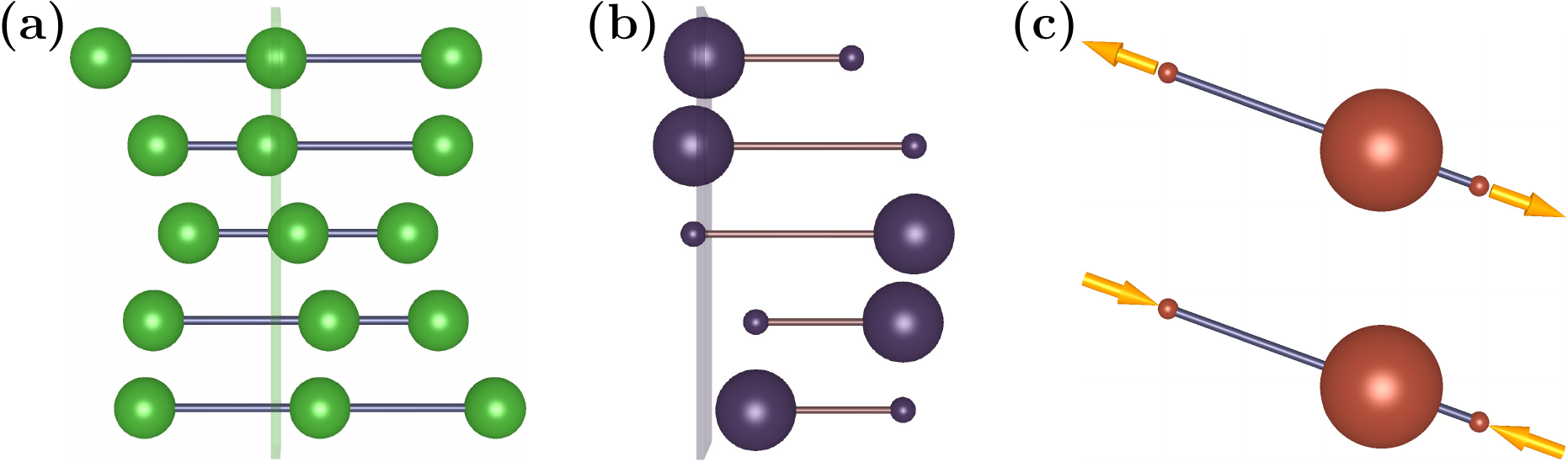}}
\caption{Illustration of different self-propulsion mechanisms for swimming at low Reynolds numbers. Panel (a) shows the three-linked-sphere swimmer suggested in Ref.~\cite{najafi2004simple}. Its deformation cycle (snapshots from top to bottom) is nonreciprocal. Due to hydrodynamic interactions, contraction or expansion of one bond leads to a different net displacement depending on whether the other bond is in the contracted or expanded state. The vertical bar is added for clarity to highlight that a net propulsion to the right has occurred during the cycle. Panel (b) depicts the nonreciprocal deformation cycle (again snapshots from top to bottom) of the two-linked-sphere swimmer introduced in Ref.~\cite{avron2005pushmepullyou}. In the completely contracted or expanded state of the bond, the swimmer exchanges the volume of the two spheres, for example by pumping the contents from one side to the other through the bond. Since the viscous drag is higher for the sphere of larger radius, the swimmer features a net propulsion to the right (again the vertical bar is added to make the net motion visible). Finally, panel (c) displays the minimal model swimmer suggested in Ref.~\cite{aditi2002hydrodynamic}. The outer spheres are small and experience a negligible viscous drag by the fluid. However, they represent centers of forces acting on the fluid and setting it into motion (indicated by the bright arrows). Due to the antiparallel alignment of the forces, the total applied force vanishes. In the upper case, the swimmer pushes the fluid outwards, which is why it is referred to as a ``pusher'', while in the lower case, it pulls the fluid inwards and would be called a ``puller''. The main swimmer body is asymmetrically located with respect to the force centers. Thus it experiences a net viscous drag by the self-induced fluid flow.}
\label{fig_swimmertypes}
\end{figure}

Different routes to reach this goal were pointed out. One example are non-reciprocal cycles of contraction and expansion of a straight three-linked-sphere swimmer \cite{najafi2004simple,golestanian2008analytic} as displayed in Fig.~\ref{fig_swimmertypes}~(a). 
Another example are volume changes of the spheres of a two-linked-sphere swimmer during contraction and expansion \cite{avron2005pushmepullyou}, see Fig.~\ref{fig_swimmertypes}~(b). Apart from that, the rigid structure of a swimming object can already break the symmetry by itself. Ideal candidates for this purpose are helical objects that have a certain handedness \cite{berg2003rotary,yonekura2003complete,wada2007model,ghosh2009controlled,zhang2009artificial, spagnolie2011comparative,tottori2012magnetic}. And as mentioned above, Janus particles intrinsically break the forward-backward symmetry, which can be exploited to start a net-propulsion mechanism \cite{paxton2004catalytic,howse2007self,jiang2010active,volpe2011microswimmers,golestanian2007designing, buttinoni2012active,theurkauff2012dynamic,buttinoni2013dynamical}. 

One swimming strategy for a swimmer at low Reynolds numbers is to exploit the drag resulting from its self-induced fluid flow. To understand this principle, it is most illustrative to consider two force centers as shown in Fig.~\ref{fig_swimmertypes}~(c). In the upper case, the forces applied to the fluid are oriented in a way that the swimmer pushes the fluid outwards. It is therefore called a pusher. The bottom case shows the opposite situation, in which the swimmer pulls the fluid inwards. Thus the swimmer is referred to as a puller. If the swimmer body is located asymmetrically with respect to the centers of force acting on the surrounding fluid, it will experience a drag from the self-induced fluid flow \cite{aditi2002hydrodynamic,hatwalne2004rheology,baskaran2009statistical}. Thus the swimmer will show a net propulsion. This propulsion principle can be generalized to more realistic swimmer shapes. 

We now turn to the situation of more than one swimmer suspended in the fluid. Then one of them feels the drag due to the resulting fluid flow induced by all other swimmers, and vice versa. 
The impact that these hydrodynamic interactions between active swimmers has on their net interactions and on their collective behavior is still under investigation. For the three-linked-sphere swimmer mentioned above \cite{najafi2004simple}, the pairwise hydrodynamic interaction was analyzed explicitly \cite{pooley2007hydrodynamic}. A complicated dependence on the relative distance, orientation, and phase of the shape change was obtained. Another study focused on the averaged flow fields around circle swimmers \cite{fily2012cooperative}. For particles that propel by pushing the fluid, a repulsive interaction was found on average, whereas for swimmers that pull the fluid, the average interaction was attractive. A pair of counterrotating pullers performs a net translational motion \cite{leoni2010dynamics,fily2012cooperative}. Straight-swimming pullers that were forced to swim parallel to each other were observed to repel each other \cite{gotze2010mesoscale,molina2013hydrodynamic}, whereas pushers in the same set-up were found to attract each other \cite{cisneros2007fluid,gotze2010mesoscale}. However, when released from the constraint of parallel swimming, no permanently bound state due to hydrodynamic interaction was found in spite of the initially attractive configuration \cite{gotze2010mesoscale}. 

Several cases of motion in a liquid environment were identified in which hydrodynamic effects seem not to play a crucial role. 
For bacterial cells, it was demonstrated that generally cell-cell interactions and cell-surface interactions are dominated by orientational diffusion, steric interactions, and lubrication forces, and that hydrodynamic interactions play a minor role \cite{drescher2011fluid}. Apart from that, different collective effects observed experimentally for microswimmers were successfully reproduced in simulations without including hydrodynamic interactions. Examples are the previously addressed observations of clusters and turbulent states \cite{buttinoni2013dynamical,wensink2012meso}. 

Nevertheless, there are many situations in which hydrodynamic interactions were shown to be important, both for the single-swimmer and for the collective behavior. 
For instance, hydrodynamic interactions with confining walls can modify the migration properties \cite{blake1971note,or2009dynamics,elgeti2010hydrodynamics, brotto2013hydrodynamics,li2014hydrodynamic, zottl2014hydrodynamics}. As demonstrated numerically, the nature of the propulsion mechanism can determine whether active microswimmers preferentially orient towards a confining surface, or whether they turn away and leave the wall \cite{li2014hydrodynamic}. Under narrow confinement between two walls, the propulsion mechanism can support orientations towards one of the walls or orientations parallel to their surfaces, which supports blocking or planar swimming motion, respectively \cite{li2014hydrodynamic,zottl2014hydrodynamics}. Naturally, this influences the formation of the clusters mentioned in Sec.~\ref{sec-active-driving-force} \cite{buttinoni2013dynamical,li2014hydrodynamic,zottl2014hydrodynamics}. Apart from that, a speed-up of microswimmers through hydrodynamic interactions with a flexible confining tube was predicted \cite{ledesma2013enhanced}. Microswimmers confined in a harmonic trap were found to orientationally order due to hydrodynamic interactions and induce a net fluid flow \cite{hennes2014self}. Moreover, in experiments, a transient capturing of microswimmers in circular trajectories around passive colloidal spheres was observed \cite{takagi2014hydrodynamic}.%

Generally, hydrodynamic interactions can induce the formation of vortices and swirls \cite{hernandez2005transport}. In that sense, hydrodynamic interactions can destroy long-ranged orientationally ordered collective motion, but also more localized orientational order in swarms \cite{lushi2014fluid}. Likewise, in the case of two-dimensional active traveling crystals, destabilizing effects of hydrodynamic interactions were observed numerically \cite{menzel2014active}. The self-propelling particles forming the active crystal were considered to migrate on a substrate covered by a thin fluid film. As a consequence of the resulting flow fields, the traveling single crystal can break up into different domains, or the crystalline lattice structure can vanish altogether. The stability of different lattice structures under hydrodynamic interactions was analyzed \cite{desreumaux2012active}. 

Recently, it became clear from theoretical and numerical investigations that hydrodynamic far-field and near-field interactions can differ in their influence on the orientational order and stability of the suspension. Naturally, far-field interactions prevail in dilute suspensions whereas near-field interactions become important in concentrated solutions of microswimmers. Most of these studies were performed using yet another swimmer model called ``squirmer'' \cite{lighthill1952squirming,blake1971spherical,ishikawa2006hydrodynamic}: a non-vanishing flow field is prescribed as a boundary condition for the surrounding fluid on the surface of each swimmer; the swimmers themselves are often considered as rigid objects, but the non-vanishing surface flow drives them through the liquid environment.%

Following these lines, we first consider dilute suspensions of swimmers that solely interact hydrodynamically. For those, it was demonstrated that orientationally ordered states are always unstable \cite{saintillan2008instabilities,saintillan2008pattern,saintillan2013active}. Furthermore, for elongated swimmers featuring a puller propulsion mechanism, it was found that isotropic and spatially homogeneous suspensions are stable \cite{saintillan2008instabilities,saintillan2008pattern,saintillan2013active}. In contrast to that, homogeneous isotropic suspensions of elongated pushers are unstable and show an interesting nonlinear dynamics including enhanced mixing \cite{saintillan2008instabilities,saintillan2008pattern,saintillan2013active}.%

When concentrated suspensions or the close proximity of confining boundaries are addressed, hydrodynamic near-field interactions become important. It is then often mandatory to explicitly resolve the induced flow fields in the vicinity of the swimmers. For this purpose, squirmer models are ideal candidates \cite{gotze2010mesoscale,evans2011orientational,alarcon2013spontaneous, zottl2014hydrodynamics,li2014hydrodynamic}. In the case of concentrated suspensions of spherical squirmers in the bulk, net polar orientational ordering of the swimming directions due to hydrodynamic interactions was numerically observed \cite{evans2011orientational,alarcon2013spontaneous}. This is in contrast to the above-mentioned results for dilute swimmer suspensions \cite{saintillan2008instabilities,saintillan2008pattern,saintillan2013active}. For pullers the degree of order was higher than for pushers \cite{evans2011orientational,alarcon2013spontaneous}. Likewise, as indicated above, hydrodynamic near-field interactions between two confining walls determine the relative orientation with respect to the boundaries \cite{li2014hydrodynamic,zottl2014hydrodynamics}. An enhanced cluster formation results for pusher squirmers when compared to pullers \cite{li2014hydrodynamic,zottl2014hydrodynamics}.%

Finally, hydrodynamic interactions can lead to synchronization. This particularly applies when active rotors or driven filaments are considered \cite{reichert2005synchronization,kim2006pumping,vilfan2006hydrodynamic,yang2008cooperation, polin2009chlamydomonas,kotar2010hydrodynamic,uchida2010synchronizationprl,uchida2010synchronization, uchida2011generic,uchida2011many,golestanian2011hydrodynamic,osterman2011finding, reigh2012synchronization,reigh2013synchronization,elgeti2013emergence, theers2013synchronization,theers2014effects}. It was summarized that rotor pairs can only synchronize if the system is not symmetric under exchanging the two rotors \cite{lenz2006collective,golestanian2011hydrodynamic}. Carpets of driven hydrodynamically coordinated active rotors were suggested to be used to pump or mix fluid in microfluidic devices \cite{uchida2010synchronizationprl}. Furthermore, carpets of actively driven deformable filaments are found on the surfaces of certain bacteria and can be used for self-propulsion \cite{brennen1977fluid,vilfan2006hydrodynamic,osterman2011finding}. Their coordinated motion can provide a swimming mechanism for these cells at low Reynolds numbers.

\subsection{Deformable self-propelled particles}

On our way of increasing complexity in active particle systems we now add a further degree of complication. Several self-propelled objects are not rigid. They deform as it is observed for instance for certain cells that crawl on substrates \cite{rappel1999self,maeda2008ordered,kaindl2012spatio}. Particularly obvious examples are self-propelled droplets on interfaces \cite{lee2002chemical,nagai2005mode,chen2009self} or in bulk fluid \cite{thutupalli2011swarming,kitahata2011spontaneous,kitahata2012spontaneous}. These droplets maintain chemical reactions \cite{thutupalli2011swarming,kitahata2011spontaneous,kitahata2012spontaneous} that lead to concentration gradients along their surfaces. A similar mechanism works by depositing chemicals from their inside to the outside \cite{lee2002chemical,nagai2005mode,chen2009self}. In effect, all these processes induce and maintain gradients of surface tension along their surfaces \cite{chen2009self,kitahata2011spontaneous,kitahata2012spontaneous, schmitt2013swimming,yoshinaga2014spontaneous}. These lead to stress gradients along the surfaces. On rigid substrates, such stress gradients on the droplet surface can directly push or pull on the droplet and lead to a net drive. In fluid environments, the surface gradients can induce convective flows on the inside and outside of the droplet surface. These convective flows in turn propel the droplets \cite{kitahata2011spontaneous,schmitt2013swimming,yoshinaga2014spontaneous}. If no special action is taken from the outside, the droplet motion has to start by spontaneous symmetry breaking; for instance by a sufficiently strong concentration fluctuation along its surface. In the ideal case, the concentration fluctuation leading to the net motion is maintained in a self-supported way by the resulting motion \cite{yoshinaga2012drift}, for example when the concentration gradients are enhanced by the induced fluid flows. It was demonstrated that self-propelled droplets can be relatively robust deformable objects. In an experiment, they were observed to deform and squeeze themselves through steric barriers \cite{chen2009self} and recover afterwards. 

\begin{figure}[t]
\centerline{\includegraphics[width=4.cm]{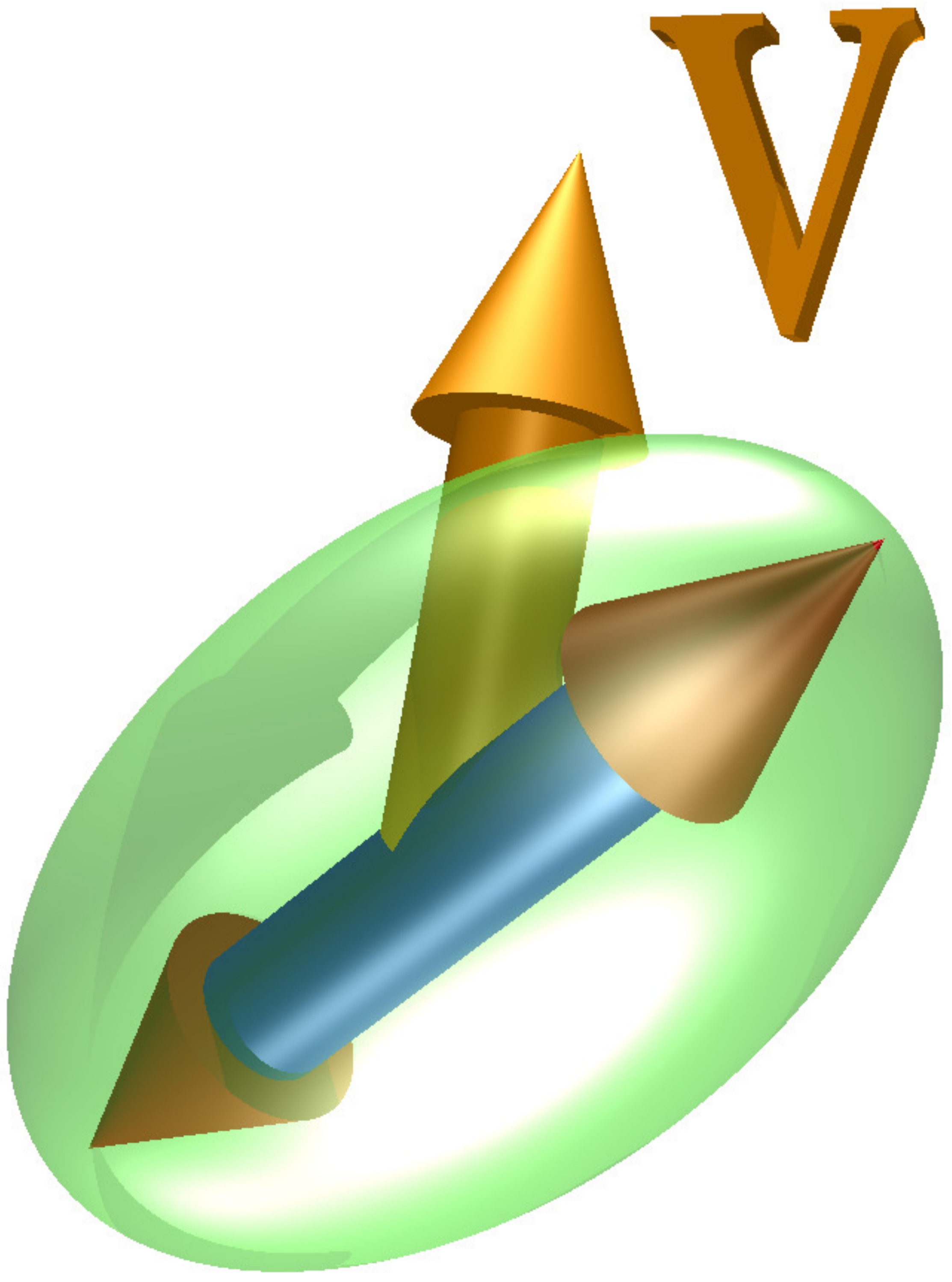}}
\caption{A self-propelled particle that deviates from its spherical ground state by axially symmetric elliptic deformations. The elliptically deformed state is described by the degree of deformation $s$ and by the orientation of the resulting ellipsoid. Since there is no deformational forward-backward asymmetry along the symmetry axis of the ellipsoid, we indicate its orientation by a double-headed arrow referred to as $\mathbf{\hat{n}}$. The self-propulsion contributes to the velocity $\mathbf{v}$ of the particle. Although representing separate degrees of freedom for each particle, the velocity $\mathbf{v}$ and the deformation tensor $\mathbf{S}=s(\mathbf{\hat{n}}\mathbf{\hat{n}}-\mathbf{I}/d)$ are generally coupled to each other, see Eqs.~(\ref{eq:ohtaohkuma1}) and (\ref{eq:ohtaohkuma2}).}
\label{fig_deformable}
\end{figure}

To study the impact of deformability on the particle motion, shape deviations from a spherical undeformed state can be taken into account in a systematic way \cite{hiraiwa2010dynamics}. The lowest order deformations are of axially symmetric elliptic shape. 
Such elliptic deformations can be described by a symmetric traceless tensor $\mathbf{S}$ of second rank \cite{ohta2009deformation}. It is formally equivalent to the order parameter tensor introduced in Eq.~(\ref{eq:nematicOPtensor}) to characterize nematic liquid crystalline phases \cite{degennes1993physics}. Now, however, $s$ parameterizes the degree of deformation, while $\mathbf{\hat{n}}$ represents the orientation of the symmetry axis of the elliptic shape changes, see also Fig.~\ref{fig_deformable}. 
The relation to the nematic order parameter tensor comes from the fact that axially symmetric elliptic deformations are forward-backward symmetric. In Fig.~\ref{fig_deformable} this is indicated by the double-headed arrow that marks the symmetry axis of the ellipsoid. 

Coupled dynamic equations between the elliptic deformation tensor $\mathbf{S}$ and the velocity vector $\mathbf{v}$ of the particle were derived on symmetry grounds and evaluated \cite{ohta2009deformable}. To lowest order in the coupling terms, they read \cite{ohta2009deformable}:
\begin{eqnarray}
\frac{\mbox{d}\mathbf{v}}{\mbox{d} t} &=& \gamma\mathbf{v}-\mathbf{v}^2\mathbf{v} -a\,\mathbf{S}\cdot\mathbf{v}, 
\label{eq:ohtaohkuma1}
\\[.1cm]
\frac{\mbox{d}\mathbf{S}}{\mbox{d} t} &=& -\kappa\mathbf{S}+b\!\left(\mathbf{v}\mathbf{v}-\frac{1}{d}\mathbf{v}^2\,\mathbf{I}\right).
\label{eq:ohtaohkuma2}
\end{eqnarray}
In passive systems, the coefficient $\gamma$ would be negative and the term would correspond to linear viscous friction with the environment. On the contrary, in active systems, this coefficient can become positive, $\gamma>0$. Then, the first two terms on the right-hand side of Eq.~(\ref{eq:ohtaohkuma1}) can lead to a stationary solution of non-vanishing velocity $\mathbf{v}\neq\mathbf{0}$, i.e.\ self-propulsion. An analogous approach can be found in the famous macroscopic continuum characterization of flocks of self-propelled particles by Toner and Tu \cite{toner1995long,toner1998flocks}. The orientation of the resulting non-vanishing velocity vector $\mathbf{v}$ is not fixed a priori and results from spontaneous breaking of the continuous rotational symmetry. In the second equation, i.e.\ in Eq.~(\ref{eq:ohtaohkuma2}), the coefficient $\kappa>0$ expresses that the shape tends to relax back to the undeformed ground state for vanishing velocity $\mathbf{v}$. As before, $\mathbf{v}\mathbf{v}$ represents a dyadic product, $d$ the dimensionality of the system, and $\mathbf{I}$ the unity matrix. Most importantly, the terms with the coefficients $a$ and $b$ include the leading-order coupling terms between the dynamic variables $\mathbf{v}$ and $\mathbf{S}$. These contributions are to a big extent responsible for the peculiar behavior of deformable self-propelled particles as it is briefly summarized in the following. 

In two spatial dimensions, the deformability can induce a bifurcation from straight to circular motion with increasing active drive \cite{ohta2009deformable}. Also quasiperiodic motions as well as array-like trajectories featuring rectangular edges were observed \cite{hiraiwa2010dynamics}. Applying an external electric field to polarizable particles or an external gravitational field, further types of trajectories such as different kinds of zig-zag motions and cycloidal motions appear \cite{tarama2011dynamics}. In the absence of external fields but in three spatial dimensions, additional helical trajectories are obtained \cite{shitara2011deformable,hiraiwa2011dynamics}. 

The collective behavior in two spatial dimensions was studied extensively \cite{ohta2009deformable,ohkuma2010deformable}. First, a global coupling was introduced that forces the deformed particles to align along their globally averaged deformation axis \cite{ohta2009deformable}. This coupling can lead to chaotic motion \cite{ohta2009deformable,ohkuma2010deformable}. After that, instead of the global coupling, a pairwise alignment interaction between deformed particles was imposed \cite{itino2011collective,itino2012dynamics}. In addition to that, as a steric interaction, a soft pairwise repulsive Gaussian potential was applied. Starting from random initial conditions at low or vanishing orientational noise amplitudes, the particles collectively order their migration directions as a function of time as expected \cite{vicsek1995novel}. All particles then collectively propel into the same direction. Moreover, they also tend to locally positionally order in a hexagonal way. Interestingly, it was observed that under compression of the system, implying an increase in the particle density, the ordered collective migration breaks down at a critical density \cite{itino2011collective,itino2012dynamics}. The system becomes disordered and fluid-like. This is just the opposite effect as to that expected from the Vicsek model, in which the orientational order in the motion increases continuously with the particle density \cite{vicsek1995novel}. 
In a later study that used an anisotropic pairwise repulsive Gaussian potential, it was concluded that the disorder transition from a collectively traveling hexagonal crystal to a fluid-like state is enabled by the soft Gaussian interaction potential \cite{menzel2012soft}. The effect is well known in corresponding equilibrium systems. At low densities, these equilibrium systems are in a fluid phase, they crystallize at intermediate densities, but they return to a fluid state at high densities \cite{stillinger1976phase,prestipino2005phase, prestipino2011hexatic,prestipino2007phase}. A central ingredient for this reentrant fluidization is the boundedness of the Gaussian interaction potential, even when two particles are placed at the same position \cite{stillinger1976phase}. It is concluded that the analogous effect is observed in active systems of self-propelled particles interacting sterically by a pairwise Gaussian potential \cite{menzel2012soft} and can be supported by their deformability \cite{itino2011collective,itino2012dynamics,menzel2012soft}. 

\begin{figure}
\centerline{\includegraphics[width=4.8cm]{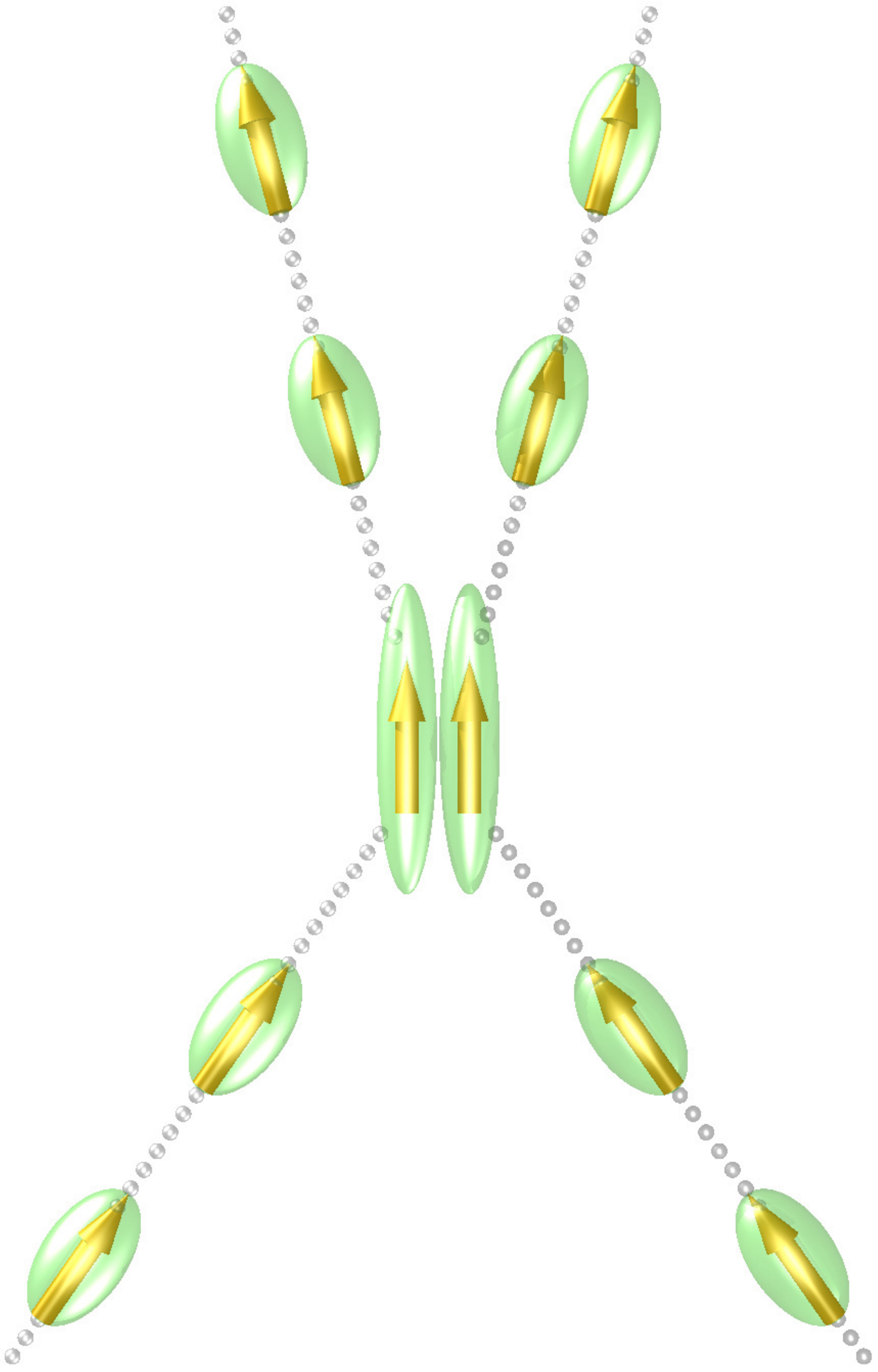}}
\caption{Aligning steric interactions between deformable self-propelled particles \cite{menzel2012soft}. Depicted is a time series from bottom to top with dotted lines indicating the particle paths. Two deformable self-propelled particles migrate towards a common point. At this point, their active drive pushes them together. Due to steric interactions, this leads to deformations. Eqs.~(\ref{eq:ohtaohkuma1}) and (\ref{eq:ohtaohkuma2}) imply that the deformations can reorient the self-propulsion directions, here towards the common axis of elongation of the particles. The outgoing angle between the particles is smaller than the ingoing angle, which means that an effective alignment mechanism is at work.}
\label{fig_deformable_alignment}
\end{figure}

Apart from that, mutual deformations of the particles due to steric interactions between them were considered \cite{menzel2012soft}. This process can provide an implicit alignment mechanism for their self-propulsion directions as illustrated in Fig.~\ref{fig_deformable_alignment} for two particles. In many-particle systems, hexagonal lattices of self-propelled particles featuring a common migration direction can emerge that represent active collectively traveling crystals. Introducing a cut-off for the steric Gaussian interaction potential, traveling crystals of rectangular lattice structure are obtained. Likewise, active resting cluster crystals with more than one particle on each lattice site are observed, as they were previously reported for equilibrium systems \cite{likos1998freezing,schmidt1999an}. In cases of strong particle deformations, laning patterns can appear from randomly initialized systems \cite{menzel2012soft}. Adding orientational noise, the Vicsek transition \cite{vicsek1995novel} from collectively ordered to disordered motion is recovered with increasing noise amplitude \cite{menzel2012soft}. 

Finally, the formation of propagating fronts was reported for two different cases. First, large systems of repulsive Gaussian interactions and reentrant fluidity become inhomogeneous at the reentrance density \cite{itino2012dynamics,tarama2014individual}. They show regions of collectively ordered and regions of disordered motion. In contrast to the traveling density bands observed in the Vicsek model \cite{chate2008collective,bertin2009hydrodynamic,peruani2011polar,menzel2012collective}, here a front of disordered motion propagates into the region of collectively ordered migration. Furthermore, the disordered region has a higher density than the area of ordered collective motion. Second, and in analogy to the results from the Vicsek model \cite{chate2008collective,bertin2009hydrodynamic,peruani2011polar,menzel2012collective}, 
traveling high-density bands of ordered collective motion emerge in systems of density-dependent active drive \cite{yamanaka2014formation}. The active drive in this case is set to increase with density. Such traveling density bands were shown to survive and penetrate through each other under repeated head-on collisions \cite{yamanaka2014formation}. 

Apart from that, the model was complexified in two different directions. On the one hand, the next-higher mode of deformation was included \cite{hiraiwa2010dynamics}. It can be described by a third-rank tensor \cite{ohta2009deformation} that is familiar from the study of liquid crystals of tetrahedratic order \cite{fel1995tetrahedral,brand2005tetrahedratic}. This uneven mode of deformation breaks the forward-backward symmetry of the particle shape. In this case, a zig-zag trajectory and chaotic motions can be obtained already for a single particle in the absence of an external field \cite{hiraiwa2010dynamics}. The description was further extended to include the lowest four modes of deformation \cite{tarama2013oscillatory}. On the other hand, in addition to self-propulsion and deformability, an active rotational (spinning) motion was experimentally observed \cite{nagai2013rotational} and theoretically considered \cite{tarama2012spinning}. 
Different kinds of circular and quasi-periodic orbital trajectories result from this modification \cite{tarama2012spinning}. Furthermore, period-doubling and chaotic behavior is observed in the trajectories and in the state variables such as in the degree of deformation \cite{tarama2012spinning}. Helical types of motion are found in three spatial dimensions, among them also one with a superhelical trajectory \cite{tarama2013dynamics}. 

\begin{figure}
\centerline{\includegraphics[width=8.5cm]{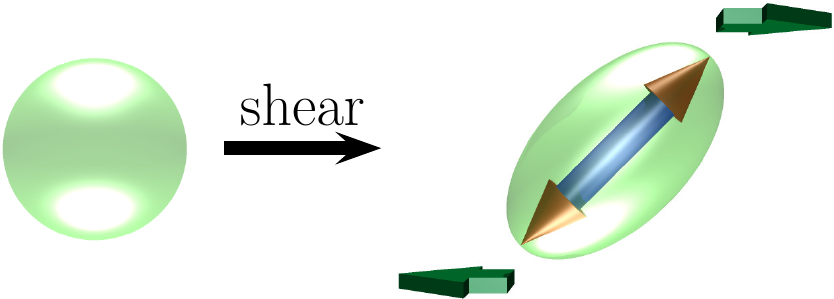}}
\caption{Elongation of a deformable particle by an externally imposed shear flow \cite{tarama2013dynamicsshear}. Initially spherical deformable particles like droplets (left) are deformed due to the elongational component of the flow (right). In addition to that, the rotational component of the shear flow tends to induce further dynamical processes, see Figs.~\ref{fig_lc_shear} and \ref{fig_vesicle_shear}.}
\label{fig_deformable_elongation}
\end{figure}

In a next step, the two-dimensional dynamics of a deformable self-propel\-led particle in a swirl flow \cite{tarama2014deformable} and in a planar linear shear flow were analyzed \cite{tarama2013dynamicsshear}. When exposed to shear, the elongational part of the flow tends to additionally deform the particle as depicted in Fig.~\ref{fig_deformable_elongation}. 
Moreover, in the case of the planar linear shear flow, the rotational contribution of the flow field tends to reorient the particle axes and can complexify the dynamic behavior. 
For this situation, an active straight motion, a winding motion, as well as different types of cycloidal motion were predicted \cite{tarama2013dynamicsshear}. An interesting issue for future studies is to find out, whether these kinds of motion represent the active analogues to the flow-aligned or tank-treading, wagging or swinging, and tumbling modes, respectively, displayed in Figs.~\ref{fig_lc_shear} and \ref{fig_vesicle_shear}. Experimentally, the shear geometry could be realized by placing self-propelling droplets on the surface of a liquid \cite{nagai2005mode,takabatake2011spontaneous} that is confined between two parallel counterpropagating walls. 

The situation gets even more complex, when an additional active rotational spinning of the particle is considered in the shear flow \cite{tarama2013dynamicsshear}. In this case, active spinning with the same sense of rotation as the externally imposed flow field must be distinguished from active spinning with the opposite sense. When the active and the externally imposed rotations counteract and balance each other, a kind of active straight trajectory is recovered. Other types of observed dynamics are periodic, quasi-periodic, different kinds of cycloidal and undulated cycloidal, different kinds of winding, as well as different kinds of chaotic motions. 

To summarize the situation, we can say that the additional degrees of freedom arising from the deformability lead to quite complex dynamics, even for isolated active particles. Combining deformability, active propulsion, active rotation, and the various types of external influences -- in our example the imposed shear flow -- opens the way to a nearly unbounded number of dynamical states. 

\subsection{Collective behavior of animals}

At the end of this review, we now reach the uppermost spring of our ladder of complexity in the field of active systems. 
By construction this is a highly non-trivial topic and we can only touch the surface of it in the present framework. The main question that needs to be answered in this context is how much the characterization of each of the following systems can be simplified. The physicist's goal is naturally to map the animal behavior onto a minimum model that captures the central ingredients but can still be evaluated efficiently. How will this work for living creatures that have their own mind and make their own decisions?  

Already for microorganisms the situation is quite difficult \cite{cates2012diffusive}. For example, reversals or changes in their migration directions are often observed \cite{wada2013bidirectional,tailleur2008statistical,berg1972chemotaxis, macnab1972gradient,polin2009chlamydomonas,min2009high}. The latter can lead to a type of run-and-tumble motion \cite{berg1972chemotaxis,macnab1972gradient,polin2009chlamydomonas,min2009high,bennett2013emergent,cates2013active}. On the one hand, this behavior may result from the complicated propulsion mechanism \cite{bennett2013emergent}. On the other hand, it can be the manifestation of an ``active decision'' \cite{yi2000robust}. If the change in migration direction occurs in response to an external stimulus, the reaction is generally referred to as ``taxis'' or ``tactic'' behavior. For instance, microorganisms can react in a chemotactic way when they detect chemicals that indicate nutrition, or they use chemicals for communication between each other without physical contact \cite{berg1972chemotaxis,macnab1972gradient,cluzel2000ultrasensitive, miller2001quorum,waters2005quorum,an2006quorum,taktikos2012collective}. Further examples concern the reaction to light (phototaxis) \cite{hader1987polarotaxis,suematsu2011localized,garcia2013light}, gravitation (gravitaxis) \cite{yoshimura2003gravitaxis,roberts2006mechanisms}, flow fields (rheotaxis) \cite{marcos2012bacterial}, adhesion gradients (haptotaxis) \cite{kolmakov2010designing,kolmakov2011designing}, and other external stimuli. Nevertheless, aspects of the collective behavior 
can often be described by simple particle descriptions in the form of extended Vicsek approaches \cite{czirok1996formation,szabo2006phase} or by minimum continuum models \cite{aranson2007model,wensink2012meso,dunkel2013fluid,svensek2013collective}. 

Grasshoppers (or locusts) form a class of insects the collective dynamics of which has attracted attention for at least one reason: cannibalism is observed for these insects and seems to influence their individual and collective behavior \cite{simpson2006cannibal,hansen2011cannibalism,bazazi2011nutritional}. On the one hand, an individual tries to rush after grasshoppers ahead of itself. On the other hand, it tries to flee from individuals behind itself. This was demonstrated by restricting the visual field through paintings on part of their eyes \cite{bazazi2008collective}. A restricted ability to detect approaching members in a group reduces motion and promotes cannibalism \cite{bazazi2008collective}, which indicates that cannibalism supports migration. The situation was modeled in a pursuit-escape approach by finite-size random walkers that experience friction with the substrate and social interactions \cite{romanczuk2009collective,romanczuk2012swarming}: other individuals moving away in the front are hunted after, whereas they are fled from if they approach from behind. In particular, the pursuit interaction tends to increase the orientational order in the collective motion. It increases the mean migration speed, which is also the case for the escape interaction at higher densities. A further study analyzed the radial distribution of other grasshoppers around one individual \cite{buhl2012using}. It revealed an isotropic shape in contrast to the anisotropic one obtained from minimum pursuit-escape approaches. The situation might still be more complex. 

Another group of animals repeatedly studied in the context of collective motion are fish. Data can be obtained by video-tracking the individuals of a swarm in artificial water tanks \cite{hensor2005modelling,becco2006experimental,gautrais2009analyzing,herbert2011inferring, katz2011inferring,gautrais2012deciphering,tunstrom2013collective}. It was observed that fish swarms were of smaller size when food sources were detected \cite{hoare2004context} and of larger size in reaction to predator attacks \cite{hoare2004context}. 
The swarms are generally elongated and denser at the front than at the back \cite{katz2011inferring,hemelrijk2012schools}. Increasing the mean density of fish in a container, a transition to cooperative collective motion could be detected at a threshold density \cite{becco2006experimental,cambui2012density} similarly to the predictions of the Vicsek model \cite{vicsek1995novel}. However, the interaction rules deduced for individuals appear to be quite different from those assumed in the Vicsek model: direct orientational alignment seems to be weak \cite{herbert2011inferring,katz2011inferring}, interactions are restricted to the one single nearest neighbor \cite{herbert2011inferring} or at least are not a simple average over all pairwise interactions \cite{katz2011inferring}, and variations of swimming speed are important \cite{herbert2011inferring,katz2011inferring}. 
In a recent study, three modes of collective swimming motion were identified \cite{tunstrom2013collective}: a swarming state of relatively low order, a polarized state, and a milling state of high global rotation. Transitions were frequently observed. 
These forms of collective behavior should be compared to the motion of single isolated fish. In the latter case, relatively continuous trajectories without sharp bends or kinks were observed, corresponding to persistently turning walkers rather than conventional random walks
\cite{gautrais2009analyzing,gautrais2012deciphering}. 

While the swimming motion of small fish can still be effectively confined to a quasi two-dimensional geometry in the lab, the study of fish and bird swarms in nature must typically consider three-dimensional textures. An exception are linear string-like and V-shaped arrangements (skeins) of, e.g., migrating geese, the structure and dynamics of which was recorded and analyzed \cite{hayakawa2010spatiotemporal,hayakawa2012group}. To empirically study real three-dimensional flocks of birds, however, new tools had to be developed to effectively reconstruct the individual positions within the swarm from recorded two-dimensional image data \cite{cavagna2008starflag1,cavagna2008starflag2,cavagna2013diffusion}. 
In contrast to the observations made for fish, starling swarms were denser at their outer boundary than in the center \cite{ballerini2008empirical,cavagna2013diffusion} without a systematic front-back asymmetry \cite{ballerini2008empirical}. A correlation between the density and number of individuals in the swarm could not be identified \cite{ballerini2008empirical}. 
On average, the probability to find a nearest-neighboring bird in the direction of motion was markedly depleted compared to other directions \cite{ballerini2008empirical,ballerini2008interaction}. 
Maybe most importantly, the interactions between the individuals were observed to follow topological rather than metric rules \cite{ballerini2008interaction,bialek2012statistical}: the interactions occur with a fixed number of nearest neighbors instead of all neighbors within a fixed interaction distance. 
Qualitative differences arise from this subtlety, for example the swarms are more cohesive under predator attacks \cite{ballerini2008interaction}. 
Furthermore, in the famous Vicsek model \cite{vicsek1995novel}, a change to non-metric topological interactions renders the order-disorder transition for collective motion continuous, without the emergence of density bands \cite{ginelli2010relevance}. This example demonstrates the importance of intensified recording and comparing to empirical data -- a task that is highly non-trivial for such complex systems as animal swarms.

\newpage
\section{Conclusions}

In the above, we took a tour through various subtopics concerning the physics of soft matter out of its equilibrium ground state. The field of soft matter has grown far too broad for all its aspects to be touched in such a brief account. Only selected issues could be included. We were guided by two central aspects characteristic for soft matter systems: they typically show large responses to external stimuli; and they are easily driven out of equilibrium. The degree of non-equilibrium served to order the different topics, where in each case we considered systems of increasing complexity. 

Static external fields can switch the state of a system from its initial equilibrium ground state to a new static equilibrium state. We considered external electric fields that reorient the director in cells of low-molecular-weight liquid crystals as well as in swollen or prestretched liquid crystalline elastomers. In the latter materials, also mechanical fields can be applied to change the director orientation. Furthermore, the electric fields can induce mechanical deformations in swollen liquid crystalline elastomers, which makes these materials candidates for the use as soft actuators. The same is true for ferrogels or magnetic elastomers when they are exposed to external magnetic fields. Apart from that, the external fields can be used to tune the mechanical properties. For example, the elastic moduli can be reversibly adjusted in a non-invasive way by applying an external field. 

After that, we addressed externally imposed shear flows that lead to steady and dynamic states of motion. In such shear flows, the director of nematic liquid crystals can be observed to ``flow align'', i.e.\ to take a steady inclination angle with respect to the flow within the shear plane; it can oscillate in a ``wagging'' motion; or it can describe continuous full ``tumbling'' rotations. In certain cases, these states can also be observed successively with decreasing shear rate. Orientational effects under shear are further found in bulk-filling, but spatially periodically modulated systems. We considered the example of micro-phase-separated block copolymer melts or solutions. Especially under large-amplitude oscillatory shear, reorientations of the structures with respect to the shear directions occur as a function of the shear rate and amplitude. Furthermore, for localized objects such as vesicles that are composed of closed bilayer membranes a picture similar to the one for the director of nematic liquid crystals emerges under steady shear flow. The steady state of constant inclination angle is called ``tank-treading'', with the bilayer membrane running around the interior of the vesicle; an oscillating inclination angle is referred to as ``swinging'', ``trembling'', or ``vacillating breathing''; and, again, full rotations occur in the ``tumbling'' mode. A non-vanishing viscosity contrast between the liquid on the inside and on the outside of the vesicle is necessary to obtain these states successively with decreasing shear rate. 

Finally, we turned to systems composed of constituents that themselves are considered to be active. They feature a mechanism of self-propulsion and possibly also of active rotations. Our focus was on the collective non-equilibrium states arising from the interactions between these constituents. Examples are ordered collective motion, traveling density bands, laning, resting and traveling crystals, or non-equilibrium clustering. We increased the complexity from point-like particles to finitely-sized objects including steric interactions; from particles propelling with constant velocity to those featuring a constant strength of active drive; from dry systems to those interacting hydrodynamically; from rigid particles to those being deformable; and, eventually, by briefly addressing recent studies on living creatures. Quite remarkably, basic aspects, such as the formation of flocks of individuals that order their migration directions at high enough density, can be found on all levels of complexity. This supports the idea that it is possible to develop basic underlying concepts and systematic methods to describe also such kinds of genuinely non-equilibrium systems. 

As a concluding remark, we would like to append a central argument in favor of this field. Although often emerging from topics of daily life or biology, many of its subareas are relatively young. The boundaries are rapidly being pushed forward, thus continuously providing new interesting problems of research. Far from equilibrium, i.e.\ inherently for active systems, the tools of understanding still need to be established, although often an intuitive picture can straightforwardly be developed. Therefore, especially for young researchers, soft matter physics provides a promising opportunity in a meanwhile well-recognized environment.

\newpage
\section{Acknowledgments}
In the first place, the author thanks Hartmut L\"owen for his continuous support. 
Apart from him, the author thanks 
Takeaki Araki, G\"unter Auernhammer, Stefan Bohlius, Alexander B\"oker, Dmitry Borin, Helmut Brand, Peet Cremer, Burkhard D\"unweg, Heino Finkelmann, Paul Goldbart, Nigel Goldenfeld, Steve Granick, Hisao Hayakawa, Marco Heinen, Sascha Hilgenfeldt, Christian Holm, J\"urgen Horbach, Thomas Ihle, Shoichi Kai, Andreas Kaiser, Toshihiro Kawakatsu, Ken Nagai, Stefan Odenbach, Takao Ohta, Yoshitsugu Oono, Harald Pleiner, Miha Ravnik, Eric Roeben, Lisa Roeder, Takahiro Sakaue, Masaki Sano, Annette Schmidt, Matthias Schmidt, Michael Schmie\-de\-berg, Maksim Sipos, Thomas Speck, Torsten St\"uhn, Daniel Sven\v{s}ek, Yuka Tabe, Kazumasa Take\-uchi, Mitsusuke Tarama, Nariya Uchida, Kenji Urayama, Axel Voigt, Hirofumi Wada, Bo Wang, Rudolf Weeber, Rik Wensink, Fangfu Ye, Natsuhiko Yoshinaga, and many others 
for scientific discussions and interactions. 
Over the past years, support leading to the above presentation was received from the Deutsche Forschungsgemeinschaft through the FOR 608 ``Nonlinear dynamics of complex continua'', a DFG research fellowship, the Japanese-German core-to-core collaboration program ``Non-equilibrium phenomena in Soft Matter'', the SPP 1681 ``Feldgesteuerte Partikel-Matrix-Wechselwirkungen: Erzeugung, skalen\"ubergreifende Mo\-del\-lie\-rung und Anwendung magnetischer Hybridmaterialien'', and the SPP 1726 ``Microswimmers -- from single particle motion to collective behavior''.


\end{document}